\documentclass{aa}  

\newcommand{\fer}{{\it Fermi}}
\newcommand{\mywidth}{7.0cm}

\usepackage{graphicx}
\usepackage{longtable}
\usepackage{mdwlist}
\usepackage{lscape}
\usepackage{natbib}
\usepackage{multirow}
\usepackage{upgreek}
\usepackage[colorlinks=true, pdfstartview=FitV, linkcolor=blue,citecolor=blue, urlcolor=blue]{hyperref}
\usepackage{wasysym}
\usepackage{txfonts}
\usepackage{gensymb}
\usepackage{morefloats}
\usepackage{threeparttable}
\usepackage{rotating}
\usepackage{graphicx}

\begin{document} 
\title{Optical spectroscopic observations \\ of low-energy counterparts of \fer-LAT $\gamma$-ray sources}

\titlerunning{Optical counterparts of $\gamma$-ray sources}
\authorrunning{H. A. Pe\~na-Herazo et al.}

\author{H. A. Pe\~na-Herazo\inst{1,2,3,4}  \and
          R. A. Amaya-Almaz{\'a}n\inst{2} \and
          F. Massaro\inst{1,3,4,5} \and
          R. de Menezes \inst{1,6} \and 
          E. J. Marchesini\inst{1,4,7,8,9} \and
          V. Chavushyan\inst{2} \and
          A. Paggi\inst{4,5} \and
          M. Landoni\inst{10,11} \and
          N. Masetti\inst{9,12} \and
          F. Ricci\inst{13} \and
          R. D'Abrusco\inst{14} \and
          C.C. Cheung\inst{15} \and
          F. La Franca\inst{16} \and
          H. A. Smith\inst{14} \and
          D. Milisavljevic\inst{17} \and
          E. Jim{\'e}nez-Bail{\'o}n\inst{18} \and
          V. M. Pati\~no-\'Alvarez\inst{2,19} \and
          G. Tosti\inst{20} 
       }

\institute{
             Dipartimento di Fisica, Universit\`a degli Studi di Torino, via Pietro Giuria 1, I-10125 Torino, Italy.  
             \and
             Instituto Nacional de Astrof\'{i}sica, \'Optica y Electr\'onica, Apartado Postal 51-216, 72000 Puebla, M\'exico.  
             \and
             Istituto Nazionale di Fisica Nucleare, Sezione di Torino, I-10125 Torino, Italy.         
             \and
             INAF-Osservatorio Astrofisico di Torino, via Osservatorio 20, 10025 Pino Torinese, Italy.  
             \and
             Consorzio Interuniversitario per la Fisica Spaziale (CIFS), via Pietro Giuria 1, I-10125, Torino, Italy.   \and
             Universidade de S\~{a}o Paulo, Departamento de Astronomia, S\~{a}o Paulo, SP 05508-090, Brazil.
             \and
             Facultad de Ciencias Astron\'{o}micas y Geof\'{\i}sicas, Universidad Nacional de La Plata, La Plata, Argentina.         
             \and
             Instituto de Astrof\'{\i}sica de La Plata, CONICET-UNLP, CCT La Plata, La Plata, Argentina.         
             \and
             INAF-Osservatorio di Astrofisica e Scienza dello Spazio, via Gobetti 93/3, I-40129, Bologna, Italy.         
             \and
             INAF-Osservatorio Astronomico di Cagliari, via della Scienza , 5, Selargius CA
             \and
             INAF-Osservatorio Astronomico di Brera, Via Emilio Bianchi 46, I-23807 Merate, Italy.         
             \and
             Departamento de Ciencias F\'isicas, Universidad Andr\'es Bello, Fern\'andez Concha 700, Las Condes, Santiago, Chile.         
             \and
             Instituto de Astrof\'{\i}sica and Centro de Astroingenier\'{\i}a, Facultad de F\'{\i}sica, Pontificia Universidad Catolica de Chile, Casilla 306, Santiago 22, Chile.         
             \and
             Center for Astrophysics | Harvard \& Smithsonian, 60 Garden Street, Cambridge, MA 02138, USA.
             \and
             Naval Research Laboratory, Space Science Division, Code 7650, Washington, DC 20375, USA.         
             \and
             Dipartimento di Matematica e Fisica, Universit{\`a} degli Studi Roma Tre, Via della Vasca Navale 84, I-00146, Roma, Italy.         
             \and
             Department of Physics and Astronomy, Purdue University, 525 Northwestern Avenue, West Lafayette, IN 47907, USA.    
             \and
             Instituto de Astronom{\'i}a, Universidad Nacional Aut{\'o}noma de M{\'e}xico, Apdo. Postal 877, Ensenada, 22800 Baja California, M{\'e}xico.
             \and
             Max-Planck-Institut f{\"u}r Radioastronomie, Auf dem H{\"u}gel 69, 53121 Bonn, Germany.
             \and
             Dipartimento di Fisica, Universit{\`a} degli Studi di Perugia, 06123 Perugia, Italy.
             }
             
\date{Received  ... ; accepted  ... }

 \abstract
{A significant fraction of all $\gamma$-ray sources detected by the Large Area Telescope aboard the \fer\ satellite is still lacking a low-energy counterpart. In addition, there is still a large population of $\gamma$-ray sources  with associated low-energy counterparts that lack firm classifications. In the last 10 years we have undertaken an optical spectroscopic campaign to address the problem of unassociated/unidentified $\gamma$-ray sources (UGSs), mainly devoted to observing blazars and blazar candidates because they are the largest population of $\gamma$-ray sources associated to date.}
{Here we describe the overall impact of our optical spectroscopic campaign on sources associated in \fer-LAT catalogs, coupled with objects found in the literature. In the literature search, we kept track of efforts by different teams that presented optical spectra of counterparts or potential counterparts of \fer-LAT catalog sources. Our summary includes an analysis of an additional 30 newly-collected optical spectra of counterparts or potential counterparts of \fer-LAT sources of previously unknown nature.
}
{New spectra were acquired at the Blanco 4-m and OAN-SPM 2.1-m telescopes, and those available in the Sloan Digital Sky Survey (data release 15) archive. }
{All new sources with optical spectra analyzed here are classified as blazars. Thanks to our campaign, we altogether discovered and classified 394 targets with an additional 123 objects collected from a literature search. 
We began our optical spectroscopic campaign between the release of the second and third \fer-LAT source catalogs (2FGL and 3FGL, respectively), and classified about 25\% of the sources with uncertain nature and discovered a blazar-like potential counterpart for $\sim$10\% of UGSs listed therein.  
In the 4FGL catalog, about 350 \fer-LAT sources are classified to date thanks to our campaign.
}
{The most elusive class of blazars are found to be BL Lacs since the largest fraction of \fer-LAT sources targeted in our observations showed a featureless optical spectrum. The same conclusion applied to the literature spectra. 
Finally, we confirm the high reliability of mid-IR color-based methods to select blazar-like candidate counterparts of unassociated/unidentified $\gamma$-ray sources.
}


%\keywords{Galaxies: Active -- BL Lacertae objects: general -- quasars: emission lines -- galaxies: luminosity function}

\maketitle

\section{Introduction}
\label{sec:intro}
With the launch in 2008 of the \fer\ Large Area telescope (\fer-LAT), a new era in $\gamma$-ray astronomy began \citep{atwood09}.
One of the most challenging key scientific objectives of the \fer-LAT mission, highlighted prior to launch, is to determine ``the type of object(s) and the mechanisms for gamma-ray emission from the unidentified gamma-ray sources'' \footnote{\url{https://fermi.gsfc.nasa.gov/science/resources/aosrd/}}.

Amongst all associated $\gamma$-ray sources listed in the \fer-LAT catalogs \citep[1FGL, 2FGL, 3FGL and 4FGL,][respectively]{abdo10a,nolan12,acero15,abdollahi20}, blazars constitute the largest known population \citep[see also][]{mirabal09,hartman99}.
In the most recent \fer-LAT catalog (i.e., the 4FGL), based on eight years of \fer-LAT survey data, they form the vast majority of all associated sources ($\sim$93\%, 3135 out of 3370 total). 
This statistic considered sources known as Blazars of Uncertain type (BCUs), which are blazars lacking a spectroscopic confirmation, and constituted more than 25\% of the entire 4FGL. 

Blazars are a subclass of radio-loud active galactic nuclei (AGNs), whose emission is interpreted as due to relativistic particles accelerated in a collimated jet aligned within a few degrees to our line of sight \citep{blandford78}. There are  two main classes of blazars: BL Lac objects and flat spectrum radio quasars (FSRQs). The {former} is characterized by an almost featureless optical spectrum, or present only weak {emission/absorption lines (with equivalent width, EW, less than 5 \AA)} mainly due to their host galaxy \citep{stickel91,stocke91}, making {redshift ($z$) estimates challenging \citep[see e.g.,][]{landoni14,landoni15b}. }
On the other hand, {the optical spectra of FSRQs appear as typical quasars \citep{stickel91}, but additionally exhibit flat radio spectra and highly polarized emission from radio to optical frequencies} \citep[see. e.g.,][]{healey07,hovatta12}. 

Recent optical spectroscopic campaigns have found that the largest fraction of {both} potential low-energy counterparts of {unidentified/unassociated $\gamma$-ray sources (UGSs) and classified BCUs are identified as} BL Lacs \citep[see e.g.,][]{landoni15,massaro16,klindt17,marchesi18,desai19,paiano19}. 
These results strongly indicate that BL Lacs are the most elusive counterparts of $\gamma$-ray sources with respect to other extragalactic classes.

Discovering more BL Lacs among UGSs has immediate scientific return both in building their $\gamma$-ray luminosity function \citep[see e.g.,][]{ajello14}, necessary to achieve a better understanding of the extragalactic $\gamma$-ray background \citep{ajello15}, and searching for signatures {of attenuation in their $\gamma$-ray spectra by extragalactic background light \citep{dominguez11,ackermann12b,sandrinelli13}}. The identification of UGSs is also useful to select potential targets for future observations with the Cherenkov Telescope Array \citep{tev,arsioli15}; to obtain more stringent limits {on dark} matter annihilation in sub-halos \citep{zechlin12,berlin14}; to test new $\gamma$-ray detection algorithms \citep{abdollahi20,kerr19}, {and to} search for new $\gamma$-ray source classes \citep{massaro17,bruni18}.

{  Over the past} decade many approaches have been used to search for UGS counterparts. These methods are mainly based {on radio observations at low- \citep[i.e., at or below $\sim1$ GHz;][]{ugs3,ugs6,giroletti16,mooney19} and high-frequency} \citep[i.e., above 1 GHz;][]{hovatta12,petrov13,schinzel15}, and/or {WISE \citep{wright10}} infrared (IR) data \citep{massaro11,dabrusco12}, and/or X-ray follow-up campaigns \citep{masetti10,paggi13,stroh13,acero13,landi15,paiano17a}, as well as statistical algorithms \citep{ackermann12a,doert14,salvetti17}, or optical polarimetry \citep{mandarakas19,liodakis19}. 
Amongst these, the most powerful tools are radio {follow-up} observations \citep{petrov13,ugs6,schinzel15,giroletti16,schinzel17} and statistical analysis of mid-IR colors \citep{dabrusco19}. 
However, to ultimately confirm the blazar nature of the low energy counterparts proposed in these studies, optical spectroscopic observations are strictly necessary \citep[see e.g.,][for a recent review]{massaro16}. 

Thus since 2014, we started a spectroscopic follow-up campaign, mainly based on observations carried out at 4 {-}m class ground-based telescopes, to (i) confirm the blazar-like nature of candidate UGS counterparts selected on the basis of their mid-IR colors \citep{massaro11}; (ii) verify if BCUs were blazars and classify them according to the possible presence {(or lack) of features} in their optical spectra, and (iii) observe any BL Lac whose redshift was still uncertain, aiming to observe them in a {low-flux} (i.e. quiescent) state \citep[see e.g.,][]{crespo16a}.

In the current paper, we describe the overall impact of our optical campaign{  , carried out since 2014, highlighting the fraction of new associations and identifications obtained in each of the \fer-LAT catalog releases. In our summary, we include the results of 30 newly-acquired optical spectra of BCUs and potential low-energy counterparts of UGSs. We obtained the majority of the new spectra (22/30) in 2018 and 2019 using the} Blanco and OAN-SPM telescopes, while we collected the remaining from an archival {search of the} Sloan Digital Sky Survey (SDSS) Data Release 15 \citep{aguado19}.

The manuscript is organized as {follows. In} Section~\ref{sec:class}, we briefly review the blazar classifications adopted during our campaign in comparison with {those} used in the \fer-LAT catalogs. Then in Section~\ref{sec:newspec}, we present the results achieved {for} the 30 new optical spectra, {with all the details on the data reduction and analysis, along with the images given in} Appendix~\ref{sec:appA}. We discuss the overall impact of our optical spectroscopic campaigns on sources classified and associated in the \fer-LAT catalogs in {Section~\ref{sec:impact} (summary tables provided in Appendix~\ref{sec:appB}), including {all the newly-derived results from the} spectra presented here.} In Section~\ref{sec:reliability}, we compare the mid-IR based classifications, mostly used to select targets during our spectroscopic campaign, with {those determined by our optical spectroscopic} observations. Finally, we {present our summary, conclusions, and future perspectives in Section~\ref{sec:summary}}. 

%We use \textit{cgs} units unless otherwise stated. 
%{  We adopt a} $\Lambda$CDM cosmology with $\Omega_{\rm m} = 0.286$, and Hubble constant $H_{0} = 69.6$ km s$^{-1}$Mpc$^{-1}$ \citep{bennett14}.
%It is worth highlighting that the entire discussion in the paper refers to the 4FGL version as available in the published paper, thus based on the first eight years of \fer-LAT observations in the 4FGL catalog \citep{abdollahi20}, even when not explicitly specified. 
{  Given our usage of a large number of acronyms and abbreviations,} mostly due to different classifications and telescopes {involved, we provide a summary} in Table~\ref{tab:acronyms}.
\begin{center}
\begin{table*}[!th] 
\label{tab:acronyms}
\tiny
\caption{Acronyms and Abbreviations used in the text.}
\begin{tabular}{ll}
\hline
Acronym {or} Abbreviation & Description \\
\noalign{\smallskip}
\hline 
\noalign{\smallskip}
AGN & Active Galactic Nuclei \\
AGU & AGN of Uncertain type \\
BCU & Blazar of Uncertain type \\
BLL & BL Lac object \\
FSRQ &Flat Spectrum Radio Quasar \\
UGS & Unidentified/Unassociated $\gamma$-ray Source \\
\noalign{\smallskip}
\hline
\noalign{\smallskip}
BZB & Roma-BZCAT label for BL Lacs \\
BZQ & Roma-BZCAT label for Flat Spectrum Radio Quasars \\
BZG & Roma-BZCAT label for blazars with optical spectra dominated by their host galaxy \\
\noalign{\smallskip}
\hline
\noalign{\smallskip}
1FGL & \fer-LAT First Source Catalog \\
2FGL & \fer-LAT Second Source Catalog \\
3FGL & \fer-LAT Third Source Catalog \\
4FGL & \fer-LAT Fourth Source Catalog \\
FL8Y & Preliminary \fer-LAT 8-year Point Source List \\
\noalign{\smallskip}
\hline
\noalign{\smallskip}
Roma-BZCAT & Roma-BZCAT multifrequency catalogue of Blazars \\
KDEBLLACS & Catalog of KDE-selected candidate BL Lacs; {Kernel Density Estimation (KDE)}\\
WIBRaLS & WISE Blazar-like Radio-Loud Sources \\
\noalign{\smallskip}
\hline
\noalign{\smallskip}
Blanco & Victor Blanco 4-m Telescope \\
Copernico & Copernico 182 cm Telescope \\
DSS & Digital Sky Survey \\
Fermi-LAT & Fermi Large Area Telescope \\
HET & Hobby-Eberly Telescope \\
Keck & W. M. Keck Observatory \\
KPNO & Kitt Peak National Observatory \\
GemN & Gemini North Observatory \\
Magellan & Magellan Telescope \\
MMT & {MMT Observatory at Mount Hopkins} \\
NOT &Nordic Optical Telescope \\
NTT & New Technology Telescope \\
OAGH & Guillermo Haro Astrophysical Observatory \\
OAN-SPM & Observatorio Astronomico Nacional San Pedro M{\'a}rtir \\
Palomar & Hale 200 inch Telescope at Palomar \\
SALT &  Southern African Large Telescope \\
SDSS & Sloan Digital Sky Survey \\
SOAR & Southern Astrophysical Research Telescope \\
TNG & Telescopio Nazionale Galileo \\
WISE & Wide-field Infrared Survey Explorer \\
6dF & The Six-degree Field Galaxy Survey \\
%\noalign{\smallskip}
%\hline
%\noalign{\smallskip}
%KDE & Kernel Density Estimation \\
%SNR & Signal-to-Noise Ratio \\
%EW & Equivalent Width \\
\noalign{\smallskip}
\hline
\end{tabular}
\end{table*}
\end{center}

\section{Blazar classifications}
\label{sec:class}
Since we are comparing \fer-LAT catalogs with sources that could be included in future releases of the Roma-BZCAT \citep{massaro09,massaro15b}, it is necessary to describe how sources are classified in these catalogs and the correspondences between the two. 

{  For the} results of our spectroscopic campaign we adopted the Roma-BZCAT {nomenclature, where sources are mainly distinguished between three types:}
\begin{enumerate}
\item BL Lac objects, labelled as BZBs, are those sources with a featureless optical spectrum, or {show} a blue continuum with absorption lines due to the host {galaxy, or weak} and narrow emission lines with equivalent width less than {5 \AA;}
\item Flat spectrum radio quasars, labelled as BZQs, {are characterized by optical spectra} showing broad emission lines and 
{  non-thermal properties at other wavelengths (flat radio spectra indicating an optically-thick jet and highly polarized emission from radio to optical);}
\item BL Lac {galaxy-dominated} sources (BZGs), {are} sources usually reported as BL Lac objects in the literature, but {their spectral energy distributions are} dominated by the host galaxy emission overwhelming the nuclear one \citep{massaro12}.  {All BZGs have a radio counterpart that, coupled with the blue excess in their optical spectra, indicate the likely presence of a jet in their nuclei, thus potentially being blazar-like sources.} {BZGs are not all expected to be genuine BZBs observed while their nuclei are in a quiescent state. Instead, some BZGs could be moderately bright AGNs, such as radio galaxies, with non-thermal emission that is not highly relativistically beamed.
}
\end{enumerate} 
In addition some objects observed during our spectroscopic campaign were simply indicated as ``quasars'', labeled as QSO, {because} the lack of additional multifrequency observations {(e.g., presence of a flat radio spectrum), prevent us from firmly establishing} their blazar nature. All sources classified as BZQs, {BZGs, or labelled as QSOs have a firm redshift measurement, while this is case-dependent in BZBs given their weak spectral features.}

{  The \fer-LAT catalogs have used a related set of labels that have evolved with their different releases, and the similarities and differences with our adopted Roma-BZCAT nomenclature are described here.}

In the latest 4FGL release, BL Lac objects are indicated as {BLLs, and} flat spectrum radio quasar as FSRQ. Then the \fer-LAT catalogs have a class {named Blazar of Uncertain type (BCU) for sources that} appear to share some blazar properties but lack an optical classification that confirms their nature. Additional sources are then indicated simply as {AGN in cases where their blazar-like nature are not formally established due to insufficient multifrequency observations. The 4FGL nomenclature was also adopted in the previous 3FGL.}
 
However, it is worth noting that in the {earlier 1FGL and 2FGL catalogs, $\gamma$-ray sources} associated with a known BL Lac or a known FSRQ were indicated as BZB and BZQ, respectively, even if they were not part of Roma-BZCAT. {These classifications were removed} in the subsequent versions of the \fer-LAT catalogs {because of the confusion generated. Importantly,} this implies that not all BZBs listed in both 1FGL and 2FGL respect the same classification criteria adopted in the Roma-BZCAT {that we} used during our spectroscopic campaign. This explains why {the first two lines in Table~\ref{tab:optcamp1} report the} BZB/BLL label for the \fer-LAT class of each catalog. 

{  Thus all Roma-BZCAT sources classified as BZBs and BZQs are indicated, respectively, as BLLs and FSRQs in the \fer-LAT catalog releases from the 3FGL onward. }
Additionally, the BCU classification was adopted starting with the 3FGL in place of AGNs of uncertain type (AGUs) used in both 1FGL and 2FGL; the class of AGN remained unchanged through all the data releases. All details about the nomenclature adopted in {the} different \fer-LAT catalogs can be found therein. A summary of the various classes used in the four Fermi-LAT data releases can be found in Table \ref{tab:optcamp3}.

\section{New optical spectra: sample selection and classification results}
\label{sec:newspec}
The strategy adopted during our optical spectroscopic campaign consists of observing small samples of targets {in} each run to minimize the impact on telescope schedules and mainly driven by visibility constraints. 

The new spectra presented here {were} collected for targets selected from the following lists: 
\begin{enumerate} 
\item BCUs {already assigned as counterparts to 4FGL catalog sources, but} whose blazar nature {were} still uncertain; 
\item radio and X-ray sources located within the $\gamma$-ray positional uncertainty of UGSs \citep{marchesini20}; 
\item BL Lacs {whose optical spectrum are} not available in the literature, or {do} not have a redshift estimate; 
\item UGSs having a WISE source with blazar-like mid-IR colors lying within {their} positional uncertainty region, most of them being part of the WISE Blazar-like Radio-Loud Sources (WIBRaLS) and KDEBLLACS catalogs \citep[][]{dabrusco19}. 
\end{enumerate} 
 In particular, the WIBRaLS {catalog sources have radio counterparts and were} selected to {all have} mid-IR colors similar to those of \fer-LAT detected {blazars, while the KDEBLLACS catalog includes} only BL Lac candidates selected using the Kernel Density Estimation (KDE) technique in the WISE W2-W3 $vs$ W1-W2 color-color diagram.

The current sample of 30 new targets includes:
\begin{itemize} 
\item Nineteen sources classified as BCUs in the 4FGL;
\item {Three} targets classified as UGSs in the preliminary version of the 4FGL (i.e., the FL8Y), all having a WISE blazar-like source lying within their $\gamma$-ray positional uncertainty region.
\item Seven BL Lacs and one FSRQ, {with their classifications provided by us to the \fer-LAT team during the preparation of the 4FGL catalog, thus appearing as such in the published 4FGL}. Our sample also includes 4FGL J1704.5-0527 and 4FGL J2115.2+1218, two known BL Lacs in the literature for which other groups published their spectra while our data analysis was in progress.
\end{itemize}

All our newly-observed BCUs are now classified as BZBs, four of them {resulting in redshift measurements}, with the only exceptions of 4FGL J1640.9+1143 and 4FGL J1858.3+4321 that {now appear} to be BZGs. The {three} UGSs analyzed here all have a BZB lying within their $\gamma$-ray positional uncertainty region, {with} the one potentially associated with 4FGL J1637.5+3005 {at $z=0.0786$.} Then all BL Lacs were confirmed as BZBs, two with redshift estimates {(4FGL J1035.6+4409 at $z=0.4438$ and 4FGL J1814.0+3828 at $z=0.2754$), while the one (FSRQ 4FGL J1459.5+1527) is a BZQ at $z=$0.3711.}

Table~\ref{tab:sample} reports all parameters and observational details about our selected sources.
%as: (1) 4FGL names, (2) \fer-LAT class as listed in 4FGL, (3) assigned counterpart in the 4FGL with the exceptions of the UGSs, (4) pointed WISE counterpart, (5) right ascension and (6) declination of the WISE source, (7) telescope used to carry out the spectroscopic observation or queried survey, (9) observing dates and (10) exposure times.
 Then in Table~\ref{tab:analysis} we show all results achieved {for our source sample, including emission/absorption lines detected, and resultant classifications; the figures with the new spectra are presented} in Appendix~\ref{sec:appA}. {Note all targets classified as BZBs with undetermined redshifts are listed in Table~\ref{tab:analysis} for completeness, but without entries for their spectral parameters.}
\begin{table*}[!h] 
\tiny
\caption{Summary of the new optical spectra analyzed and presented here.}\label{tab:sample}
\begin{center}
\begin{tabular}{lllllllcc}
\hline
4FGL & 4FGL  & 4FGL        & WISE & R.A.    & Dec.    & Facility & Obs. Date & Exposure \\
name & class & counterpart & name & (J2000) & (J2000) &          & {dd/mm/yyyy}  & (sec)    \\
 {(1)} &  {(2)}  &  {(3) }  &  {(4)} &  {(5)}    &  {(6) }   &  {(7) }&  {(8)} &  {(9)} \\
\noalign{\smallskip}
\hline 
\noalign{\smallskip}
J0836.9+5833 & bcu  & NVSS J083705+583151          & J083706.00+583152.9 & 08:37:06.01 & +58:31:52.9 & SDSS DR15 & \dots & \dots    \\
J0914.8+5846 & bll    & SDSS J091451.54+584438.1  & J091451.57+584438.1 & 09:14:51.58 & +58:44:38.1 & SDSS DR15 & \dots & \dots    \\
J1019.3+5625 & ugs  &     \dots                                      & J101919.15+562428.8 & 10:19:19.16 & +56:24:28.8 & SDSS DR15 & \dots  & \dots    \\
J1035.6+4409 & bll    & 7C 1032+4424                         & J103532.12+440931.4 & 10:35:32.13 & +44:09:31.5 & SDSS DR15 & \dots & \dots   \\
J1226.0+5622 & bll    & SDSS J122602.81+562254.6  & J122602.82+562254.9 & 12:26:02.83 & +56:22:54.9 & SDSS DR15 & \dots & \dots    \\
J1238.1-4541 & bcu  & PMN J1238-4541                     & J123806.03-454129.6 & 12:38:06.04 & -45:41:29.6 & Blanco    & 12/06/2019 & 2x600 \\
J1403.4+4319 & bll    & NVSS J140319+432018          & J140319.46+432020.1 & 14:03:19.46 & +43:20:20.1 & SDSS DR15 & \dots  & \dots    \\
J1454.7+5237 & bll    & 87GB 145311.3+524904         & J145445.32+523655.4 & 14:54:45.32 & +52:36:55.5 & SDSS DR15 & \dots  & \dots   \\
J1459.5+1527 & fsrq  & MG1 J145921+1526               & J145922.16+152654.9 & 14:59:22.17 & +15:26:54.9 & OAN-SPM       & 03/07/2019 & 3x600 \\
J1545.0-6642 & ugs   &       \dots                                   & J154458.88-664146.9 & 15:44:58.89 & -66:41:46.9 & Blanco    & 12/06/2019 & 3x600 \\
J1545.8-2336 & bcu   & J1545-2339                             & J154546.58-233928.4 & 15:45:46.59 & -23:39:28.4 & Blanco    & 12/06/2019 & 2x600 \\
J1600.3-5811 & bcu   & MRC 1556-580                        & J160012.36-581102.8 & 16:00:12.37 & -58:11:02.8 & Blanco    & 12/06/2019 & 3x400 \\
J1637.5+3005 & ugs  &       \dots                                   & J163738.24+300506.4 & 16:37:38.24 & +30:05:06.5 & OAN-SPM       & 04/07/2019 & 3x900 \\
J1640.9+1143 & bcu  & TXS 1638+118                         & J164058.89+114404.2 & 16:40:58.90 & +11:44:04.2 & OAN-SPM       & 05/07/2019 & 2x1200\\
J1647.1+6149 & bcu  & RX J1647.3+6153                   & J164723.42+615347.5 & 16:47:23.42 & +61:53:47.6 & OAN-SPM       & 03/07/2019 & 3x900 \\
J1704.5-0527 & bll     & NVSS J170433-052839           & J170433.83-052840.7 & 17:04:33.84 & -05:28:40.8 & Blanco    & 12/06/2019 & 2x1200\\
J1705.4+5436 & bcu  & NVSS J170520+543700          & J170520.54+543659.8 & 17:05:20.55 & +54:36:59.8 & OAN-SPM       & 06/07/2019 & 3x900 \\
J1706.8+3004 & bcu  & 87GB 170454.3+300758         & J170650.43+300412.5 & 17:06:50.44 & +30:04:12.6 & SDSS DR15 & \dots & \dots   \\
J1744.4+1851 & bcu  & 1RXS J174420.1+185215        & J174419.81+185217.9 & 17:44:19.81 & +18:52:18.0 & OAN-SPM       & 07/07/2019 & 3x900 \\
J1810.7+5335 & bcu  & 2MASS J18103800+5335016  & J181037.98+533501.5 & 18:10:37.99 & +53:35:01.5 & OAN-SPM       & 07/07/2019 & 3x900 \\
J1814.0+3828 & bcu  & 2MASS J18140339+3828107  & J181403.43+382810.1 & 18:14:03.44 & +38:28:10.2 & OAN-SPM       & 06/07/2019 & 3x900 \\
J1838.4-6023 & bcu   & 2MASS J18382063-6025224  & J183820.63-602522.6 & 18:38:20.64 & -60:25:22.6 & Blanco    & 12/06/2019 & 3x300 \\
J1858.3+4321 & bcu  & NVSS J185813+432452          & J185813.43+432451.9 & 18:58:13.43 & +43:24:51.9 & OAN-SPM       & 04/08/2018 & 3x1200\\
J1929.4+6146 & bcu  & TXS 1928+616                          & J192935.09+614629.4 & 19:29:35.10 & +61:46:29.4 & OAN-SPM       & 07/07/2019 & 3x1200\\
J2043.7+0000 & bcu  & 2MASS J20435020+0001280  & J204350.15+000127.8 & 20:43:50.16 & +00:01:27.9 & OAN-SPM       & 02/07/2019 & 3x1200\\
J2046.8-4258 & bcu   & MRSS 285-029065                   & J204644.01-425713.2 & 20:46:44.01 & -42:57:13.2 & Blanco    & 12/06/2019 & 2x400 \\
J2115.2+1218 & bcu  & NVSS J211522+121802           & J211522.00+121802.6 & 21:15:22.00 & +12:18:02.7 & OAN-SPM       & 04/07/2019 & 3x900 \\
J2141.4+2947 & bcu  & 87GB 213913.0+293303          & J214123.89+294706.2 & 21:41:23.90 & +29:47:06.2 & OAN-SPM       & 05/07/2019 & 3x900 \\
J2208.2+0350 & bll    & SDSS J220812.70+035304.6   & J220812.70+035304.5 & 22:08:12.70 & +03:53:04.5 & OAN-SPM       & 06/07/2019 & 3x900 \\
J2235.3+1818 & bcu  & 2MASS J22352860+1816356  & J223528.60+181635.5 & 22:35:28.61 & +18:16:35.6 & OAN-SPM       & 07/07/2019 & 3x900 \\
\noalign{\smallskip}
\hline
\end{tabular}\\
\end{center}
Note: columns (1) 4FGL name; (2) \fer-LAT class as listed in 4FGL; (3) assigned counterpart in the 4FGL {(none assigned for UGSs)}; (4) {targeted} WISE counterpart; (5) right ascension and (6) declination of the WISE target; (7) telescope used to carry out the spectroscopic observation or queried survey; (8) observing dates and (9) exposure times in seconds.
\end{table*}

\begin{table*}[!h] 
\tiny
\caption{Summary of the new optical spectra analyzed and presented here.}\label{tab:analysis}
\begin{center}
\begin{tabular}{lllllrrrc}
\hline
4FGL & WISE & Class & z & Line &  EW  & $\lambda_{obs}$& Type & Ca II break \\
name & name & & &  ID & (\AA) &  & (\AA)  & \\
   {(1)} &  {(2)} &  {(3) } &  {(4)} &  {(5)} &  {(6) } &  {(7) } &  {(8)} &  {(9)}\\
\noalign{\smallskip}
\hline 
\noalign{\smallskip}
  J0836.9+5833 & J083706.00+583152.9 & bzb & \dots & \dots & \dots & \dots &  & \dots\\
  J0914.8+5846 & J091451.57+584438.1 & bzb & \dots & \dots & \dots & \dots & \dots & \dots\\
  J1019.3+5625 & J101919.15+562428.8 & bzb & \dots & \dots & \dots & \dots & \dots & \dots\\
  J1035.6+4409 & J103532.12+440931.4 & bzg & 0.4438 & [O II] & 10 & 5380 & E & 0.25\\
  \dots & \dots & \dots & \dots & H & 6 & 5730 & A & \dots\\
  \dots & \dots & \dots & \dots & K & 7 & 5680 & A & \dots\\
  J1226.0+5622 & J122602.82+562254.9 & bzb & \dots & \dots & \dots & \dots & \dots & \dots\\
  J1238.1-4541 & J123806.03-454129.6 & bzb & \dots & \dots & \dots & \dots & \dots & \dots\\
  J1403.4+4319 & J140319.46+432020.1 & bzb & \dots & \dots & \dots & \dots & \dots & \dots\\
  J1454.7+5237 & J145445.32+523655.4 & bzb & \dots & \dots & \dots & \dots & \dots & \dots\\
  J1459.5+1527 & J145922.16+152654.9 & bzb & 0.3711 & [O II] & 5 & 5110 & E & 0.06\\
  \dots & \dots & \dots & \dots & H & 3 & 5441 & A & \dots\\
  \dots & \dots & \dots & \dots & K & 3 & 5391 & A & \dots\\
  J1545.0-6642 & J154458.88-664146.9 & bzb & \dots & \dots & \dots & \dots &  & \dots\\
  J1545.8-2336 & J154546.58-233928.4 & bzb & 0.1204 & Mg I & 6 & 5800 & A & \dots\\
  \dots & \dots & \dots & \dots & Na I & 4 & 6602 & A & \dots\\
  J1600.3-5811 & J160012.36-581102.8 & bzb & \dots & \dots & \dots & \dots & \dots & \dots\\
  J1637.5+3005 & J163738.24+300506.4 &  {bzg} & 0.0786 & [O II] & 10 & 4019 & E & 0.25\\
  \dots & \dots & \dots & \dots & K & 5 & 4243 & A & \dots\\
  \dots & \dots & \dots & \dots & H & 8 & 4280 & A & \dots\\
  \dots & \dots & \dots & \dots & G & 6 & 4644 & A & \dots\\
  \dots & \dots & \dots & \dots & [O III] & 3 & 5401 & E & \dots\\
  \dots & \dots & \dots & \dots & Mg I & 3 & 5581 & A & \dots\\
  J1640.9+1143 & J164058.89+114404.2 & bzg & 0.0799 & K & 8 & 4248 & A & 0.35\\
  \dots & \dots & \dots & \dots & H & 6 & 4285 & A & \dots\\
  \dots & \dots & \dots & \dots & G & 5 & 4649 & A & \dots\\
  \dots & \dots & \dots & \dots & H$\beta$ & 3 & 5250 & A & \dots\\
  \dots & \dots & \dots & \dots & Mg I & 13 & 5586 & A & \dots\\
  \dots & \dots & \dots & \dots & Na I & 10 & 6365 & A & \dots\\
  J1647.1+6149 & J164723.42+615347.5 & bzb & 0.347 & K & 3 & 5298 & A & 0.11\\
  \dots & \dots & \dots & \dots & H & 4 & 5349 & A & \dots\\
  \dots & \dots & \dots & \dots & G & 5 & 5798 & A & \dots\\
  J1704.5-0527 & J170433.83-052840.7 & bzb & \dots & \dots & \dots & \dots & \dots & \dots\\
  J1705.4+5436 & J170520.54+543659.8 & bzb & \dots & \dots & \dots & \dots & \dots & \dots\\
  J1706.8+3004 & J170650.43+300412.5 & bzb & \dots & \dots & \dots & \dots & \dots & \dots\\
  J1744.4+1851 & J174419.81+185217.9 & bzb & \dots & \dots & \dots & \dots & \dots & \dots\\
  J1810.7+5335 & J181037.98+533501.5 & bzb & \dots & \dots & \dots & \dots & \dots & \dots\\
  J1814.0+3828 & J181403.43+382810.1 & bzb & 0.2754 & K & 6 & 5017 & A & 0.23\\
  \dots & \dots & \dots & \dots & H & 5 & 5059 & A & \dots\\
  \dots & \dots & \dots & \dots & G & 5 & 5492 & A & \dots\\
  J1838.4-6023 & J183820.63-602522.6 & bzb & 0.120 & H$\beta$ & 3 & 5447 & A & \dots\\
  \dots & \dots & \dots & \dots & Mg I & 4 & 5795 & A & \dots\\
  \dots & \dots & \dots & \dots & Na I & 3 & 6602 & A & \dots\\
  J1858.3+4321 & J185813.43+432451.9 & bzg & 0.1356 & [O II] & 5 & 4233 & E & 0.26\\
  \dots & \dots & \dots & \dots & K & 3 & 4468 & A & \dots\\
  \dots & \dots & \dots & \dots & H & 3 & 4506 & A & \dots\\
  \dots & \dots & \dots & \dots & G & 6 & 4888 & A & \dots\\
  \dots & \dots & \dots & \dots & Na I & 4 & 6692 & A & \dots\\
  J1929.4+6146 & J192935.09+614629.4 & bzb & 0.2117 & [O II] & 1 & 4516 & E & 0.17\\
  \dots & \dots & \dots & \dots & K & 5 & 4767 & A & \dots\\
  \dots & \dots & \dots & \dots & H & 4 & 4808 & A & \dots\\
  \dots & \dots & \dots & \dots & G & 5 & 5215 & A & \dots\\
  \dots & \dots & \dots & \dots & Mg I & 3 & 6271 & A & \dots\\
  J2043.7+0000 & J204350.15+000127.8 & bzb & \dots & \dots & \dots & \dots & \dots & \dots\\
  J2046.8-4258 & J204644.01-425713.2 & bzb & \dots & \dots & \dots & \dots & \dots & \dots\\
  J2115.2+1218 & J211522.00+121802.6 & bzb & $>0.498$ & Mg II & 2 & 4188 & A & \dots\\
  \dots & \dots & \dots & \dots & Mg II & 4 & 4199 & A & \dots\\
  J2141.4+2947 & J214123.89+294706.2 & bzb & \dots & \dots & \dots & \dots & \dots & \dots\\
  J2208.2+0350 & J220812.70+035304.5 & bzb & \dots & \dots & \dots & \dots & \dots & \dots\\
  J2235.3+1818 & J223528.60+181635.5 & bzb & \dots & \dots & \dots & \dots & \dots & \dots\\

\noalign{\smallskip}
\hline
\end{tabular}\\
\end{center}
Note: columns (1) 4FGL name; (2) {targeted} WISE counterpart; (3) {classification based on our analysis of the collected spectra}; (4) {redshift measurement when multiple emission/absorption lines are present, with a lower limit derived in one case from the Mg II doublet} ; (5) emission/absorption line identified; (6) {corresponding} equivalent width (EW); 
(7) observed wavelength of the emission/absorption line listed in col. (5); 
(8) type of spectral feature/line: $E=$ emission, $A=$ absorption;
(9) Ca II break intensity.
\end{table*}

\section{Impact of optical spectroscopic observations on the \fer-LAT source catalogs}
\label{sec:impact}
Here we summarize all results achieved to date, distinguishing those obtained thanks to our optical spectroscopic campaign from those found in the literature. Our summary is presented separating each \fer-LAT catalog release available to date to highlight the evolution of the impact of our campaign on \fer-LAT associations. However, this also implies that sources listed in more than one \fer-LAT catalog are counted in each of them.  

In Appendix~\ref{sec:appB} we also report both summary lists including all details obtained thanks to our observations as well as those found in the literature search. 
 {We remark that sources spectroscopically identified both during our follow-up campaign and found in the literature reported in different versions of the \fer-LAT catalogs are not independent.}

\subsection{Optical Spectroscopic Campaigns}
During our optical spectroscopic campaigns we analyzed 441 observations and we found 394 optical spectra that allowed us to clearly classify the {targets;} this is our ``clean'' sample reported in Appendix~\ref{sec:appA}. These spectra include also those found in the SDSS and 6dF databases and analyzed as part of our campaign. 

For the 47 optical spectra not used, {27 observations} revealed an ``incorrect'' target {(e.g.,} a star lying within the positional uncertainty region of a {UGS), while an} additional seven spectra {achieved only low signal-to-noise ratios (SNR) thus were} not reported in our summary table presented here. {The remaining 13 cases were targets observed} with more than one telescope.

In our ``clean'' sample, 237 sources out of 394 were previously unclassified and had no optical spectrum present in the literature. We observed 306 targets with ground based telescopes while 88 spectra were collected from archival {SDSS and 6dF observations.} In the clean sample, 121 of them lie in the northern hemisphere while the remaining 116 in the southern one. 
{  Our most widely-used facilities} were OAN-SPM in the northern hemisphere and SOAR telescope at southern declinations. 

During our campaign we also {observed} a total of 128 Roma-BZCAT sources, mostly BL Lac candidates or sources with uncertain classification or lacking a $z$ {estimate, in order to} provide updated information in the next release of the blazar catalog. Then there are also 59 targets that, while we carried out our campaign, were also analyzed in other papers.
 
Results on the impact of our optical spectroscopic observations, carried out since the first release of the \fer-LAT source catalog, on the association of $\gamma$-ray sources listed therein are all summarized as follows. 

In Table~\ref{tab:optcamp1} we report the number of \fer-LAT sources observed in each catalog as classified therein. For example in the 2FGL we analyzed 76 spectra of sources listed therein as AGN or AGU. 
\begin{table}[!h] 
\caption{$\gamma$-ray classification of targets observed during our spectroscopic campaign as reported in each \fer-LAT catalog.}
\begin{center}
\label{tab:optcamp1}
\begin{tabular}{lrrrrr}
\hline

\fer-LAT class & 1FGL & 2FGL & 3FGL & 4FGL & FL8Y \\
 {(1)} &  {(2)} &   {(3) }&   {(4)} &   {(5)} &   {(6)} \\
\noalign{\smallskip}
\hline 
\noalign{\smallskip}
BZB/BLL      &   32 &   55 &   80 &  212 &   1  \\ 
BZQ/FSRQ      &    5  &  10 &   10 &   36 &   0  \\ 
AGN/AGU/BCU  &   10 &   76 &  139 &   86 &  10  \\
UGS /UNK     &   58 &   52 &   57 &   13 &  15  \\
\noalign{\smallskip}
\hline
\noalign{\smallskip}
Total         &  105 &  193 &  286 &350$^*$& 26  \\
\noalign{\smallskip}
\hline
\end{tabular}\\
\end{center}
\footnotesize{Note: {col. (1)} lists the \fer-LAT class, in particular BZB and BZQ were mainly used {in the} 1FGL and 2FGL together with {the} AGU classification, then they were removed in the {later 3FGL and 4FGL catalogs.} The UNK class {of sources with unknown nature were} all counted together with UGSs in our summary; {cols. (2,3,4,5,6) list} the number of sources classified according to col. (1) {for} each \fer-LAT catalog. \\ ($^*$) The total number of 4FGL sources is 350 instead of 347 since we observed three targets lying within the positional uncertain regions of three sources associated to one pulsar and two radio galaxies. Results of our campaign confirmed the 4FGL classification and in particular for the pulsar we only found a quasar within its $\gamma$-ray positional uncertainty ellipse.}
\end{table}

Table~\ref{tab:optcamp2} {summarizes} the fraction of uncertain sources, {e.g.,} BCUs and UGSs in the 3FGL or similar in {the} other catalogs, for which we provide a firm optical classification with respect to the total number reported in the original \fer-LAT catalog. Since we started our campaign after the 2FGL release and during the 3FGL preparation, the impact of our observations {was} larger on these two {catalogs. In particular,} we were able to classify $\sim$25\% of the uncertain AGNs and found potential blazar-like low-energy counterparts for 9\% and 6\% of UGSs listed therein.

\begin{table}[!h] 
\caption{Fractions of \fer-LAT sources with uncertain nature (i.e., AGUs, BCUs and UGSs) analyzed during our campaign and computed with respect to the total number in each \fer-LAT catalog.}
\label{tab:optcamp2}
\begin{center}
\begin{tabular}{lrrrr}
\hline

\fer-LAT class &  1FGL &  2FGL &  3FGL &  4FGL \\  
 {(1) }&   {(2)} &   {(3)} &   {(4)} &  { (5)} \\  
\noalign{\smallskip}
\hline 
\noalign{\smallskip}
AGN/AGU/BCU  &  0.08 &  0.27 &  0.24 &  0.06 \\
UGS/UNK      &  0.09 &  0.09 &  0.06 &  0.01 \\
\noalign{\smallskip}
\hline
\end{tabular}\\
\end{center}
\footnotesize{Note: {col. (1)} lists the \fer-LAT class, in particular {the} AGU classification was used only in {the} 1FGL and 2FGL, then replaced by BCU {in the later 3FGL and 4FGL catalogs. The UNK class of sources with unknown nature were} all counted together with UGSs in our summary; {cols. (2,3,4,5)} report the fraction of sources in each class computed over the whole number of sources belonging to that class listed in each \fer-LAT catalog.}
\end{table}

In Table~\ref{tab:optcamp3} we show for each sample of sources listed in all \fer-LAT catalogs the {classifications} we obtained. For example in the 3FGL we observed 103 sources listed therein as AGN or BCU for which their optical spectra clearly {indicate them to be BL Lacs. Moreover,} as reported in parenthesis {in} the BZB column, 20 out of these 103 targets have also a firm redshift {measurement}. We also classified {an} additional 14 AGNs or BCUs {from} the 3FGL as BZQ, plus {seven} more simply as ``quasars'' since {they lack radio spectral information, thus the entry is} reported as 14[+7] in Table~\ref{tab:optcamp3}. Finally, 15 more AGNs or BCUs {from} the 3FGL were classified as BZGs, all with a firm $z$ {measurement, contributing to the total of 139 sources indicated} in Table~\ref{tab:optcamp1}.  
\begin{table}[!h]
\caption{Classification results achieved thanks to our optical spectroscopic campaign split with respect to sources listed in each \fer-LAT catalog.}
\label{tab:optcamp3}
\begin{center}
\begin{tabular}{lrrrr}
\hline
\fer-LAT class &   BZB   &   BZQ   &   BZG   \\
 {(1)} &    {(2)}   &    {(3)}   &    {(4)}   \\
\noalign{\smallskip}
\hline 
\noalign{\smallskip}
in 1FGL & & & \\        
\noalign{\smallskip}
\hline 
\noalign{\smallskip}
BZB      &   32(1) &    \dots   &    \dots    \\
BZQ      &    1    &    4    &    \dots    \\
AGN/AGU  &    7(1) &    1    &    2    \\
UGS      &   48(2) &    2[+5]&    3    \\
\noalign{\smallskip}
\hline 
\noalign{\smallskip}
in 2FGL & & & \\        
\noalign{\smallskip}
\hline 
\noalign{\smallskip}
BZB      &   54(3) &    \dots   &    1    \\
BZQ      &    3    &    7    &    \dots   \\
AGN/AGU  &   54(6) &   13[+2]&    7    \\
UGS      &   36(4) &   3[+12]&    1    \\
\noalign{\smallskip}
\hline 
\noalign{\smallskip}
in 3FGL & & & \\        
\noalign{\smallskip}
\hline 
\noalign{\smallskip}
BLL      &   79(3) &    \dots    &    1    \\
FSRQ     &    \dots    &    9[+1]&    \dots    \\
AGN/BCU  &  103(20)&   14[+7]&   15    \\
UGS      &   40(5) &   4[+12]&    1    \\
\noalign{\smallskip}
\hline 
\noalign{\smallskip}
in 4FGL & & & \\        
\noalign{\smallskip}
\hline 
\noalign{\smallskip}
BLL      &  193(18)&    1    &   18    \\
FSRQ     &    2(2) &   30[+4]&    \dots   \\
AGN/BCU  &   66(23)&    1[+10]&   9    \\
UGS/UNK  &    8(1) &    -[+5]&   \dots    \\
\noalign{\smallskip}
\hline 
\noalign{\smallskip}
in FL8Y & & & \\        
\noalign{\smallskip}
\hline 
\noalign{\smallskip}
BLL      &    1    &    \dots  &    \dots    \\
FSRQ     &   \dots   &    \dots   &    \dots    \\
AGN/BCU  &    6(5) &    3    &    1    \\
UGS      &   15(3) &    \dots   &    \dots    \\
\noalign{\smallskip}
\hline
\end{tabular}\\
\end{center}
\footnotesize{Note: {col. (1)} lists the \fer-LAT class according to labels reported in each \fer-LAT catalog; {cols. (2,3,4) indicate} the number of BZBs, BZQs and BZGs classified thanks to our observations, respectively. Number in parenthesis for col. (2) corresponding to BZB classification indicates those BL Lacs having a firm $z$ estimate, while that in col. (3) for BZQs indicate additional sources classified as quasars for which information about their radio spectral shape were not found.}
\end{table}

\begin{figure*}{}
%\label{fig:campaign}
\begin{center}
$\begin{array}{cc}
\includegraphics[width=9.2cm]{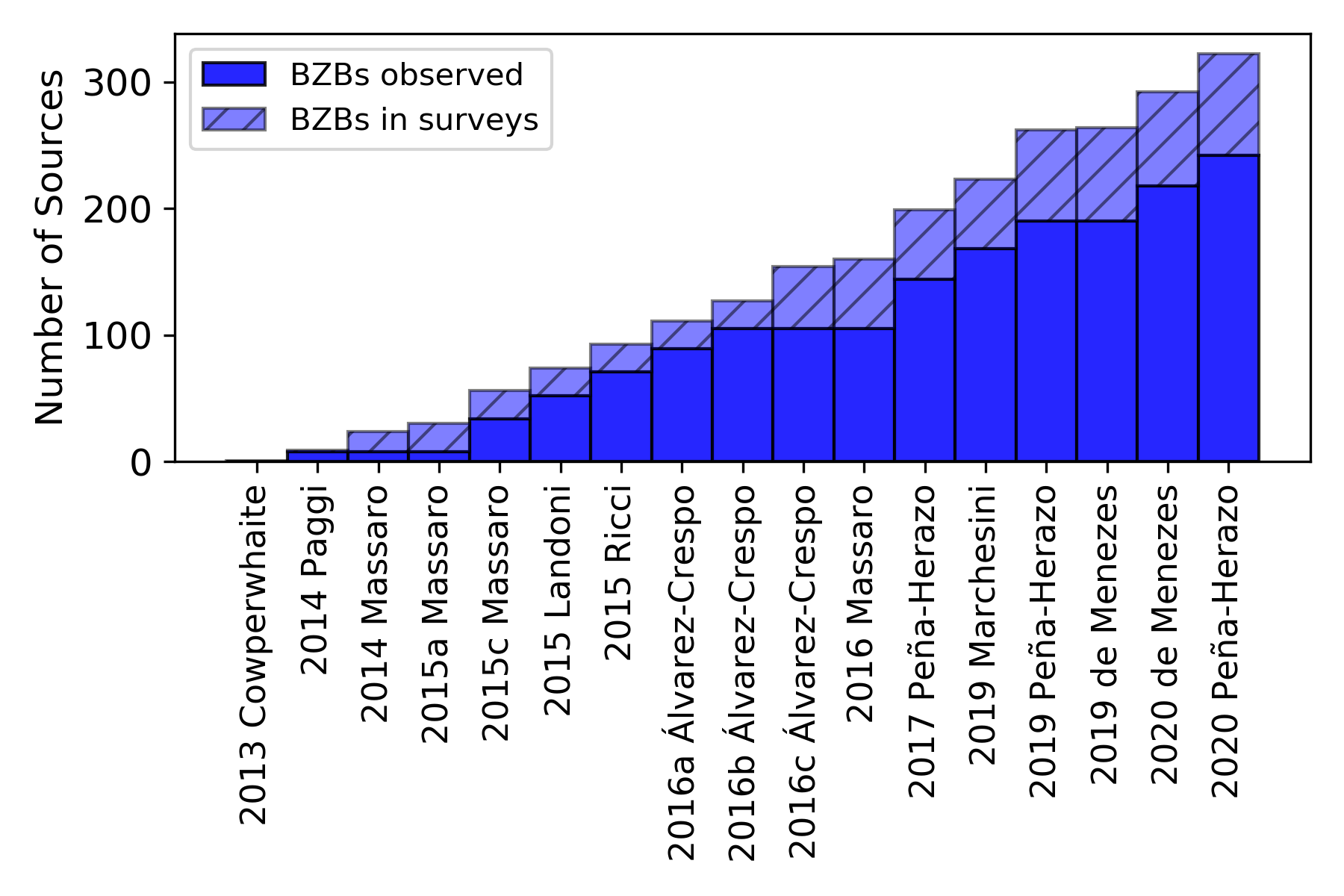}&
\includegraphics[width=8.75cm]{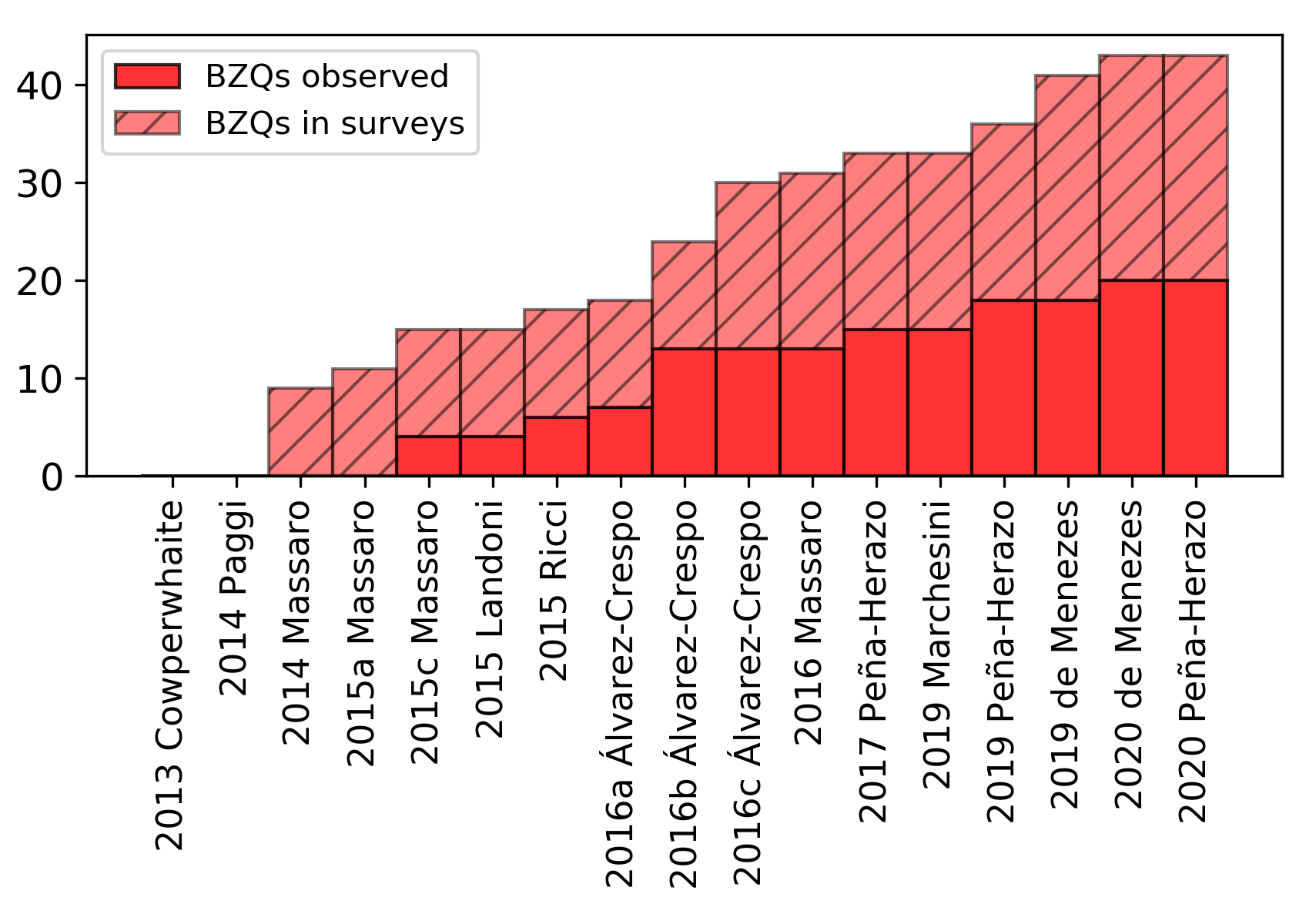}\\

\end{array}$
\end{center}
\caption{
Cumulative distributions of all sources observed during our spectroscopic campaign (non-hatched bars) and classified as BZBs (\textit{left panel} in blue) or BZQs (\textit{right panel} in red) in addition to those found {in our archival searches of other} spectroscopic surveys (hatched bars). 
}\label{fig:campaign}
\end{figure*}

Finally, Figure~\ref{fig:campaign} shows the cumulative distributions of all confirmed blazars observed during our spectroscopic campaign in each published {paper. Sources classified as BZBs and BZQs are indicated separately, and we further distinguished between those observed by us} and those discovered thanks to archival searches in major surveys (i.e., SDSS and 6dF).

\subsection{Literature search}
During our optical spectroscopic campaign, other teams worldwide were presenting optical spectra of sources listed in the \fer-LAT catalogs as UGSs or BCUs. Thus, {we continuously kept track of these works to avoid duplicating targets and provided our summaries of the literature results to our \fer-LAT collaboration colleagues while they were preparing each new release of the $\gamma$-ray source catalogs}. Thus here we report all information retrieved from the literature {from that timespan} that met the same criteria adopted to classify sources in our campaign and those of the Roma-BZCAT.

In the literature we found a total of 123 sources, {with 67 northern hemisphere objects and 56 in the southern hemisphere.} These observations {include} (i) 23 known blazars listed in the latest release of the Roma-BZCAT and (ii) 19 targets collected from data available in the SDSS {or 6dF} surveys.

{  Following Table~\ref{tab:optcamp1} from the previous section, here, Table~\ref{tab:literature1} provides all information from the literature on} the total number of sources analyzed for each \fer-LAT catalog since the 1FGL release. {It is evident that most of efforts reported in the literature were} focused on 3FGL sources. This was because {most} observational campaigns carried out in parallel with {our work started later, with the main exception being the observations performed by \citet{shaw13}.}
\begin{table}[!h] 
\caption{$\gamma$-ray classification of sources analyzed in the literature as reported in each \fer-LAT catalog.}
\label{tab:literature1}
\begin{center}
\begin{tabular}{lrrrr}
\hline
\fer-LAT class & 1FGL & 2FGL & 3FGL & 4FGL\\
 {(1)} & { (2)} &  {(3) }&  {(4)} &  {(5)}\\
\noalign{\smallskip}
\hline 
\noalign{\smallskip}
BZB/BLL          &    9 &   11 &   14 &   53 \\ 
BZQ/FSRQ         &    1 &    \dots &    \dots &    5 \\ 
AGN/AGU/BCU  &    6 &   23 &   58 &   53 \\
UGS/UNK          &   22 &   25 &   51 &    7 \\
\noalign{\smallskip}
\hline
\noalign{\smallskip}
Total         &   38 &60$^*$&  123 &  118 \\
\noalign{\smallskip}
\hline
\end{tabular}\\
\end{center}
Note: {col. (1)} lists the \fer-LAT class, in particular BZBs and BZQs were mainly used {in the} 1FGL and 2FGL together with {the} AGU classification, then they were removed in the {later 3FGL and 4FGL catalogs where the UNK class was introduced} for sources with unknown nature that were all counted together with UGSs in our summary; {cols. (2,3,4,5) lists} the number of sources classified according to col. (1) {for} each \fer-LAT catalog. ($^*$) {In the 2FGL,} the total number of sources is 60 instead of 59 {because we found in the literature that one 2FGL source classified as a pulsar was later classified as BCU in both the 3FGL and 4FGL.}\\
\end{table}

Then in Table~\ref{tab:literature3} we show for each sample of sources listed in all \fer-LAT catalogs the classification reported in the literature.
\begin{table}[!h]
\caption{Classification results collected from the literature distinguishing sources belonging to each \fer-LAT catalog.}
\label{tab:literature3}
\begin{center}
\begin{tabular}{lrrrr}
\hline
\fer-LAT class &   BZB   &   BZQ   &   BZG   \\
 {(1)} &   {(2)}   &    {(3) }  &   { (4)}   \\
\noalign{\smallskip}
\hline 
\noalign{\smallskip}
in 1FGL & & & \\        
\noalign{\smallskip}
\hline 
\noalign{\smallskip}
BZB      &    9(3) &    \dots   &    \dots    \\
BZQ      &    \dots  &    [+1]&    \dots    \\
AGN/AGU  &    6(2) &    \dots    &    \dots    \\
UGS      &   22(5) &    \dots    &    \dots   \\
\noalign{\smallskip}
\hline 
\noalign{\smallskip}
in 2FGL & & & \\        
\noalign{\smallskip}
\hline 
\noalign{\smallskip}
BZB      &   11(4) &    \dots    &    \dots   \\
BZQ      &   \dots    &   \dots    &    \dots   \\
AGN/AGU  &   19(3) &    3   &    1    \\
UGS      &   23(8) &    1[+1]&  \dots    \\
\noalign{\smallskip}
\hline 
\noalign{\smallskip}
in 3FGL & & & \\        
\noalign{\smallskip}
\hline 
\noalign{\smallskip}
BLL      &   14(4) &    \dots   &   \dots   \\
FSRQ     &   \dots   &   \dots    &    \dots  \\
AGU/BCU  &   53(8) &    4    &    1    \\
UGS      &   50(22)&    [+1]&   \dots   \\
\noalign{\smallskip}
\hline 
\noalign{\smallskip}
in 4FGL & & & \\        
\noalign{\smallskip}
\hline 
\noalign{\smallskip}
BLL      &   53(15)&   \dots   &    \dots  \\
FSRQ     &    4(1) &    1    &    \dots   \\
AGU/BCU  &   49(15)&    3    &    1    \\
UGS/UNK  &    6(1) &    [+1]&    \dots  \\
\noalign{\smallskip}
\hline
\end{tabular}\\
\end{center}
\footnotesize{Note: {col. (1)} lists the \fer-LAT class according to labels reported in each \fer-LAT catalog; {cols. (2,3,4) indicate the number of BZBs, BZQs and BZGs classified in the literature, respectively. Numbers in parenthesis in col. (2) correspond to BZB classifications of BL Lacs having a firm $z$ measurement, while those in col. (3) for BZQs indicate additional sources classified as quasars but without sufficient information on their radio spectral shapes.}}
\end{table}

For {the} literature results we did not report the comparison with the FL8Y, the intermediate catalog between the 3FGL and the 4FGL {because most of our efforts} were focused on 3FGL sources, quite close to the {time of the FL8Y release. Nevertheless,} during our optical spectroscopic campaign we observed some sources listed in the FL8Y catalog only, {and} all sources found in the literature belong to one of the major releases (3FGL or 4FGL) and thus the comparison with the FL8Y it is not relevant.

\section{Comparing mid-IR color predictions with optical campaign and literature results}
\label{sec:reliability}
In this section we present the comparison between the predicted {classifications based on mid-IR colors with those ultimately provided by optical spectroscopic observations, distinguishing between sources collected during our campaign and those found in the literature. }

 {We classified sources based on the Ca II break feature in the optical spectra. The break is defined as $C = (F_+-F_-)/F_+$, with $F_+$ and $F_-$ being the fluxes measured respectively at rest-frame wavelengths of 3750-3950 \AA\, and 4050-4250 \AA\, \citep{landt02}. The division line for BZGs is $C\geq0.25$, otherwise they are classified as BZB.}

A large fraction of targets selected for our optical spectroscopic observations was based on the statistical analysis of mid-IR colors as presented in the WIBRaLS catalog \citep{wibrals}, for which the second version, WIBRaLS2, has been recently released together with the KDEBLLACS \citep{dabrusco19}. The main description of these two catalogs is summarized as follows. 
\begin{itemize}
\item  {The WIBRaLS2 catalog contains 9541 candidate blazars, selected among WISE sources detected in all four W1, W2, W3, and W4 filters (3.4, 4.6, 12, and 22 $\mu$m, respectively), and whose colors are similar to those of confirmed $\gamma$-ray emitting blazars. 
As shown in our previous analyses \citep{wibrals,massaro16,dabrusco19}, Galactic extinction affects only the W1-band for sources lying at low Galactic latitudes ($|b|<$20$^\circ$) at levels of only $2-5\%$ of the magnitude, thus the WISE magnitudes were not corrected for extinction.
The selection is performed in the three-dimensional Principal Component (PC) space generated by the distribution of W1-W2, W2-W3 and W3-W4 colors for a sample of bona fide $\gamma$-ray blazars (the locus). Differently from the direct color space, in the PC space the region occupied by the locus can be modeled with coaxial cylinders. This method also distinguishes among BZB-like, BZQ-like, or MIXED candidates based on the WISE colors, hereafter labelled as WBZB, WBZQ, and MIXED types, respectively. The MIXED-type sources have colors consistent with both blazar classes. WISE-selected sources are further required to have a radio counterpart that is determined to be radio-loud (according to the $q_{22}$ parameter; see below). }

\item  {The KDEBLLACS catalog includes 5579 BL Lac candidates selected among WISE sources that are not detected in the W4 band. The colors of these candidates lie within the region of the W1-W2 vs W2-W3 color-color diagram occupied by a set of confirmed BL Lacs with WISE counterparts not detected in W4. The region of the color-color plane used for the selection is defined as the area enclosed by the 5\% contour of the 2D density distribution of the confirmed BLLacs, determined using the KDE method. As for the WIBRaLS, the final members of the KDEBLLACS catalog are selected to have a radio-loud counterpart (according to the $q_{12}$ parameter; see below). }
\end{itemize}

 {Both the $q_{22}$ and $q_{12}$ parameters are simply modified definitions of the so-called $q$ parameter \citep{helou85}, defined as the logarithm of the ratio of far-IR to radio flux densities ($S_{radio}$).
Applied to the WISE flux densities in the W4 ($S_{\rm 22\mu\,m}$) and W3 ($S_{\rm 12\mu\,m}$) bands, the definitions of radio-loudness were
$q_{22} = log(S_{22\mu\,m}/S_{radio})$ for WIBRaLS and
$q_{12} = log(S_{12\mu\,m}/S_{radio})$ for  KDEBLLACS.} 

\begin{figure*}{}
\begin{center}
\includegraphics[width=13cm]{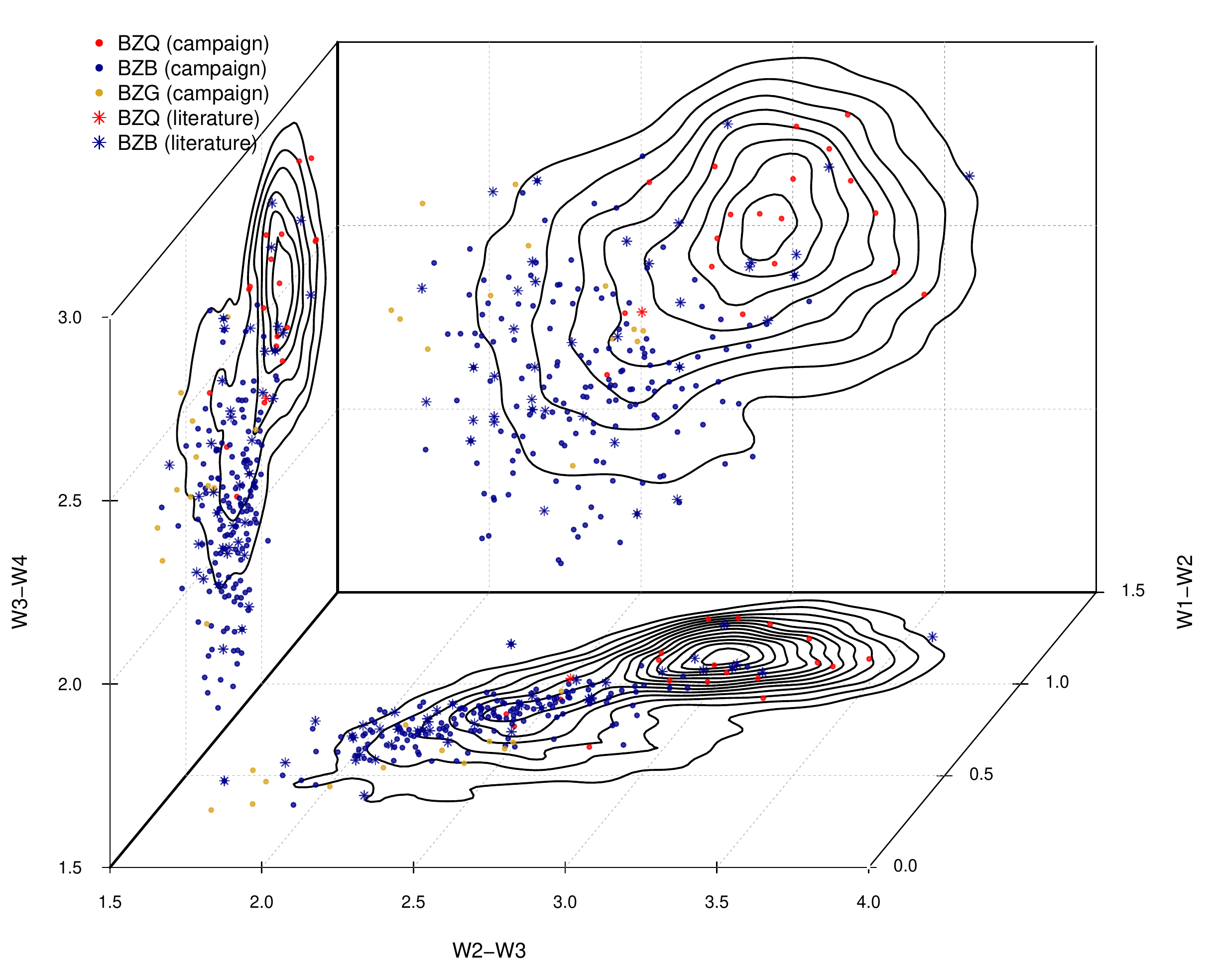} 
\end{center}
\caption{ {Projections on the three WISE color-color planes of {spectroscopically-identified} sources discussed in this paper with counterparts in WIBRaLS. Red, blue and yellow points indicate, respectively, candidates classified as BZQ, BZB and BZG sources based on their optical spectra. Circles and stars are associated with sources observed in this campaign and sources for which spectra are available in the {literature, respectively.} The black lines on the three planes are the isodensity contours of the 2D projections of the 3D color distribution of locus sources (not plotted for clarity) used to define the WIBRaLS 3D model in the WISE colors space.}}
\label{fig:wise3dplot}
\end{figure*}

Comparing our 394 sources with those listed in WIBRaLS we found 212 {total matches, consisting of 178 WBZBs, 22 WBZQs, and 12 MIXED. In Table~\ref{tab:compare1}, we report how they were classified based on our} spectroscopic observations. In this table we adopted the same nomenclature of Table~\ref{tab:optcamp3} to indicate those targets classified as quasars but lacking radio spectral information and thus are not qualified to be BZQ according to our criteria.  {We also present in Figure~\ref{fig:wise3dplot} the projections of the three WISE color-color planes, for the sources in this work that have a WIBRaLS counterpart.}

\begin{table}[!h] 
\caption{Comparison between the expected classification provided in WIBRaLS2 catalog with {those obtained in} our optical spectroscopic campaign.}
\label{tab:compare1}
\begin{center}
\begin{tabular}{lcrrrr}
\hline
WIBRALS Type & BZB & BZQ & BZG & Total \\ 
 {(1)} &  {(2) }&  {(3)} &  {(4)} &  {(5)} \\ 
\noalign{\smallskip}
\hline 
\noalign{\smallskip}
WBZB  &  5  &  \dots       &  \dots  &   5 \\
            & 32  &  \dots    &  1  &  33 \\
            & 78  &  3       &  4  &  85 \\
            & 46  &  [+2]  &  7  &  55 \\
\noalign{\smallskip}
\hline
\noalign{\smallskip}
WBZQ  &  \dots &  [+1]  &  \dots  &   1 \\ 
             &  1  &  5 [+1]  &  \dots  &   7 \\
             &  1  &  6 [+1]  &  \dots  &   8 \\ 
             &  \dots  &  5 [+1]  &  \dots  &   6 \\
\noalign{\smallskip}
\hline
\noalign{\smallskip}
MIXED &  \dots  &  \dots      &  -  &   \dots \\
             &  4  &   [+1]  &  \dots &   5 \\ 
             &  4  &  2 [+1]  &  \dots  &   7 \\
             &  \dots  &  \dots       &  \dots  &   \dots \\
\noalign{\smallskip}
\hline
\end{tabular}\\
\end{center}
Note: {col. (1) indicates the class/type as listed in the WIBRaLS2 catalog; cols. (2,3,4) list the} classification results provided by our optical spectroscopic observations; col. (5) indicates total number of sources observed for each type and subclass.\\
\end{table}

Overall we found that only {8 out of 212 WIBRaLS2 sources} are classified as quasars. Then, we confirmed the nature of 161 out of 178 (i.e., 90\%) of those targets expected to be BZBs, {and} $\sim$70\% of the WBZQs and the largest fraction of those indicated as MIXED in WIBRaLS2 are BZBs. It is worth noting that only 12 sources out of 212 are classified as BZGs and they were all expected to be BZBs according to their mid-IR colors in WIBRaLS2. 

On the other hand, {only 43 spectroscopic identifications were achieved out of 394 sources selected from the KDEBLLACS catalog. In all these cases, they were expected to be BZBs based on the mid-IR colors, and the spectroscopic observations confirmed 40 as BZBs, with only 3 better classified as BZGs.}

This comparison strongly supports both the reliability of methods based on mid-IR colors to select candidate counterparts for UGSs and that there is a high chance to confirm the blazar-like nature of {BCUs and that they are most likely classified as BZBs.}

{  Similar conclusions could be drawn for the literature results where we compared WIBRaLS2 with sources analyzed by other groups and} found 37 matches out of 123 targets (see Table~\ref{tab:compare2}). All WIBRaLS2 sources were confirmed as BZBs with only one {exception,} 3FGL J0644.3-6713 that is a BZQ. {In the KDEBLLACS catalog, we found} 35 out of 123 sources listed among those classified in the literature and again all of them are BZBs.

\begin{table}[!h] 
\caption{Comparison between the expected classification provided in WIBRaLS2 catalog with {those obtained in} our literature search.}
\label{tab:compare2}
\begin{center}
\begin{tabular}{lcrrr}
\hline
WIBRALS Type & BZB & BZQ & Total \\ 
 {(1)} &  {(2)} &  {(3)} & { (4)} \\ 
\noalign{\smallskip}
\hline 
\noalign{\smallskip}
WBZB  &  0  &  \dots      &   1 \\
            &  6  &  \dots       &   6 \\
            & 10  &  \dots       &  10 \\
            & 13  &  \dots    &  12 \\
\noalign{\smallskip}
\hline
\noalign{\smallskip}
WBZQ  &  1  &  \dots       &   1 \\ 
            &  3  &  \dots     &   3 \\
            &  1  &  \dots      &   1 \\ 
            &  1  &  \dots      &   1 \\
\noalign{\smallskip}
\hline
\noalign{\smallskip}
MIXED &  \dots  &  \dots       &   \dots \\
            &  1  &  \dots      &   1 \\ 
            &  1  &  1       &   2 \\
            &  \dots  &  \dots       &   \dots \\
\noalign{\smallskip}
\hline
\end{tabular}\\
\end{center}
Note: {col. (1) indicates the class/type as listed in the WIBRaLS2 catalog; cols. (2,3) list the} classification results found during our literature search; {col. (4) indicates} total number of sources observed for each type and subclass.\\
\end{table}

\section{Summary and Conclusions}
\label{sec:summary}
In this work we summarized all results achieved to date thanks to the optical spectroscopic campaign we carried out to unveil the nature of \fer-LAT sources classified as BCUs and to potentially identify blazar-like sources lying within the positional uncertainty regions of UGSs.

Since the beginning of our campaign in 2014 we analyzed 394 unique targets confirming the blazar-like nature of 371 of them. These are classified as 300 BZBs (38 with a firm redshift {measurement}), 40 BZQs, and 31 BZGs. Additionally, there are 23 sources for which the lack of radio spectral information prevented us from labeling them as BZQs thus are simply indicated as quasars. 

{  Altogether, we} observed 122 targets in the northern hemisphere mostly thanks to the OAN-SPM facility and 116 targets in the southern one predominantly through observations with the SOAR telescope. We observed 306 targets out of 394 using ground-based telescopes while 88 spectra were collected from archival observations {from} large spectroscopic surveys. The selection of our targets was mainly based on mid-IR colors since 212 out of 394 also belong to the WIBRaLS2 catalog and 43 to the KDEBLLACS \citep{dabrusco19} and the expected blazar classification reported therein was also mainly confirmed by our followup observations. All these results also include 30 new spectra presented in this paper for which all details on the data reduction and analysis are available in Appendix~\ref{sec:appA}.

Here we also described results found in an extensive literature search carried out in parallel {with} our campaign to avoid observing targets already classified by other groups. {From these} literature results we found a total of 123 sources, 67 lying in the northern hemisphere while 56 in the southern one. These observations {resulted in one BZG, four BZQs, and one} quasar, while all remaining sources were classified as BZBs {(34 with firm redshift measurements). Of} these total 123 sources, only in 19 cases were spectra analyzed from archival observations available thanks to spectroscopic surveys. A significant fraction of them (i.e.,  {$\sim$59\%}) {also have} a mid-IR counterpart in one of the two main catalogs used to select targets for our campaign, namely 37 in WIBRaLS2 and 35 in KDEBLLACS. All these sources are classified as BZBs with only one exception, thus mainly confirming the expected classification reported {therein based} on WISE colors.

Finally, we conclude that $\sim$20\% of blazars currently listed in the 4FGL are classified thanks to our optical spectroscopic campaign and additional $\sim$7\% arise from literature results. Moreover the largest fraction of them are BL Lac objects thus confirming that this is the most elusive class of extragalactic $\gamma$-ray sources. 

Our observational campaign is still {ongoing, and while preparing this paper, an} additional 62 spectra have been already collected. Considering new observing runs already awarded to our group and {considering our efficiency in acquiring spectra to date}, we expect to release results for {an} additional $\sim$500 targets by around 2022.
\vspace{1cm}

%\acknowledgements
H.P.-H. and R.A.-A. acknowledge support from CONACyT program for Ph.D. studies. This work was partially supported from CONACyT research grant No. 280789.

Based upon observations acquired at the Observatorio Astron{\'o}mico Nacional San Pedro M{\'a}rtir (OAN-SPM), Baja California, M{\'e}xico.
This work is based on data acquired at Blanco CTIO telescope.
We thank the staff at the OAN-SPM and Blanco CTIO telescope for all their help during the observation runs. 
F.R. acknowledges support from FONDECYT Postdoctorado 3180506 and CONICYT project Basal AFB-170002.
The work of R.M. is supported by FAPESP (Funda\c{c}\~ao de Amparo \`a Pesquisa do Estado de S\~ao Paulo) under grants 2016/25484-9, 2018/24801-6.
Work by C.C.C. at the Naval Research Laboratory is supported by the Office of Naval Research 6.1.

This work is supported by the ``Departments of Excellence 2018 - 2022 Grant awarded by the Italian Ministry of Education, University and Research (MIUR) (L. 232/2016).
This research has made use of resources provided by the Compagnia di San Paolo for the grant awarded on the BLENV project (S1618\_L1\_MASF\_01) and by the Ministry of Education, Universities and Research for the grant MASF\_FFABR\_17\_01.
A.P. acknowledges financial support from the Consorzio Interuniversitario per la Fisica Spaziale (CIFS) under the agreement related to the grant MASF\_CONTR\_FIN\_18\_02.

This publication makes use of data products from the Wide-field Infrared Survey Explorer, which is a joint project of the University of California, Los Angeles, and the Jet Propulsion Laboratory/California Institute of Technology, funded by the National Aeronautics and Space Administration.

Funding for the Sloan Digital Sky Survey IV has been provided by the Alfred P. Sloan Foundation, the U.S. Department of Energy Office of Science, and the Participating Institutions. SDSS-IV acknowledges support and resources from the Center for High Performance Computing at the University of Utah. The SDSS web site is www.sdss.org. SDSS-IV is managed by the Astrophysical Research Consortium for the Participating Institutions of the SDSS Collaboration including the Brazilian Participation Group, the Carnegie Institution for Science, Carnegie Mellon University, the Chilean Participation Group, the French Participation Group, Harvard-Smithsonian Center for Astrophysics, Instituto de Astrof\'isica de Canarias, The Johns Hopkins University, Kavli Institute for the Physics and Mathematics of the Universe (IPMU)/University of Tokyo, the Korean Participation Group, Lawrence Berkeley National Laboratory, Leibniz Institut f{\"u}r Astrophysik Potsdam (AIP), Max-Planck-Institut f{\"u}r Astronomie (MPIA Heidelberg), Max-Planck-Institut f{\"u}r Astrophysik (MPA Garching), Max-Planck-Institut f{\"u}r Extraterrestrische Physik (MPE), National Astronomical Observatories of China, New Mexico State University, New York University, University of Notre Dame, Observat{\'o}rio Nacional/MCTI, The Ohio State University, Pennsylvania State University, Shanghai Astronomical Observatory, United Kingdom Participation Group, Universidad Nacional Aut{\'o}oma de M{\'e}xico, University of Arizona, University of Colorado Boulder, University of Oxford, University of Portsmouth, University of Utah, University of Virginia, University of Washington, University of Wisconsin, Vanderbilt University, and Yale University. 

TOPCAT8 \citep{taylor05} was extensively used in this work for the preparation and manipulation of the tabular data.

{}

\appendix{}

\section{Data reduction and analysis of new spectroscopic observations}
\label{sec:appA}
{  This Appendix describes in more detail, the results for the 30 sources analyzed in this paper (see Sect.~\ref{sec:newspec}).
The results were derived from new observations with the Blanco and OAN-SPM telescopes, and archival data collected from the SDSS.}

{  Our targets were selected} from the KDEBLLACS {and} WIBRALS2 catalogs \citep{dabrusco19}, {a sample of X-ray selected sources \citep{marchesini20}, and BL Lac objects with previously-undetermined redshifts that would benefit from new observations.}
In our sample, {seven} sources were observed at Blanco ({  2 UGSs,} 5 BCUs), 15 at OAN-SPM ({  6 UGSs, 9 BCUs),} while the remaining {eight} (i.e., all 5 BL Lacs, {1 UGS, and} 2 BCU) were found in {the SDSS DR15 archive} \citep{aguado19}. We provided the spectroscopic {classifications of these five BL Lacs to the \fer-LAT team} during the preparation of the 4FGL catalog release {thus why they were already classified therein}.

{  As in previous analyses from our spectroscopic campaign \citep[e.g.,][]{pena17,marchesini19},} all spectroscopic data sets acquired were optimally extracted and reduced following standard procedures with IRAF \citep{tody86,tody93}.
{  Thus here we provide only a brief overview of the data acquisition and our} extraction procedure of {the optical spectra; see \citet{marchesini19} for details.}
{  Details of the Blanco telescope observations are provided below as this was the first time it was used in} our campaign. 
{  Further} details regarding the OAN-SPM observations are given in {our past papers} \citep{masetti13,massaro15c}. 
{  We describe the selection of the SDSS spectra and our results from tthe measurements that utilized the DR15 pipeline-calibrated spectra.}

All spectra are de-reddened for {Galactic} absorption assuming $E(B-V)$ values from the relation presented by \citet{schlegel98}. Although the main scientific objective of our campaign does not require a precise flux calibration, we {derived relative photometry by observing} a spectrophotometric standard star during each night of the different observing runs. To detect faint spectral features, aimed at estimating redshifts, we also present normalized spectra.

{  All the following figures showing the new spectra are in the same format adopted in our previous analyses.} We claim a line identification whenever there are at least two absorption and/or emission lines present that can be identified with a redshift value. 
{  As in our previous analyses \citep[see e.g.,][]{masetti13, marchesini19}, these spectral features must be detected in more than one exposure.}

\subsection{Victor Blanco Telescope}
Southern declination sources (i.e., at Dec. $<0^{\circ}$) were observed {in visiting mode at the} Victor Blanco 4 {-}m telescope in Cerro Tololo, Chile on {2019 June 12}. We made use of the COSMOS spectrograph red grism (r2k), {center slit with $1.2\arcsec$ width, and the} OG530 filter. This setup gave a spectral range of 5515-9635 \AA\ and a dispersion of 1 \AA/pixel. We acquired Hg-Ne comparison lamp spectra on each target position for the wavelength calibration.

\subsection{OAN-SPM}
Sources in the northern hemisphere (i.e., at Dec. $>0^{\circ}$) were observed with the 2.1 m telescope of the OAN-SPM in San Pedro M\'{a}rtir (Mexico) in two observing runs {on 2018 August 4 and 2019 July 2-7.}

The telescope carries a Boller and Chivens spectrograph {with} a 1024$\times$1024 pixel E2V 4240 CCD, tuned to the 4000--8000 \AA $\,$range with a resolution of 10 \AA/pixel {and} a slit width of 2.5\arcsec. Wavelength calibration was done using CuHe-NeAr and He-Ar comparison lamps.

For each acquisition, we carried out {bias subtraction, flat-field correction, and cosmic-ray removal. To remove cosmic rays, we collected 2-3 individual exposures per target and averaged them. The multiple exposures allowed us to identify dubious spectral features using the L.A. Cosmic IRAF algorithm \citep{vandockum01}.}

\subsection{SDSS Archival Search}
To investigate those UGSs lying in the SDSS footprint and {to search for new BL Lac objects,} we adopted the same procedure as described in \citet{massaro16,pena19}.
{  In summary, we} first searched for all UGSs in the footprint of the SDSS DR15. {We then} selected UGSs having at least one source, with an available optical spectrum, lying within their $\gamma$-ray positional uncertainty region. Then, we visually inspected all these optical sources selecting those with a typical BL Lac spectrum (i.e., featureless or with emission/absorption lines {with EW $<5$ $\AA$).}

We remark that for all UGSs analyzed using the SDSS spectra, we carried out our investigation and provided {the classifications} directly to the \textit{Fermi}-LAT collaboration. Thus they appeared {originally} as UGS only in the preliminary version of the 4FGL used to select our targets. For the following we will keep considering them as UGSs, however in Table~\ref{tab:sample} we report {their new classifications in the final, published 4FGL in parenthesis}. Basically they are listed as 2 BCUs, 5 BL Lacs and 1 UGS. These eight sources appear {to all be BZBs with no redshift estimates due to their featureless spectra with the exception of 7C 1032+4424 (4FGL J1035.6+4409)} at $z=$0.444, that is classified as {a} blazar of galaxy type (BZG) according to the criteria outlined in the Roma-BZCAT \citep{massaro09,massaro15b}.

\begin{figure*}{}
\begin{center}$
\begin{array}{cc}
\includegraphics[width=\mywidth]{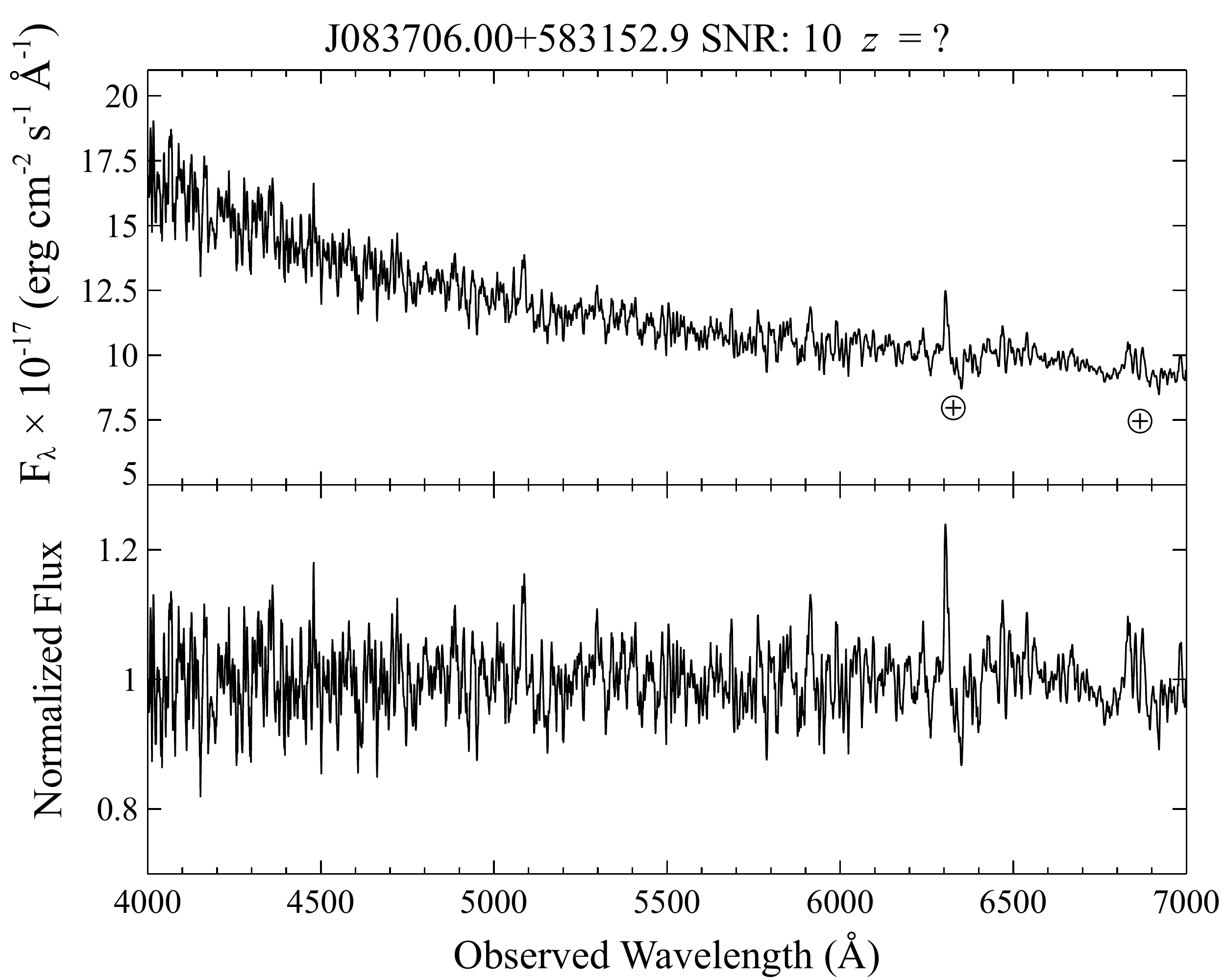} &
\includegraphics[clip=true, width=7cm]{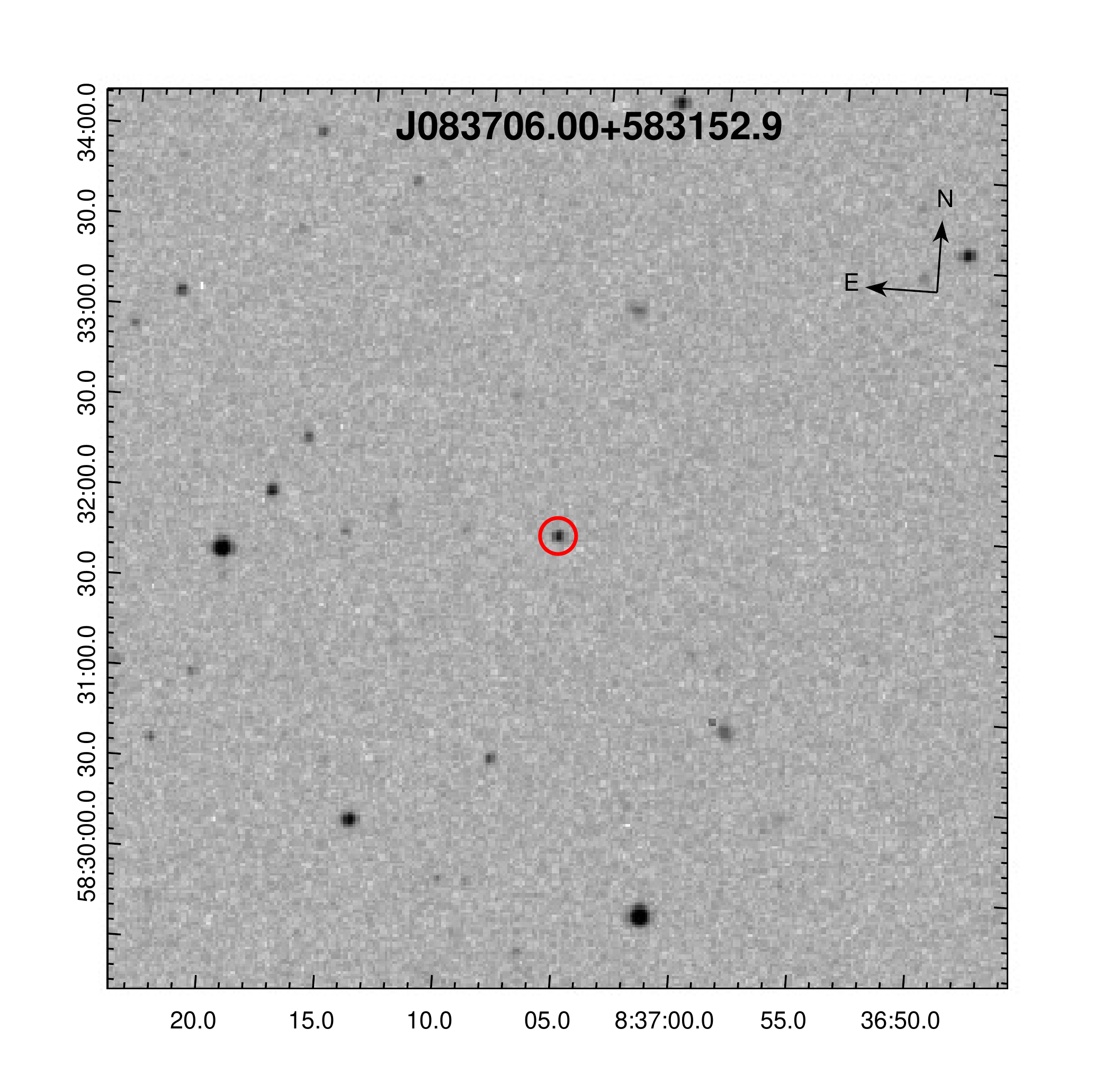} \\
J083706.00+583152.9-eps-converted-to
\end{array}$
\end{center}
\caption{(Left panels) Optical spectrum of WISE J083706.00+583152.9, the potential counterpart of 4FGL J0836.9+5833. The Signal-to-Noise Ratio (SNR) of the spectrum is indicated along the top. 
(Right panel) The finding chart ( $5'\times 5'$ ) retrieved from the Digitized Sky Survey (DSS) highlighting the location of the counterpart (red circle).}
\label{fig:J0837}
\end{figure*}

\begin{figure*}{}
\begin{center}$
\begin{array}{cc}
\includegraphics[width=\mywidth]{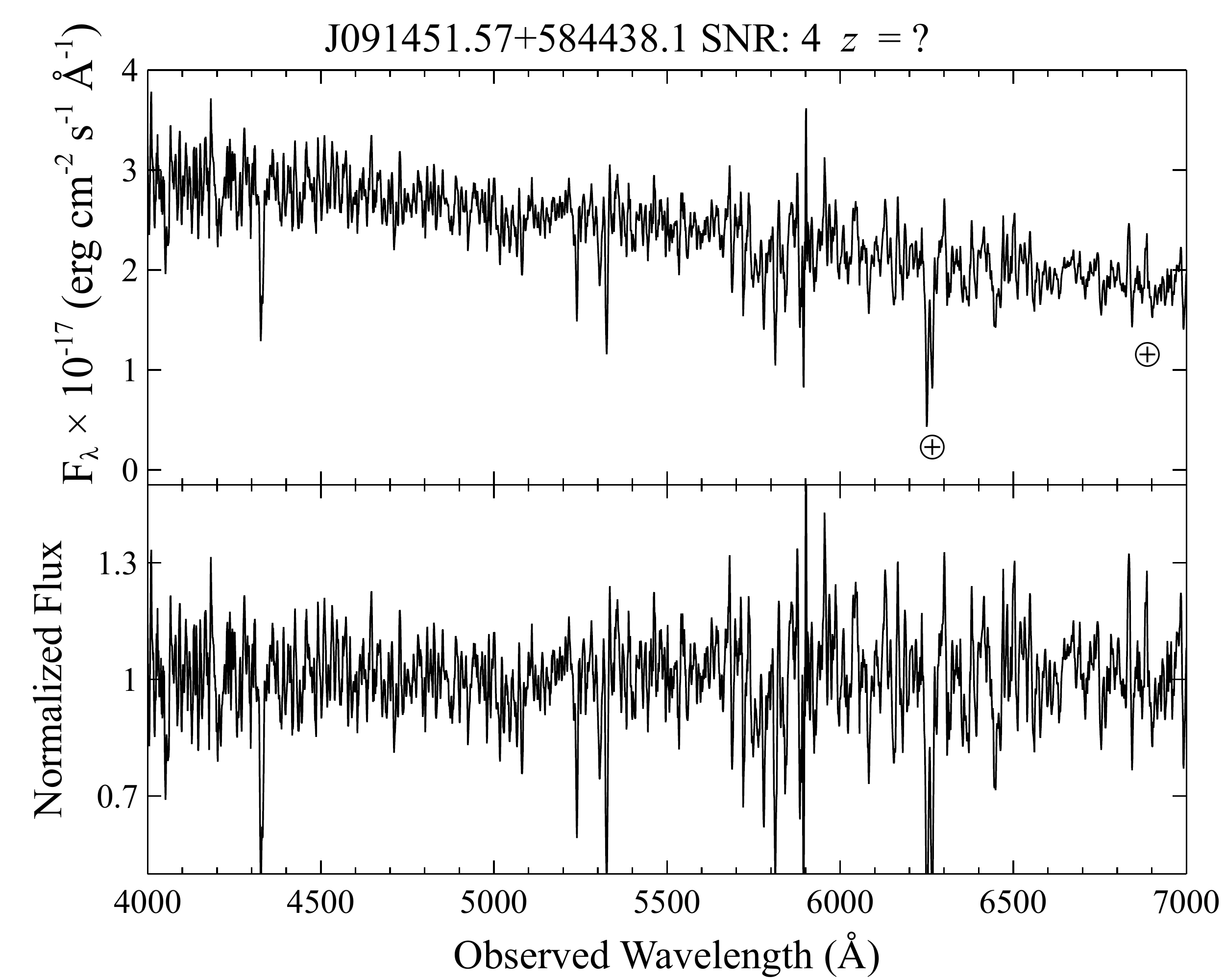} &
\includegraphics[clip=true, width=7cm]{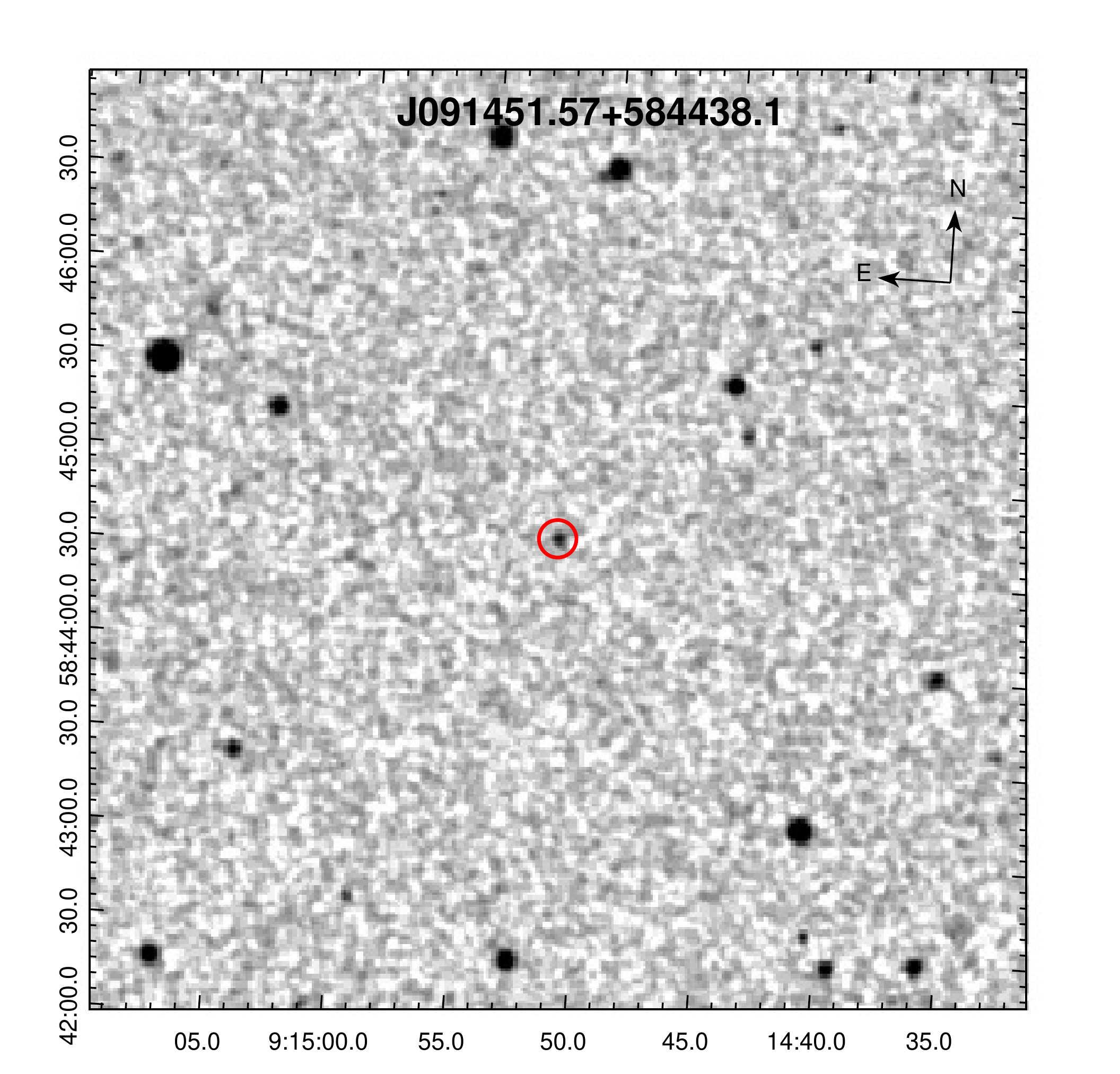} \\
\end{array}$
\end{center}
\caption{ 
As in Figure~\ref{fig:J0837} but for WISE J091451.57+584438.1, the potential counterpart of 4FGL J0914.8+5846.
}
\label{fig:J0914}
\end{figure*}

\begin{figure*}{}
\begin{center}$
\begin{array}{cc}
\includegraphics[width=\mywidth]{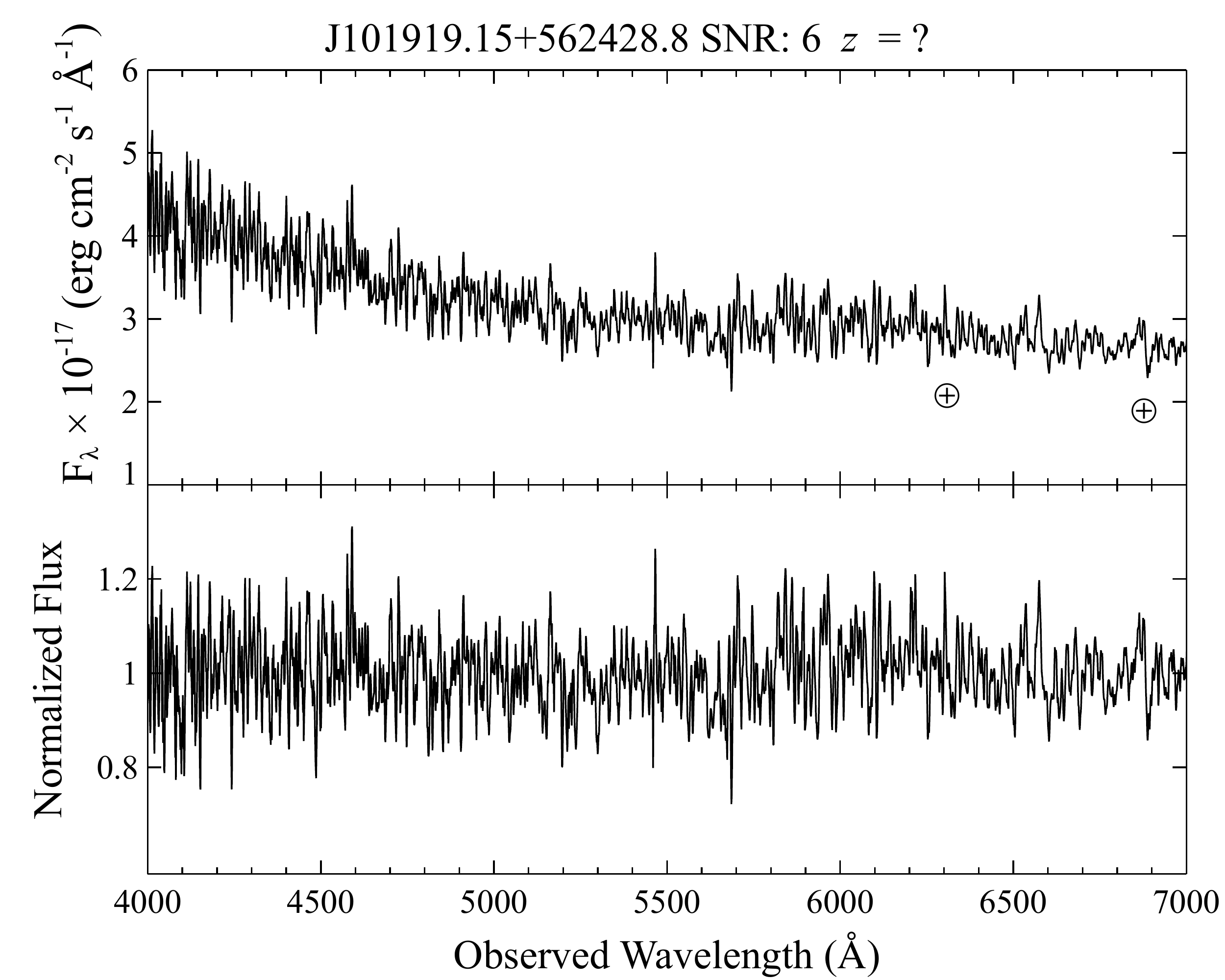} &
\includegraphics[clip=true, width=7cm]{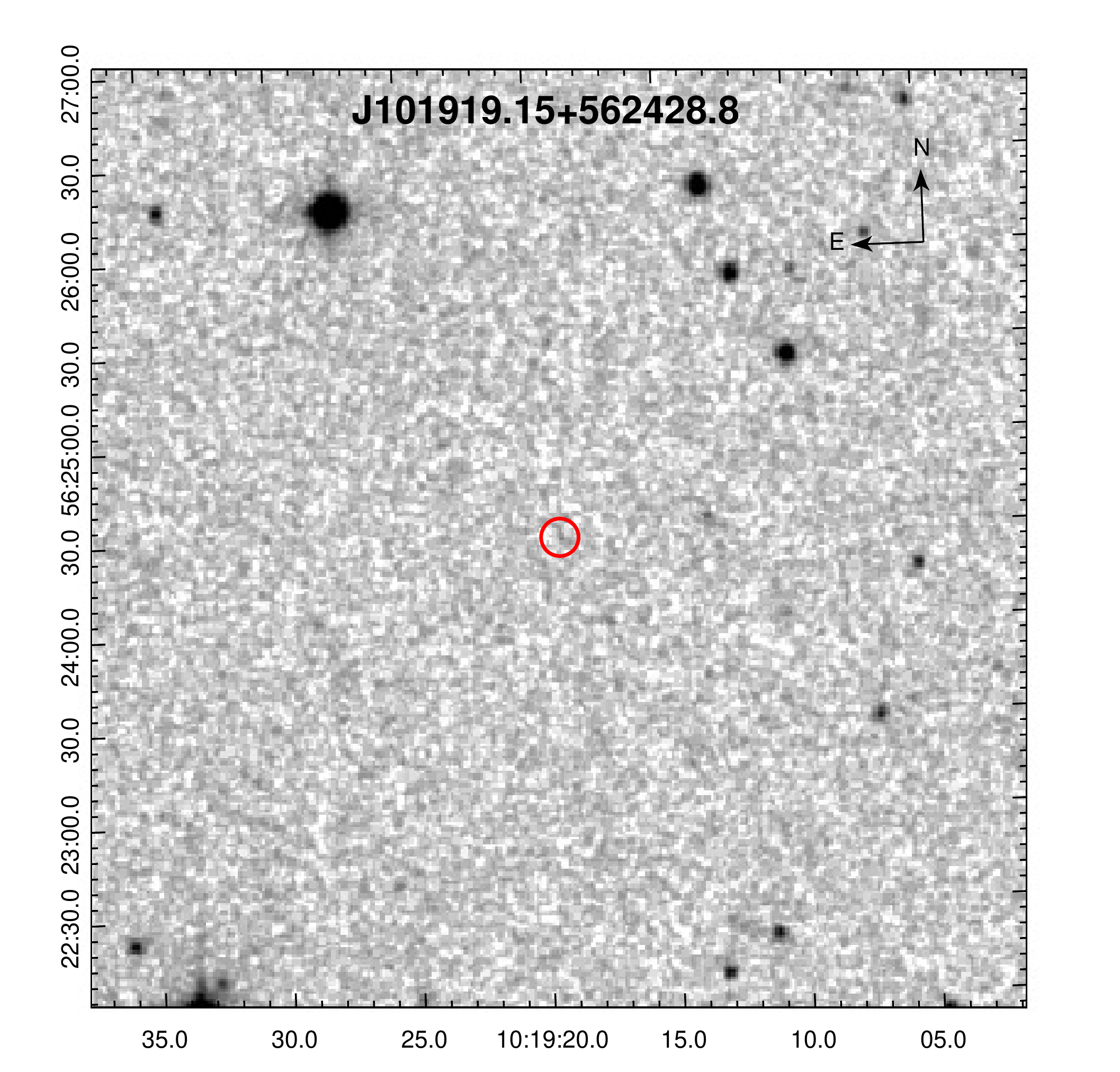} \\
\end{array}$
\end{center}
\caption{
As in Figure~\ref{fig:J0837} but for WISE J101919.15+562428.8, the potential counterpart of 4FGL J1019.3+5625.}
\label{fig:J1019}
\end{figure*}

\begin{figure*}{}
\begin{center}$
\begin{array}{cc}
\includegraphics[width=\mywidth]{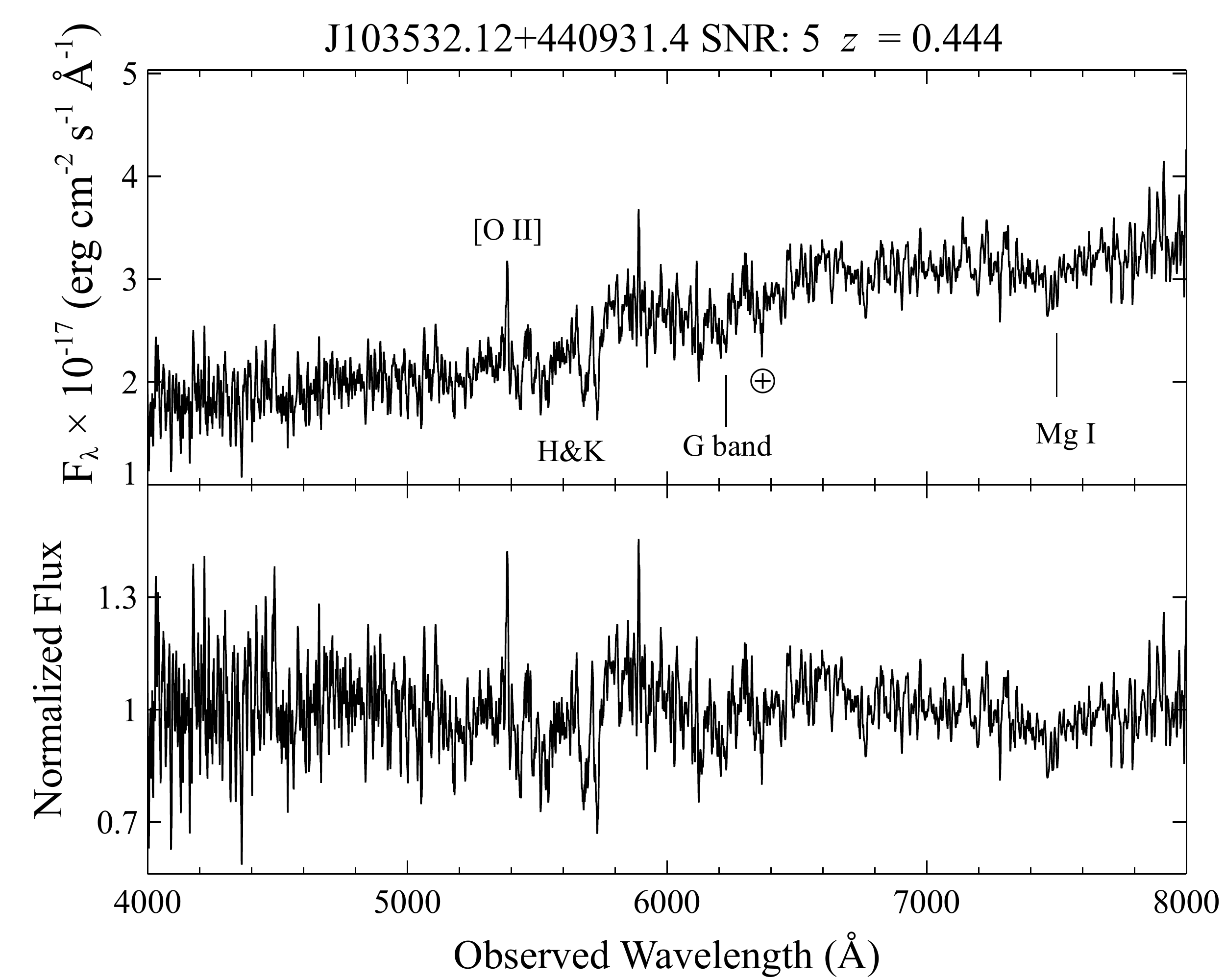} &
\includegraphics[clip=true, width=7cm]{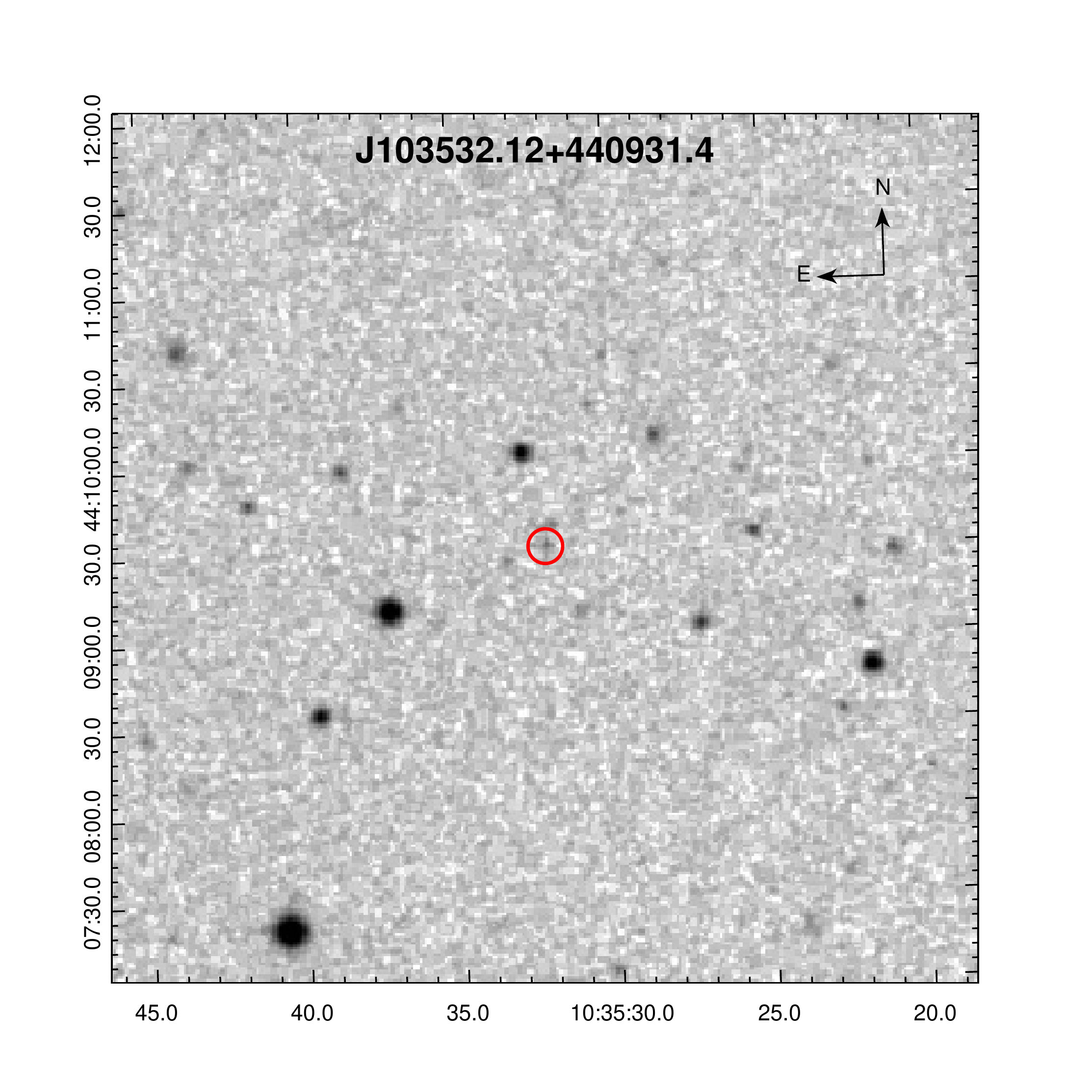} \\
\end{array}$
\end{center}
\caption{As in Figure~\ref{fig:J0837} but for WISE J103532.12+440931.4, the potential counterpart of 4FGL J1035.6+4409.}
\label{fig:J1035}
\end{figure*}

\begin{figure*}{}
\begin{center}$
\begin{array}{cc}
\includegraphics[width=\mywidth]{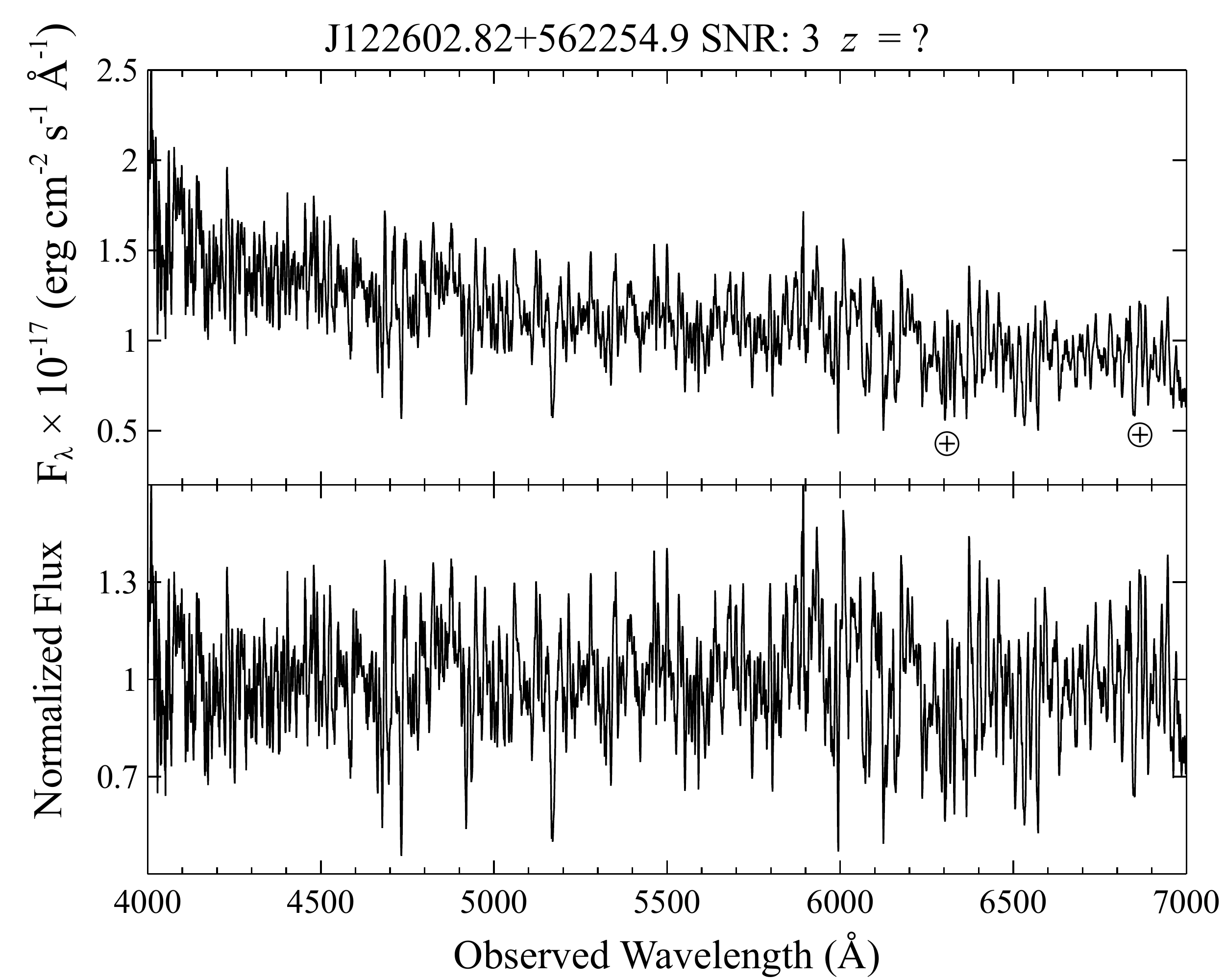} &
\includegraphics[clip=true, width=7cm]{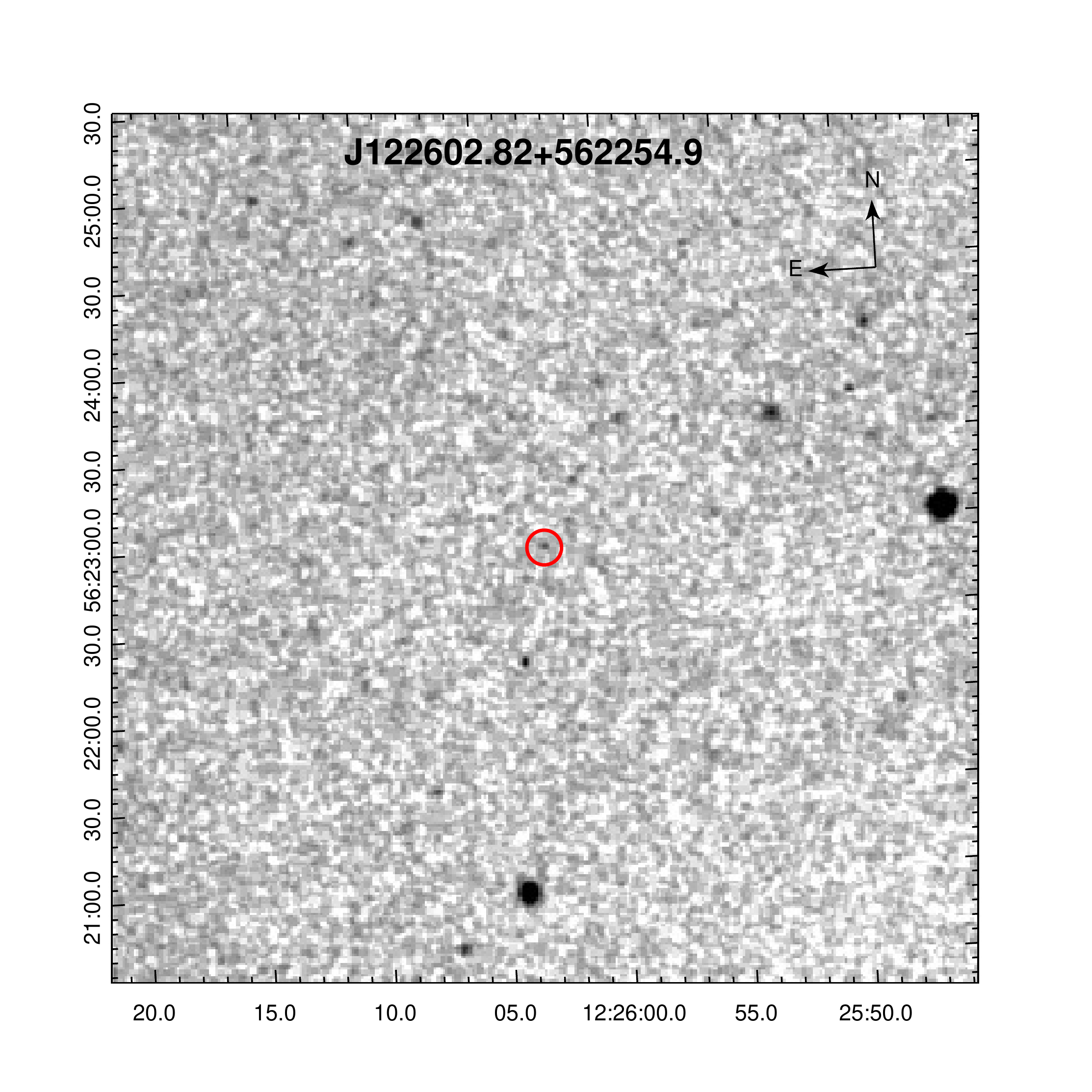} \\
\end{array}$
\end{center}
\caption{As in Figure~\ref{fig:J0837} but for WISE J122602.82+562254.9, the potential counterpart of 4FGL J1226.0+5622.}
\label{fig:J1226}
\end{figure*}

\begin{figure*}{}
\begin{center}$
\begin{array}{cc}
\includegraphics[width=\mywidth]{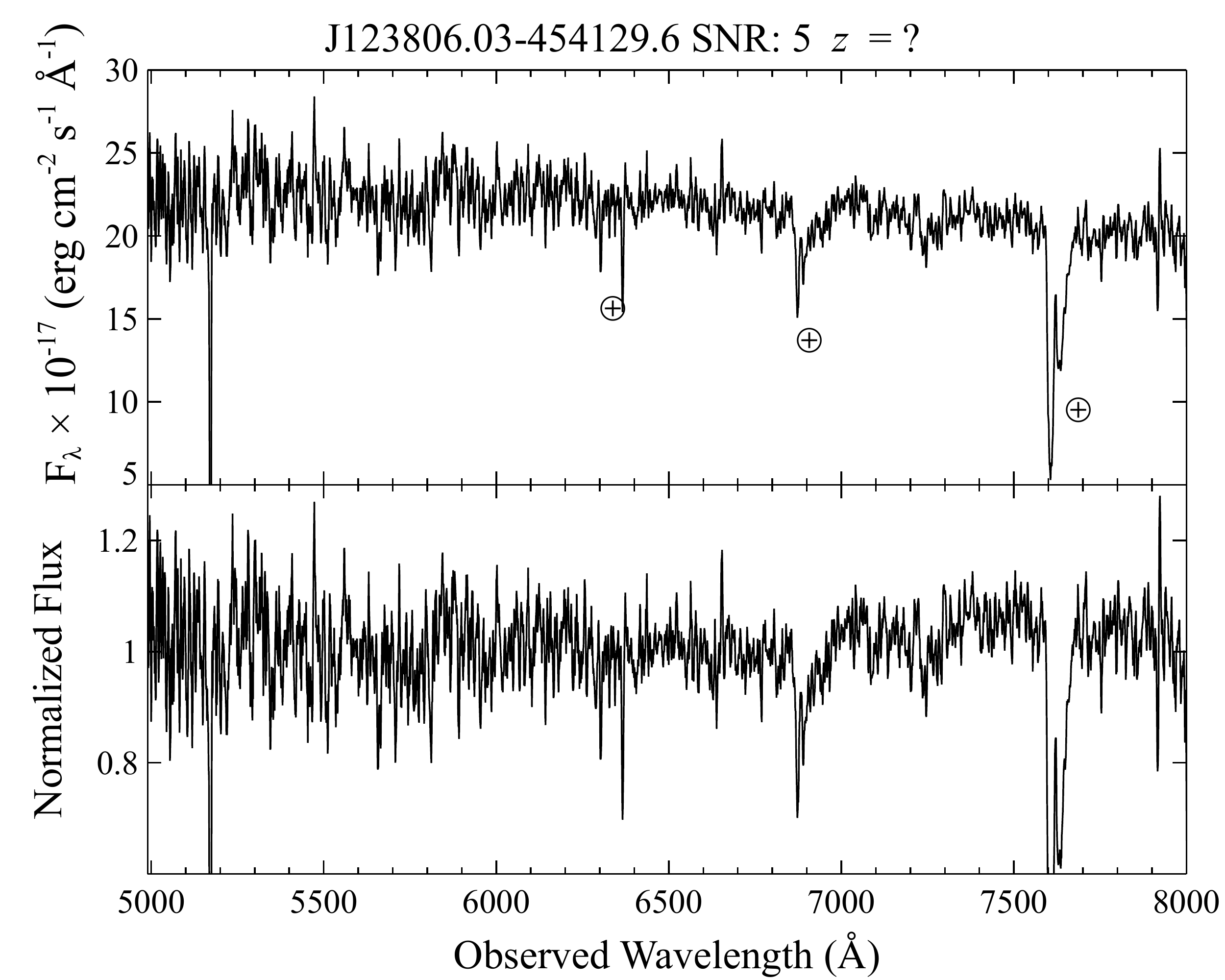} &
\includegraphics[clip=true, width=7cm]{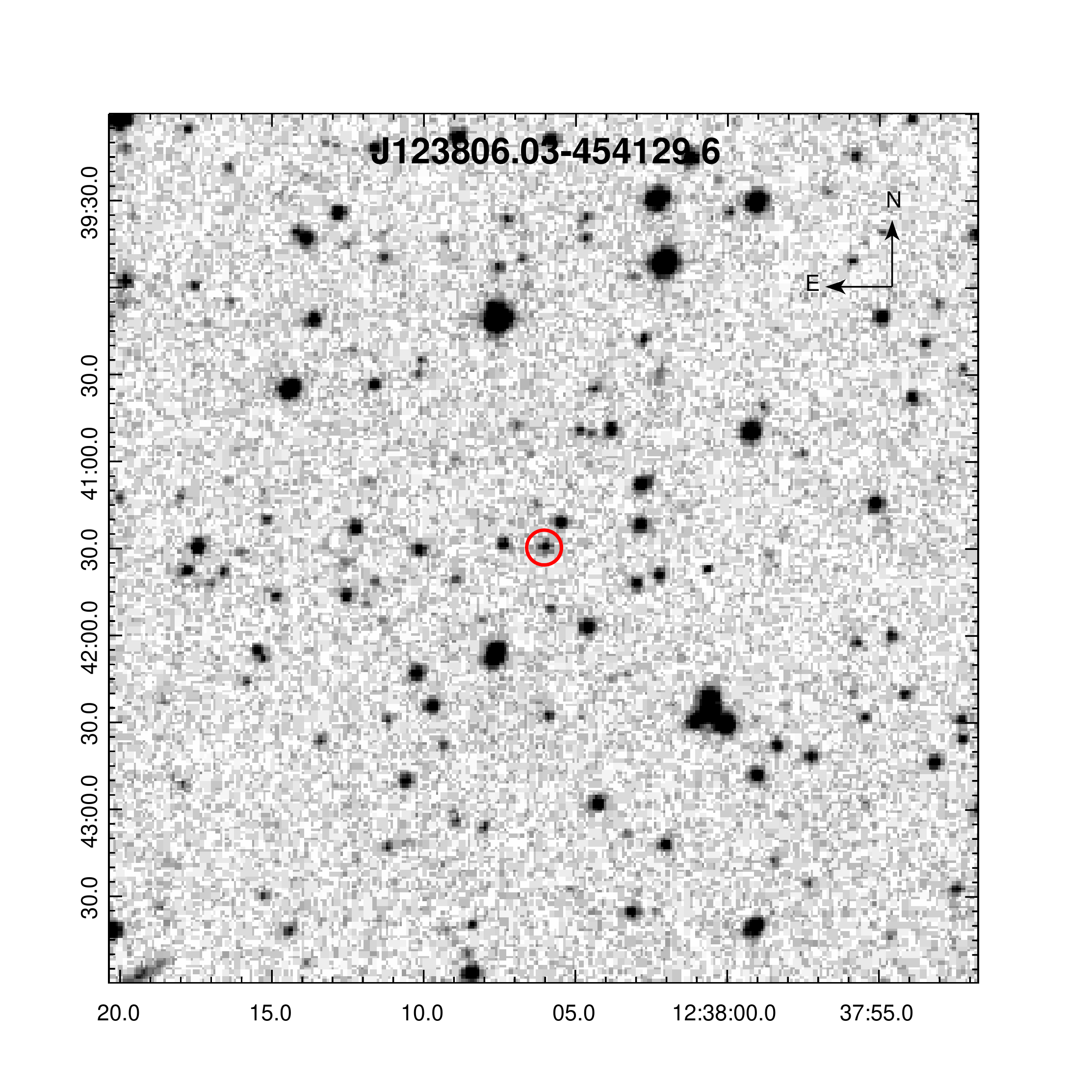} \\
\end{array}$
\end{center}
\caption{As in Figure~\ref{fig:J0837} but for WISE J123806.03-454129.6, the counterpart of 4FGL J1238.1-4541.}
\label{fig:J1238}
\end{figure*}

\begin{figure*}{}
\begin{center}$
\begin{array}{cc}
\includegraphics[width=\mywidth]{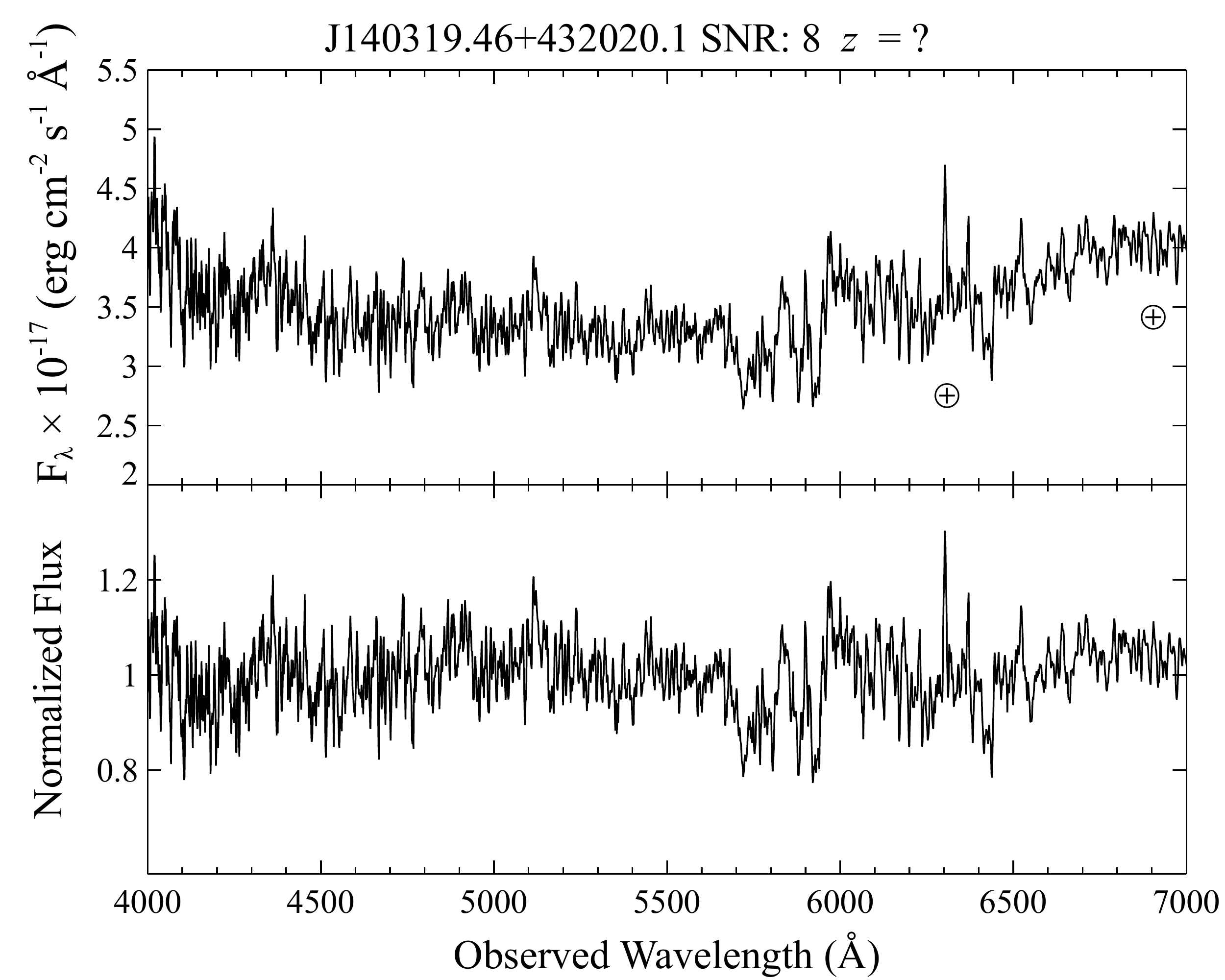} &
\includegraphics[clip=true, width=7cm]{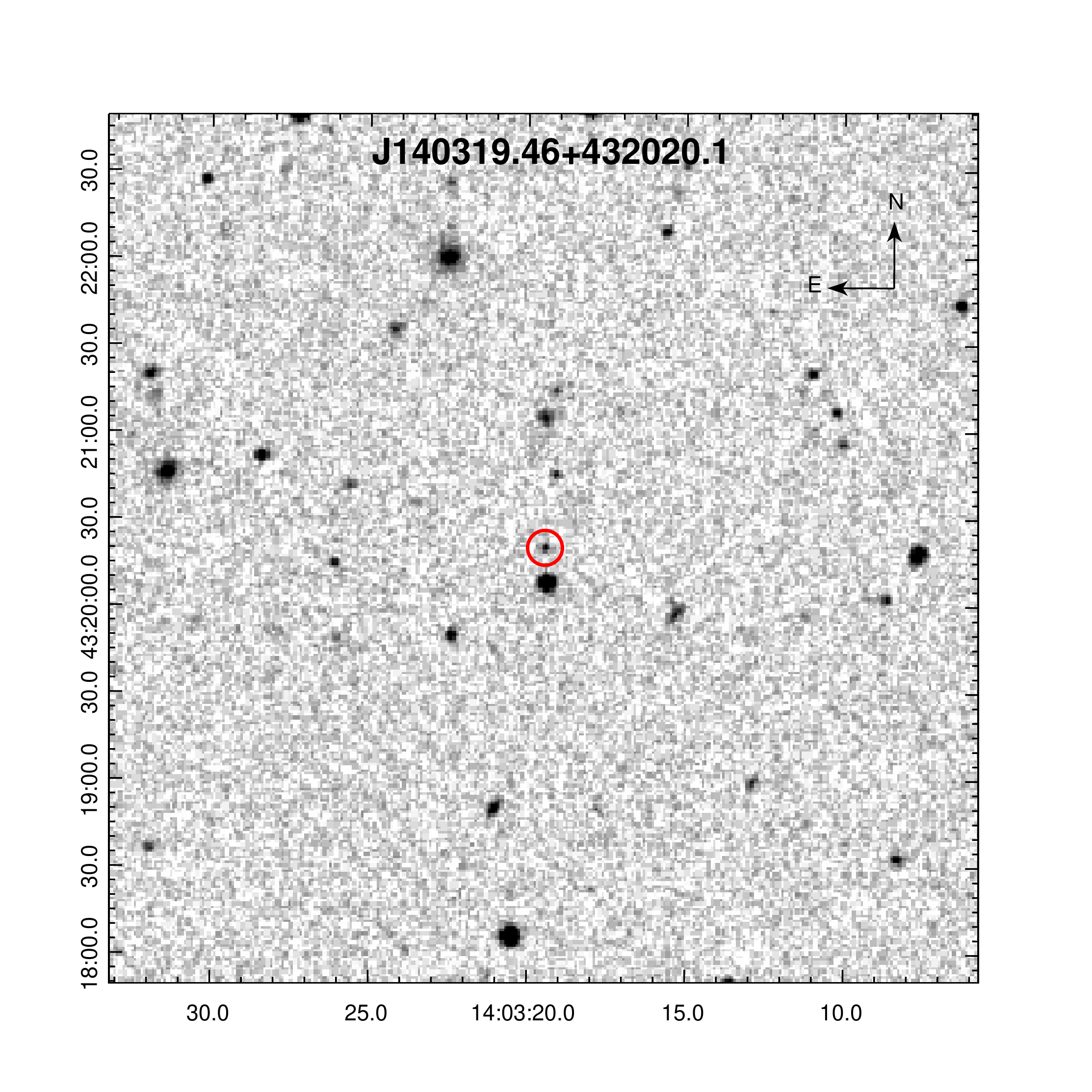} \\
\end{array}$
\end{center}
\caption{As in Figure~\ref{fig:J0837} but for WISE J140319.46+432020.1, the potential counterpart of 4FGL J1403.4+4319.}
\label{fig:J1403}
\end{figure*}

\begin{figure*}{}
\begin{center}$
\begin{array}{cc}
\includegraphics[width=\mywidth]{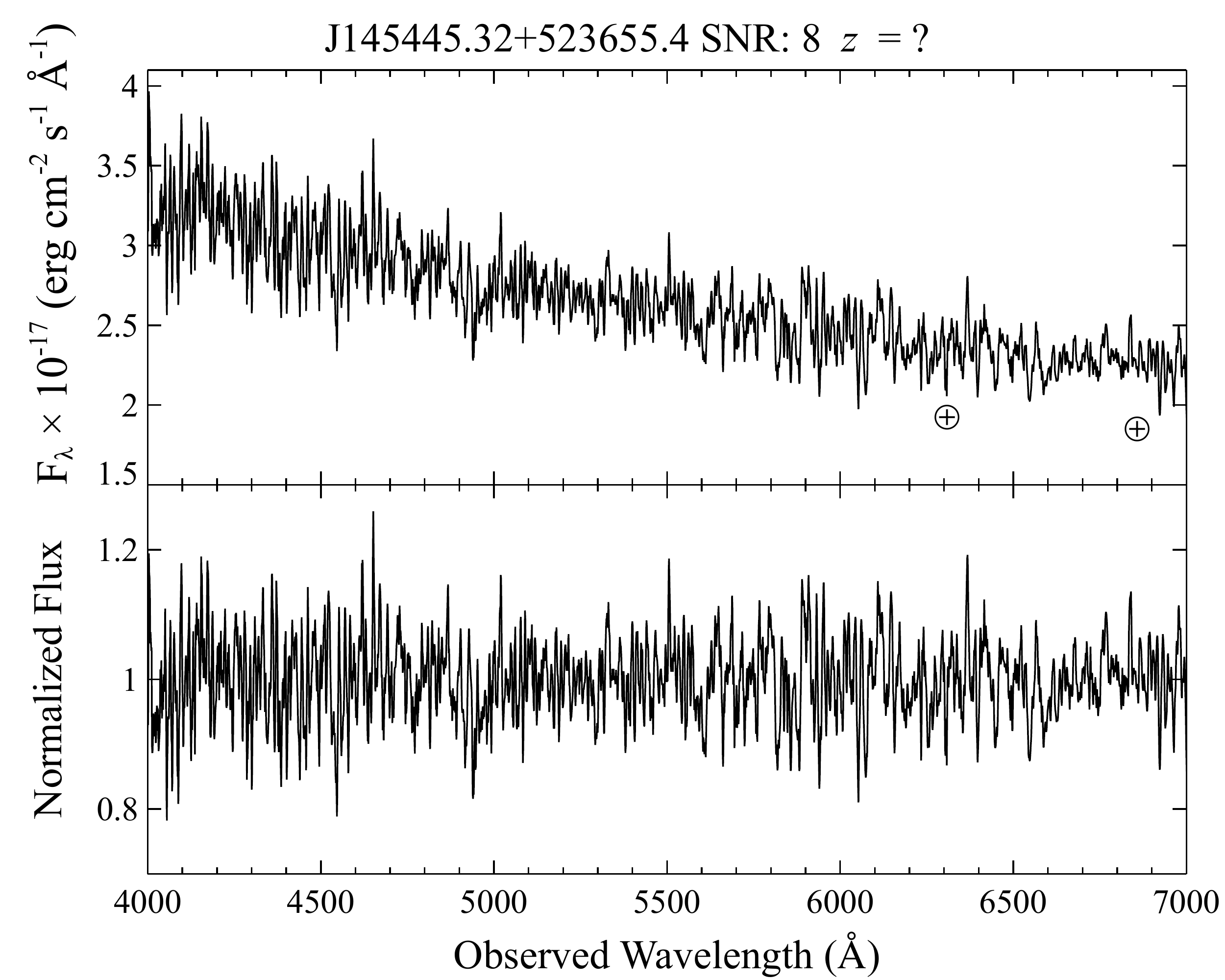} &
\includegraphics[clip=true, width=7cm]{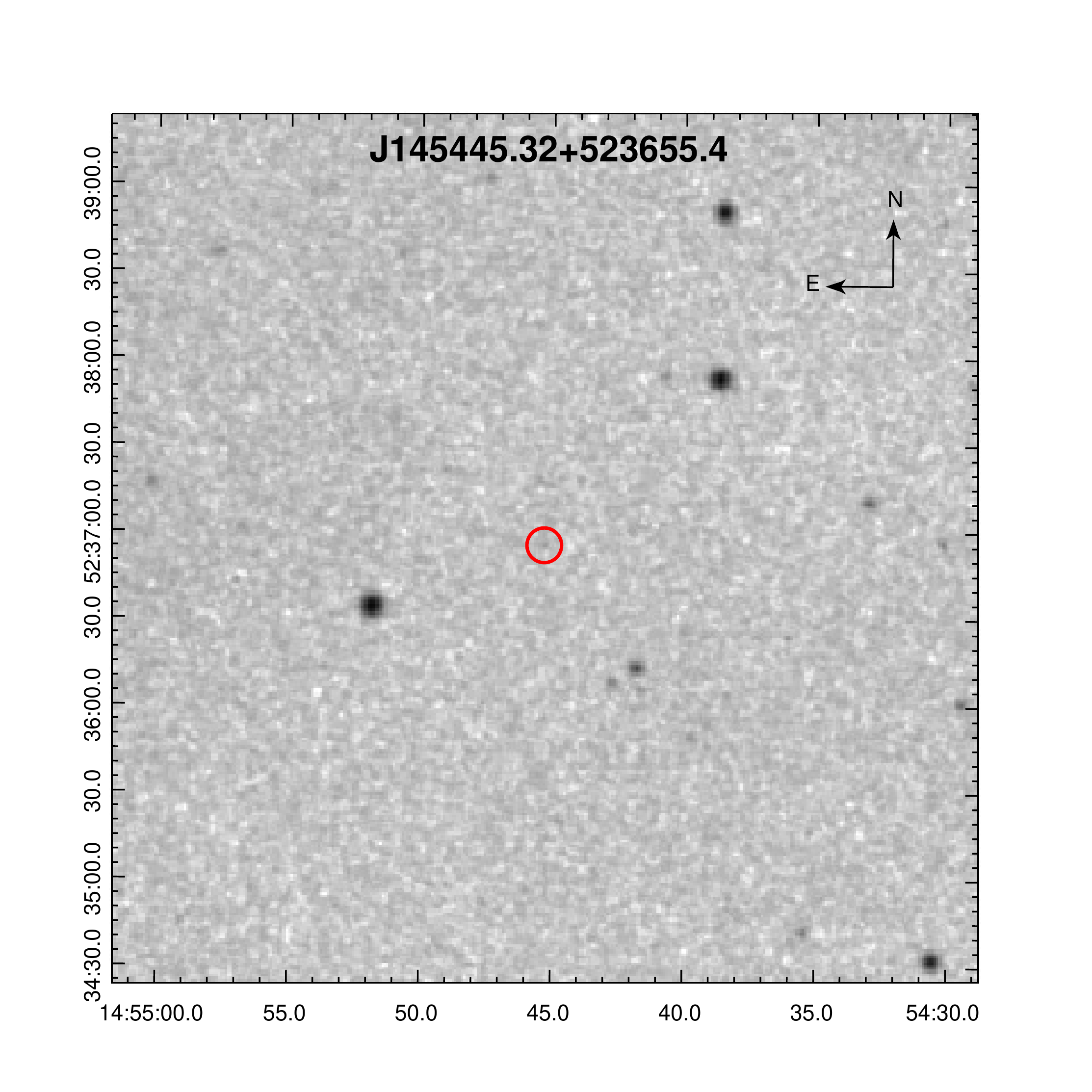} \\
\end{array}$
\end{center}
\caption{As in Figure~\ref{fig:J0837} but for WISE J145445.32+523655.4, the potential counterpart of 4FGL J1454.7+5237.}
\label{fig:J1454}
\end{figure*}

\begin{figure*}{}
\begin{center}$
\begin{array}{cc}
\includegraphics[width=\mywidth]{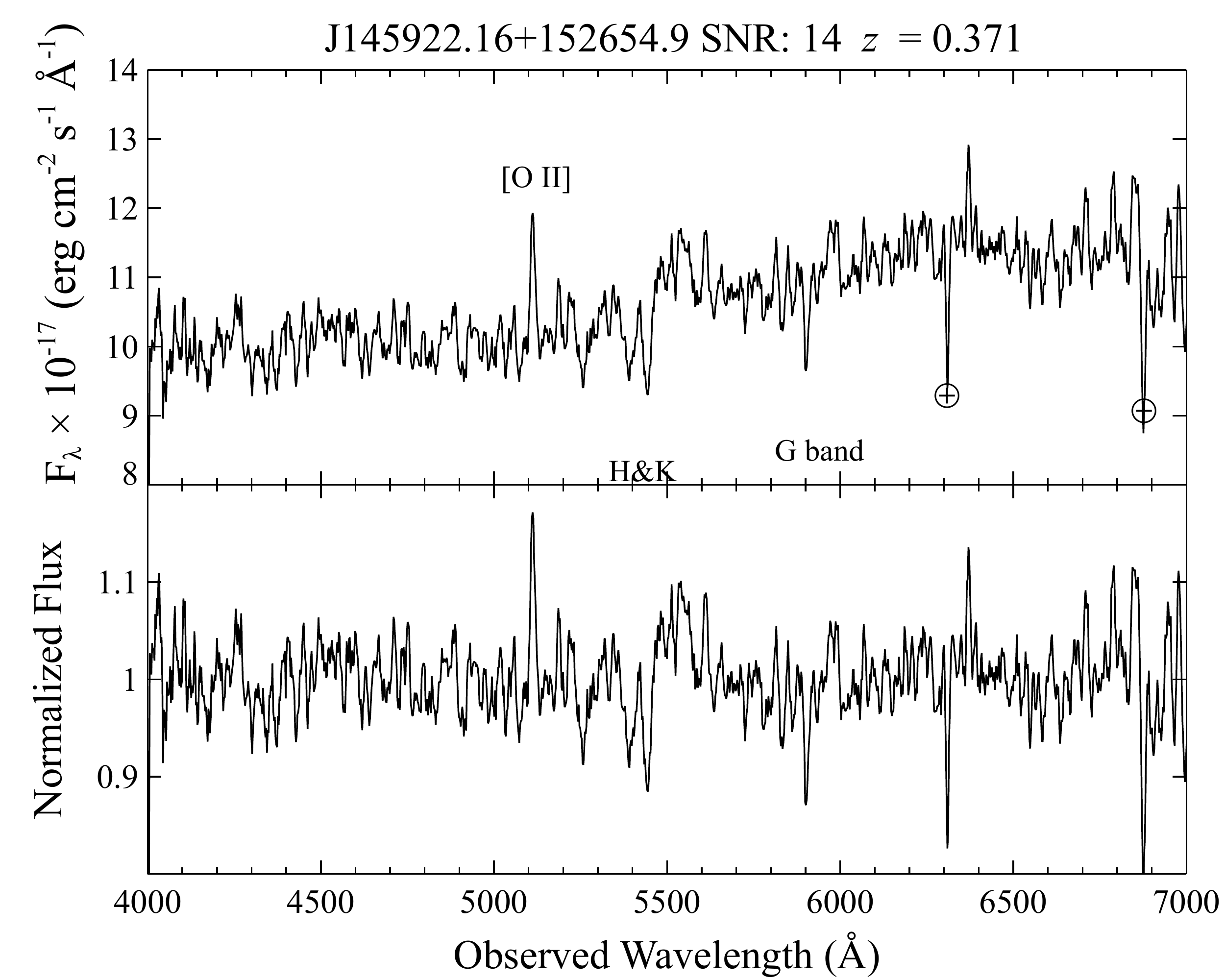} &
\includegraphics[clip=true, width=7cm]{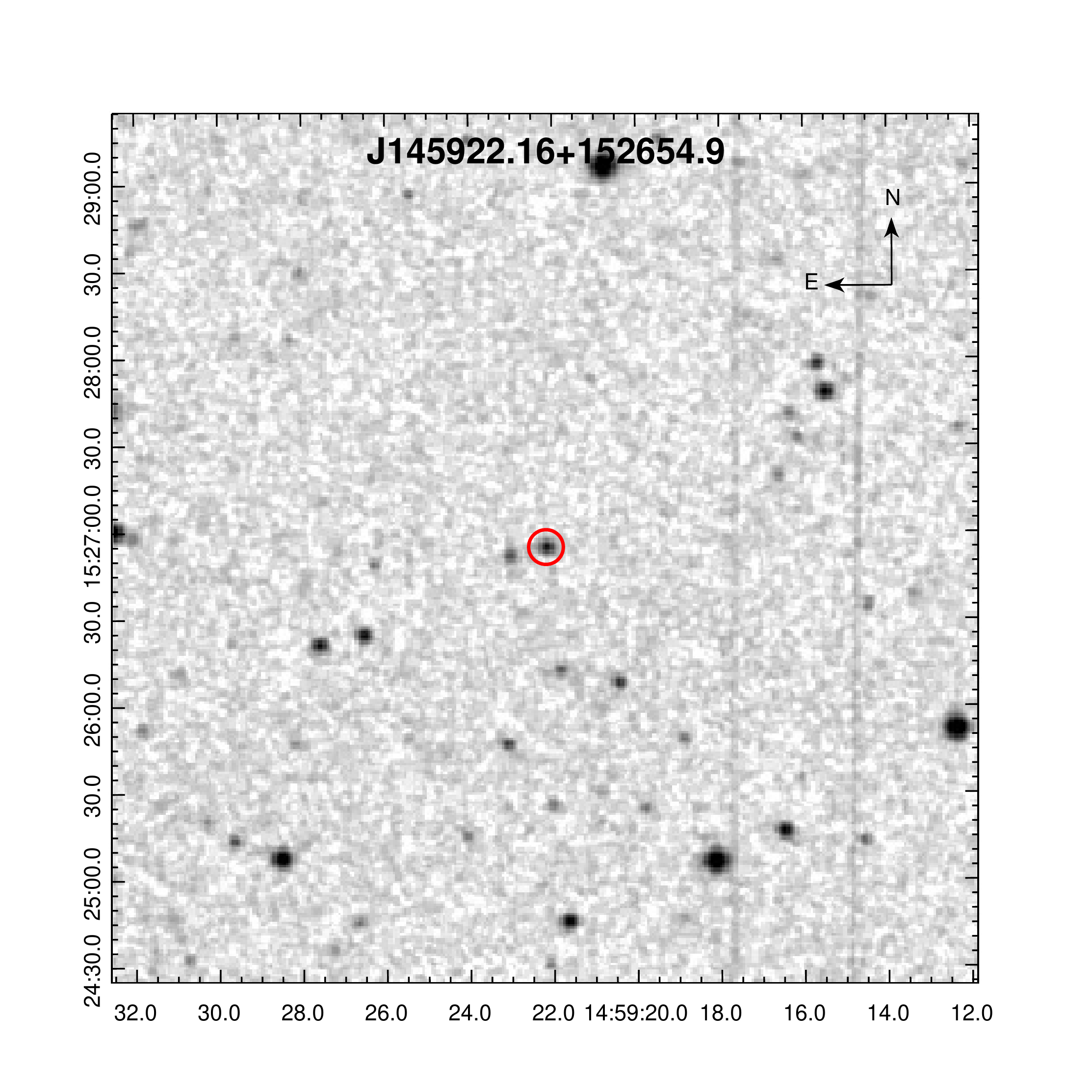} \\
\end{array}$
\end{center}
\caption{As in Figure~\ref{fig:J0837} but for WISE J145922.16+152654.9, the counterpart of 4FGL J1459.5+1527.}
\label{fig:J1459}
\end{figure*}

\begin{figure*}{}
\begin{center}$
\begin{array}{cc}
\includegraphics[width=\mywidth]{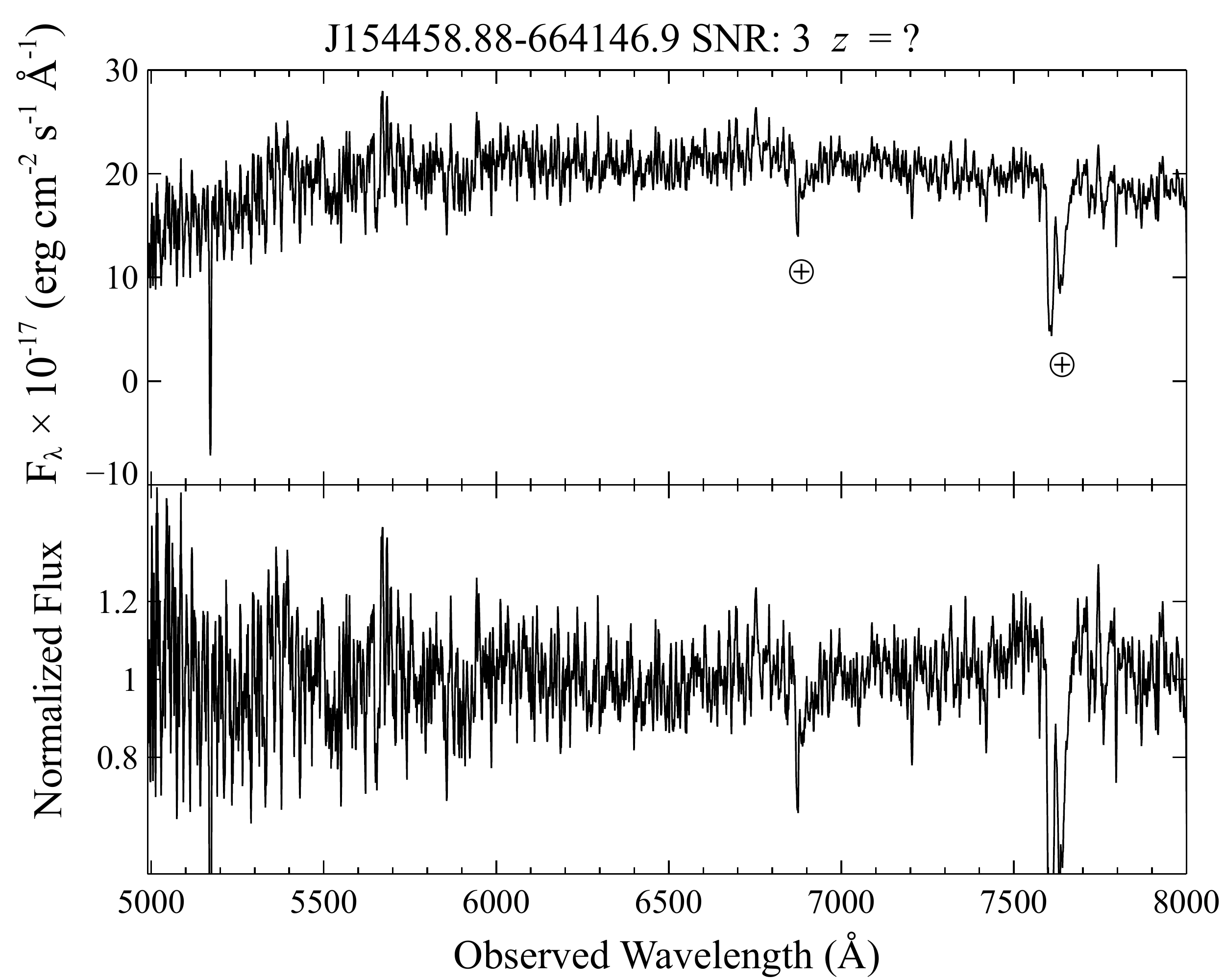} &
\includegraphics[clip=true, width=7cm]{{J145922.16+152654.9}.pdf} \\
\end{array}$
\end{center}
\caption{As in Figure~\ref{fig:J0837} but for J154458.88-664146.9, the potential counterpart of 3FGL J1545.0-6641.}
\label{fig:J1544}
\end{figure*}

\begin{figure*}{}
\begin{center}$
\begin{array}{cc}
\includegraphics[width=\mywidth]{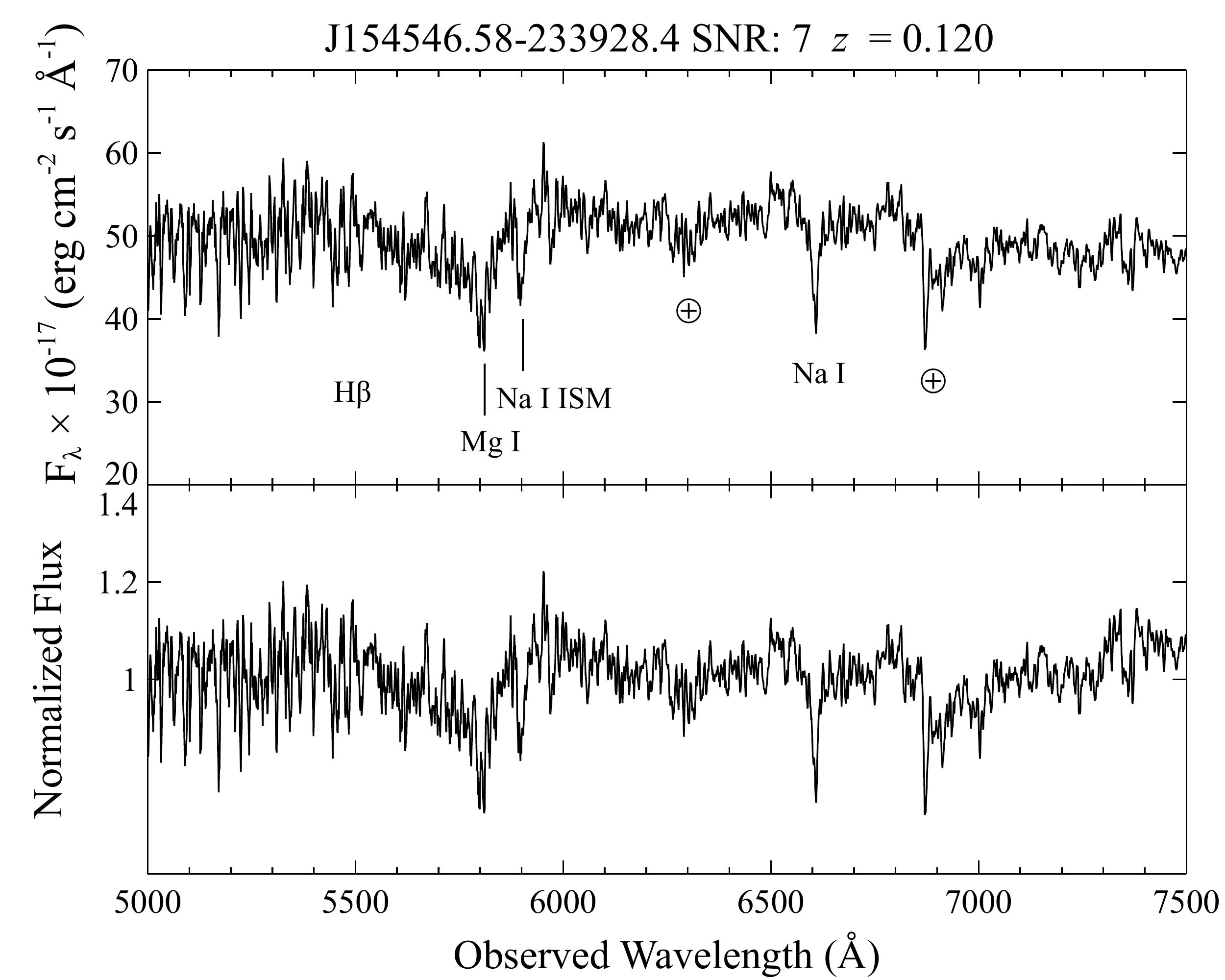} &
\includegraphics[clip=true, width=7cm]{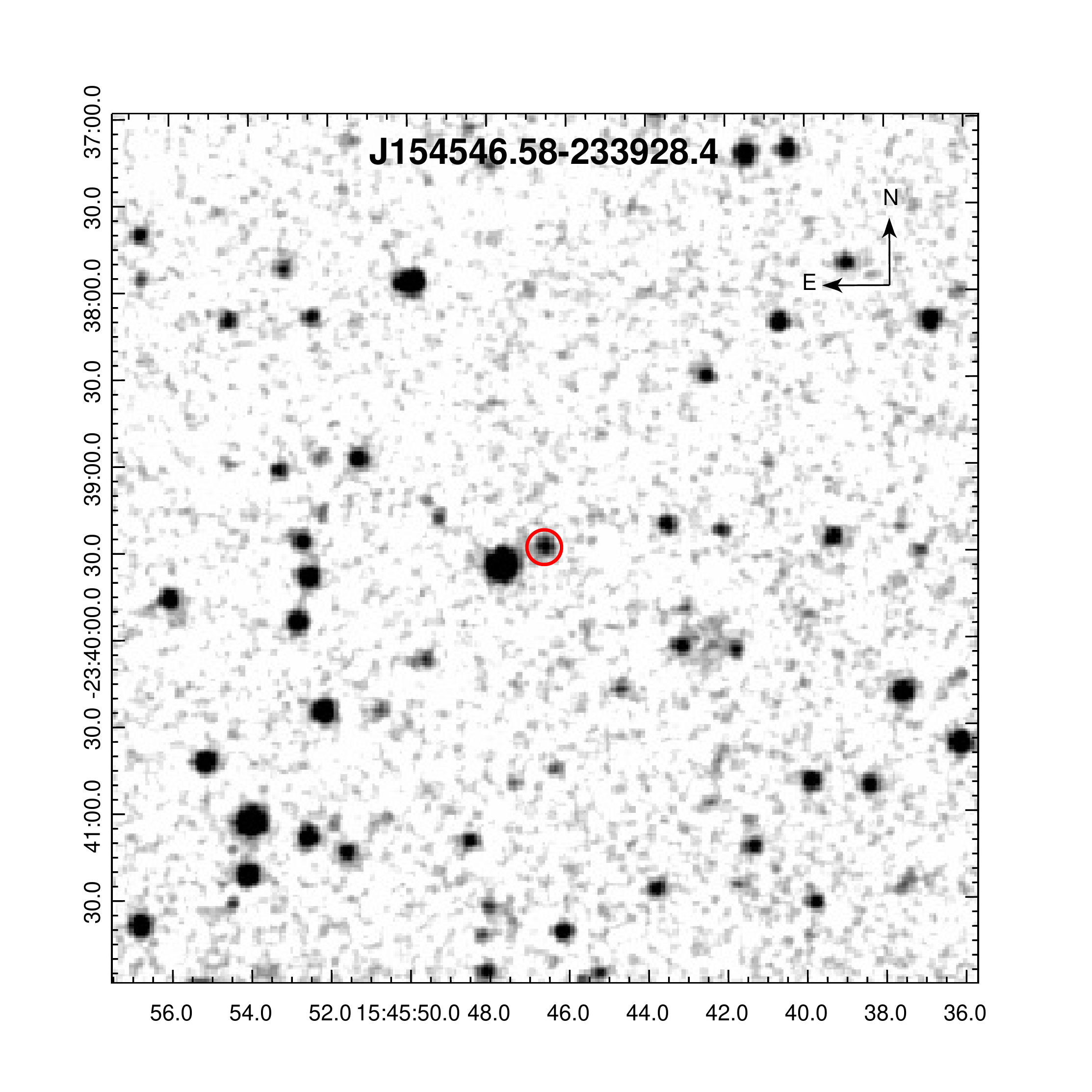} \\
\end{array}$
\end{center}
\caption{As in Figure~\ref{fig:J0837} but for WISE J154546.58-233928.4, the counterpart of 4FGL J1545.8-2336.}
\label{fig:J1545}
\end{figure*}

\begin{figure*}{}
\begin{center}$
\begin{array}{cc}
\includegraphics[width=\mywidth]{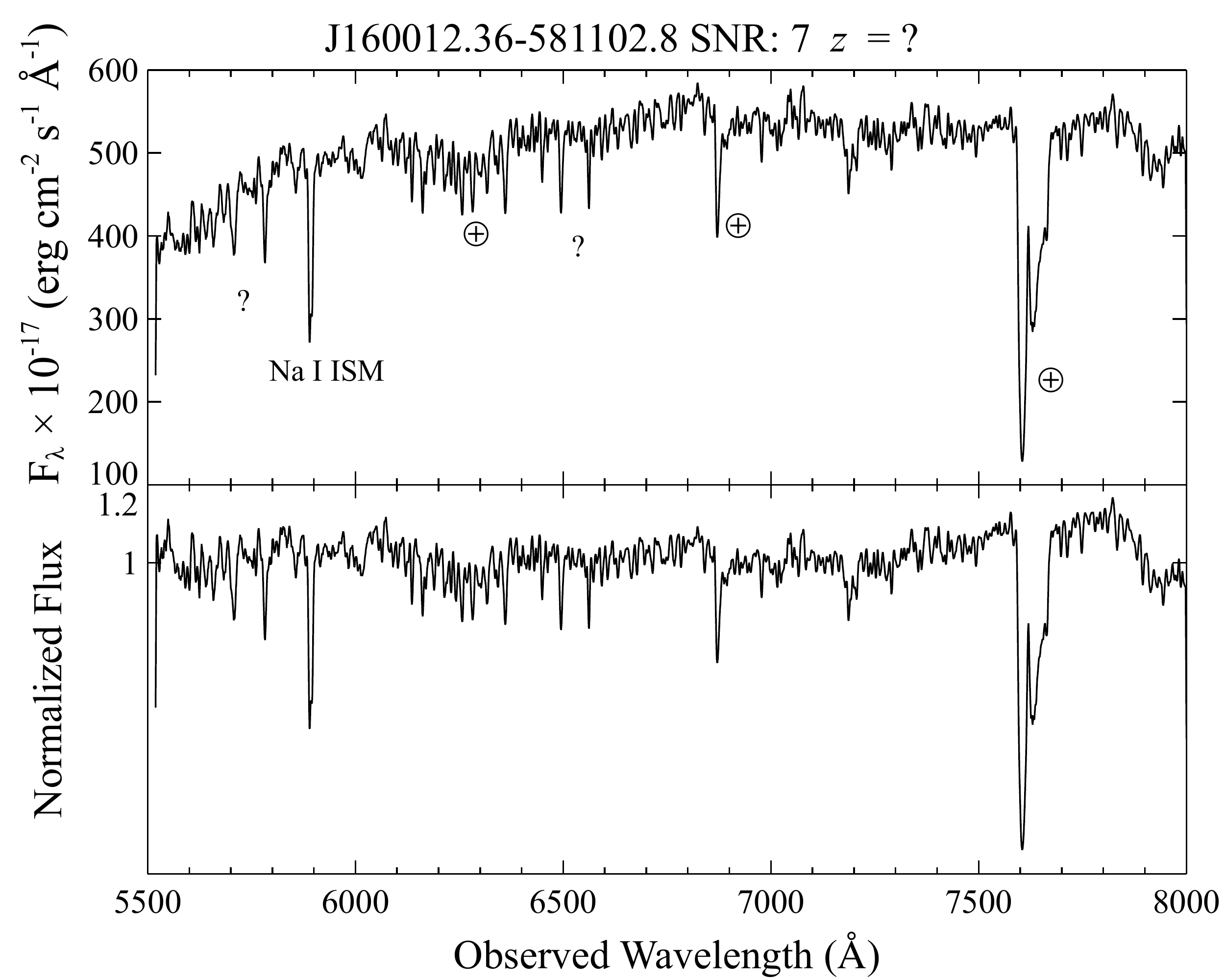} &
\includegraphics[clip=true, width=7cm]{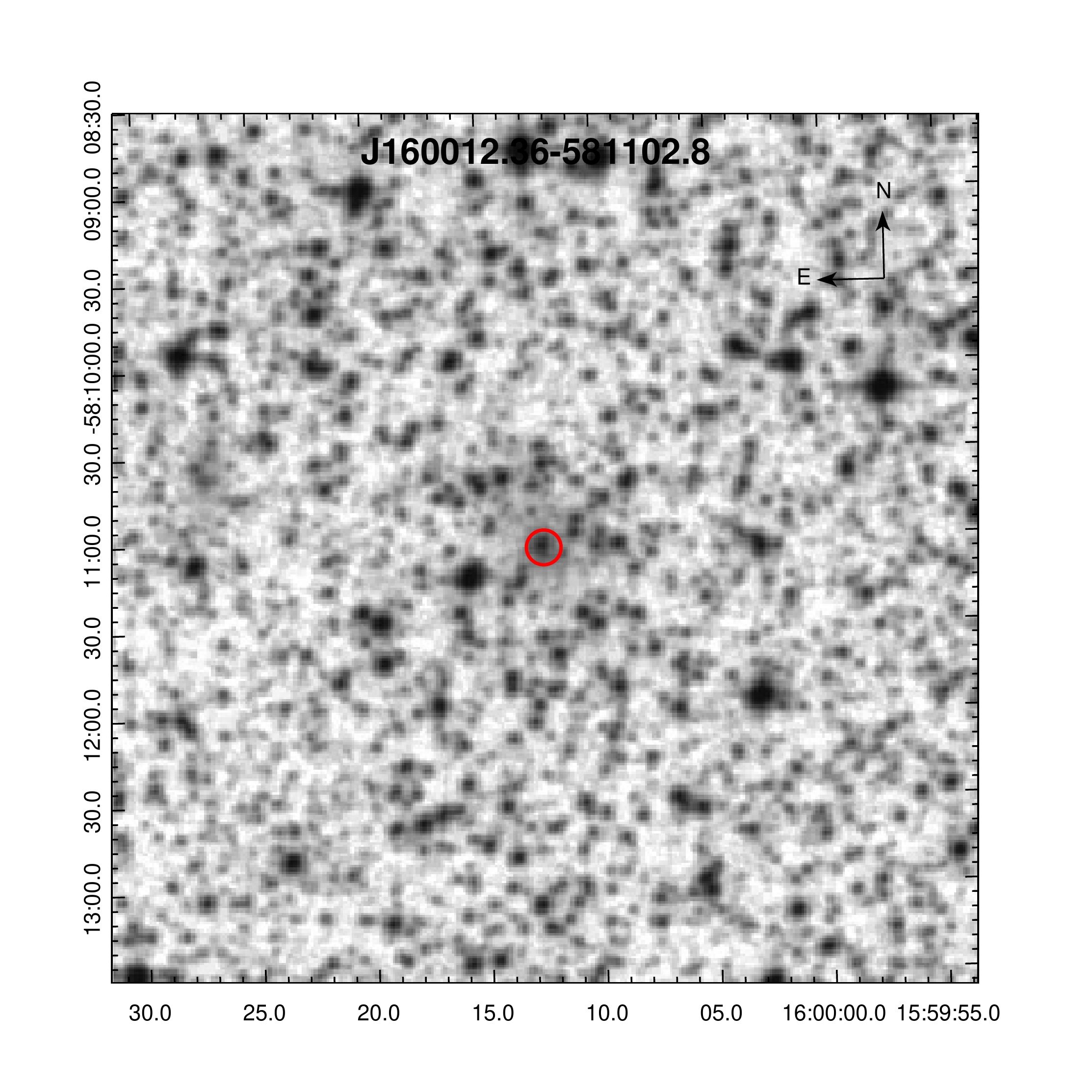} \\
\end{array}$
\end{center}
\caption{As in Figure~\ref{fig:J0837} but for WISE J160012.36-581102.8, the counterpart of 4FGL J1600.3-5811.}
\label{fig:J1600}
\end{figure*}

\begin{figure*}{}
\begin{center}$
\begin{array}{cc}
\includegraphics[width=\mywidth]{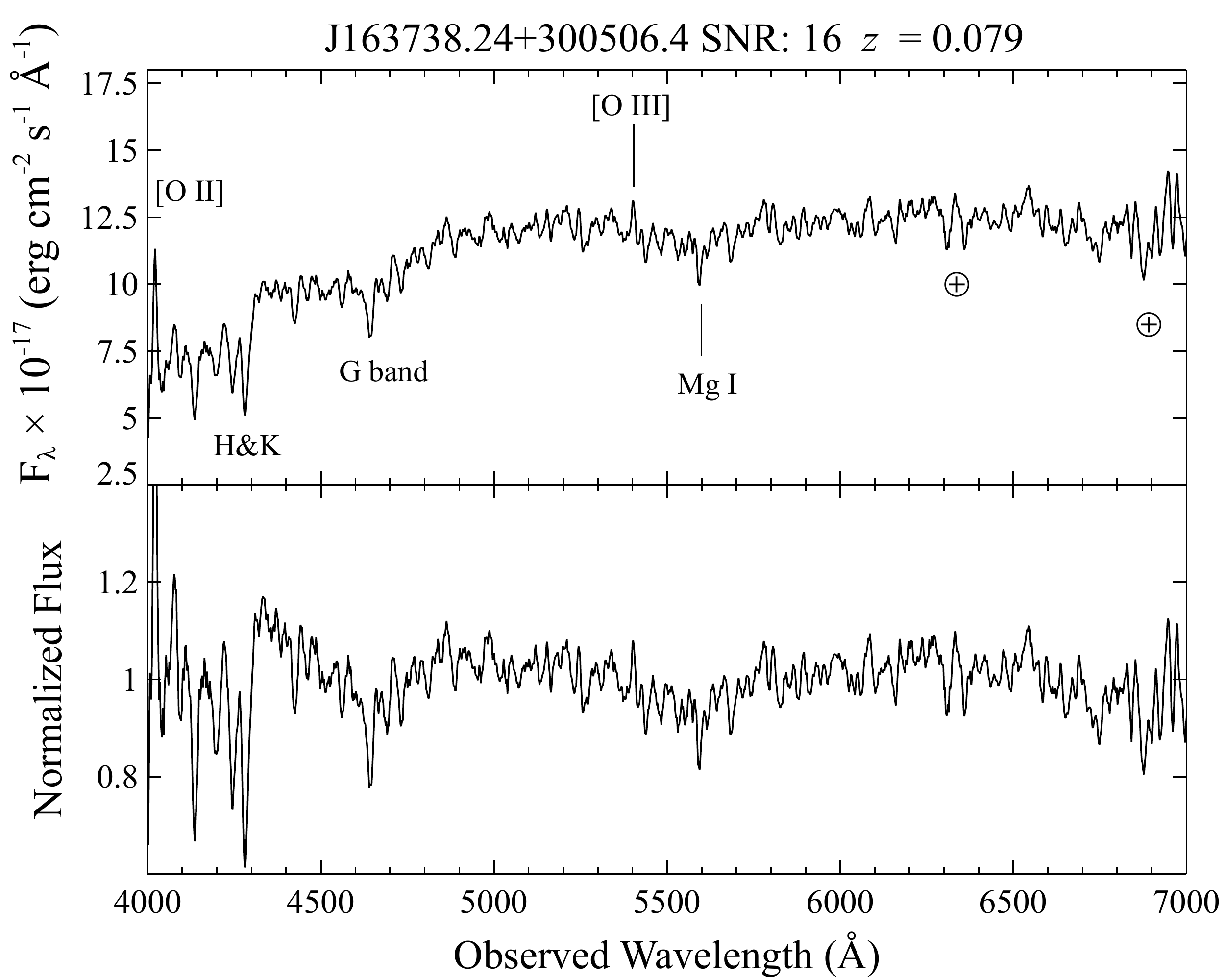} &
\includegraphics[clip=true, width=7cm]{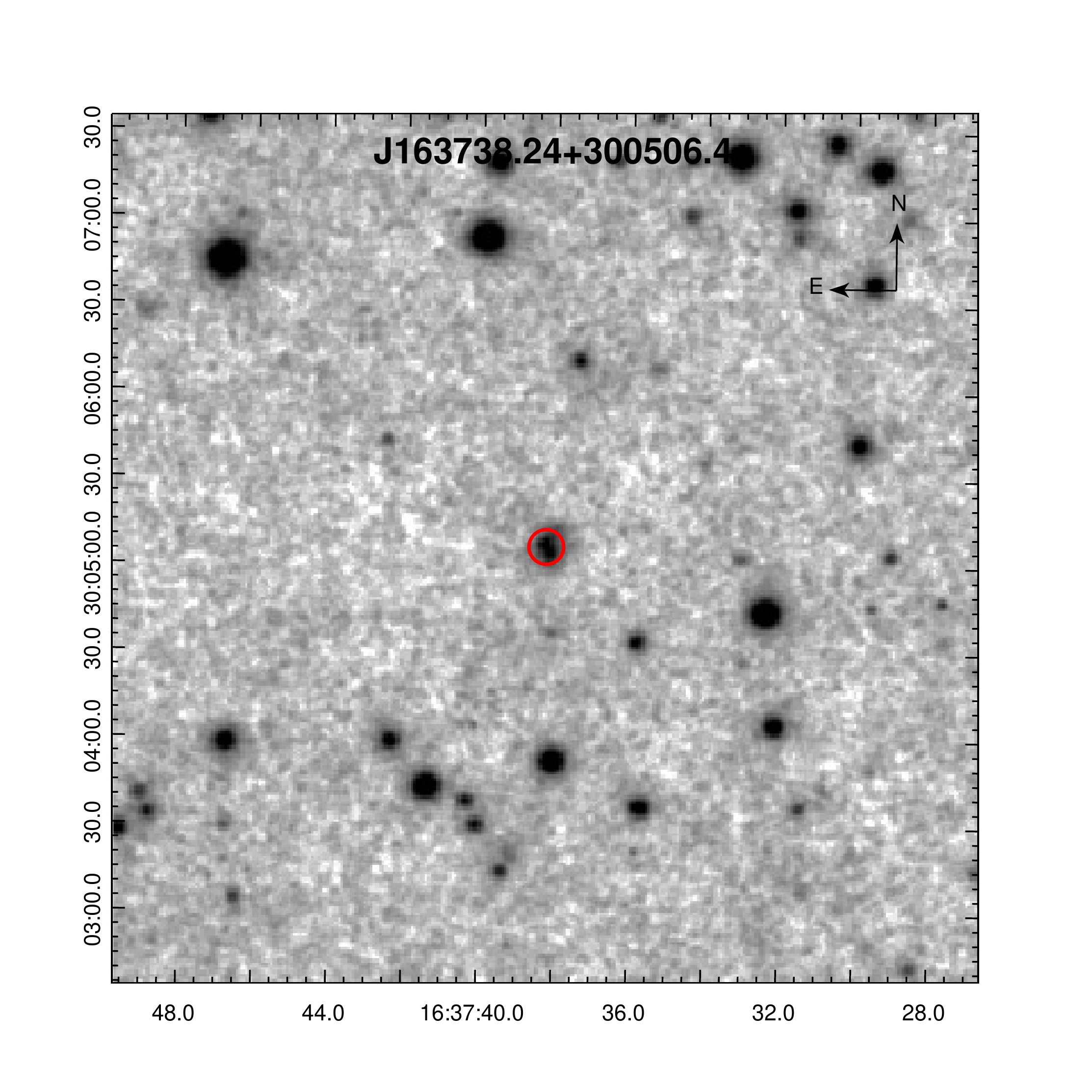} \\
\end{array}$
\end{center}
\caption{As in Figure~\ref{fig:J0837} but for WISE J163738.24+300506.4, the potential counterpart of 4FGL J1637.5+3005.}
\label{fig:J1637}
\end{figure*}

\begin{figure*}{}
\begin{center}$
\begin{array}{cc}
\includegraphics[width=\mywidth]{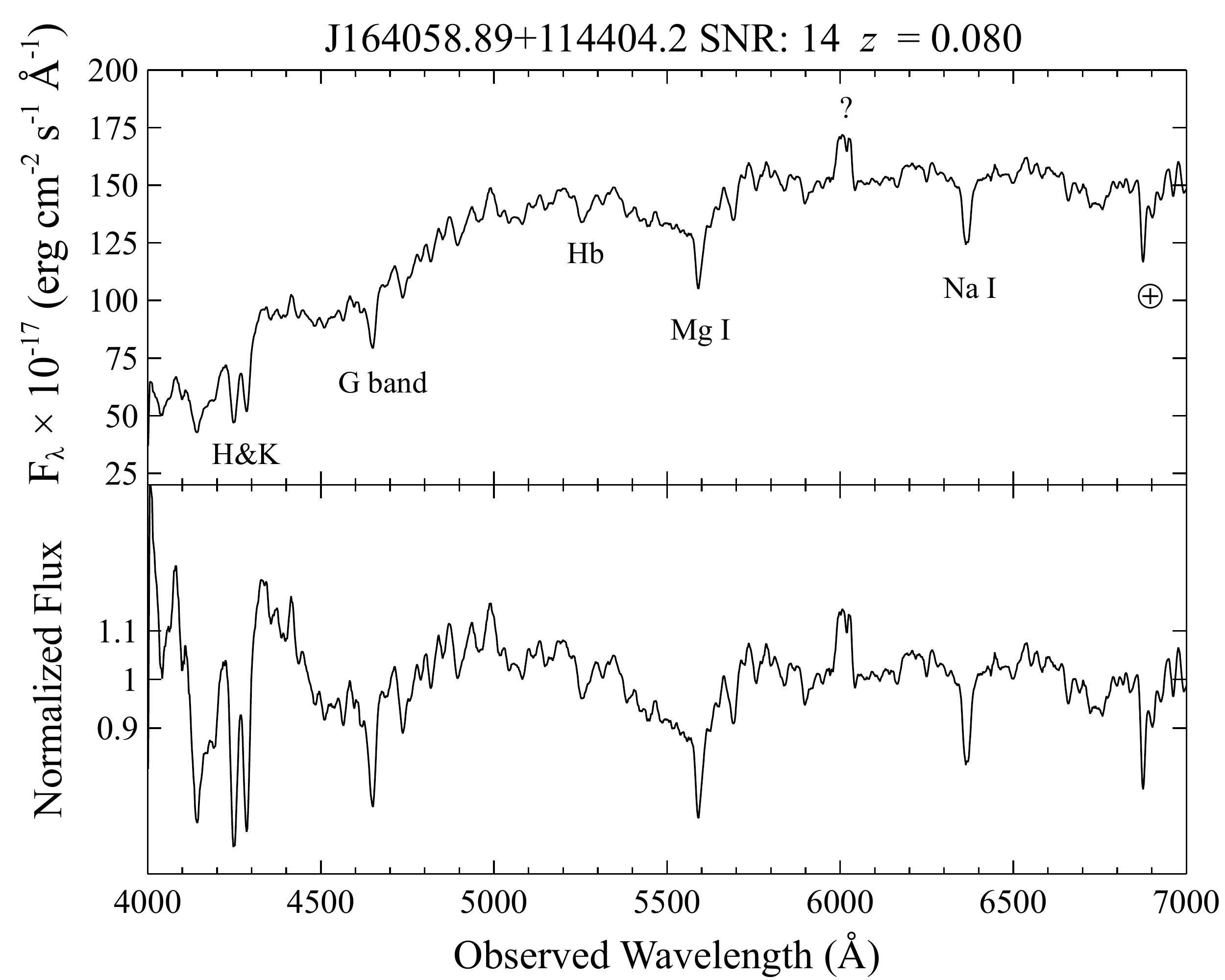} &
\includegraphics[clip=true, width=7cm]{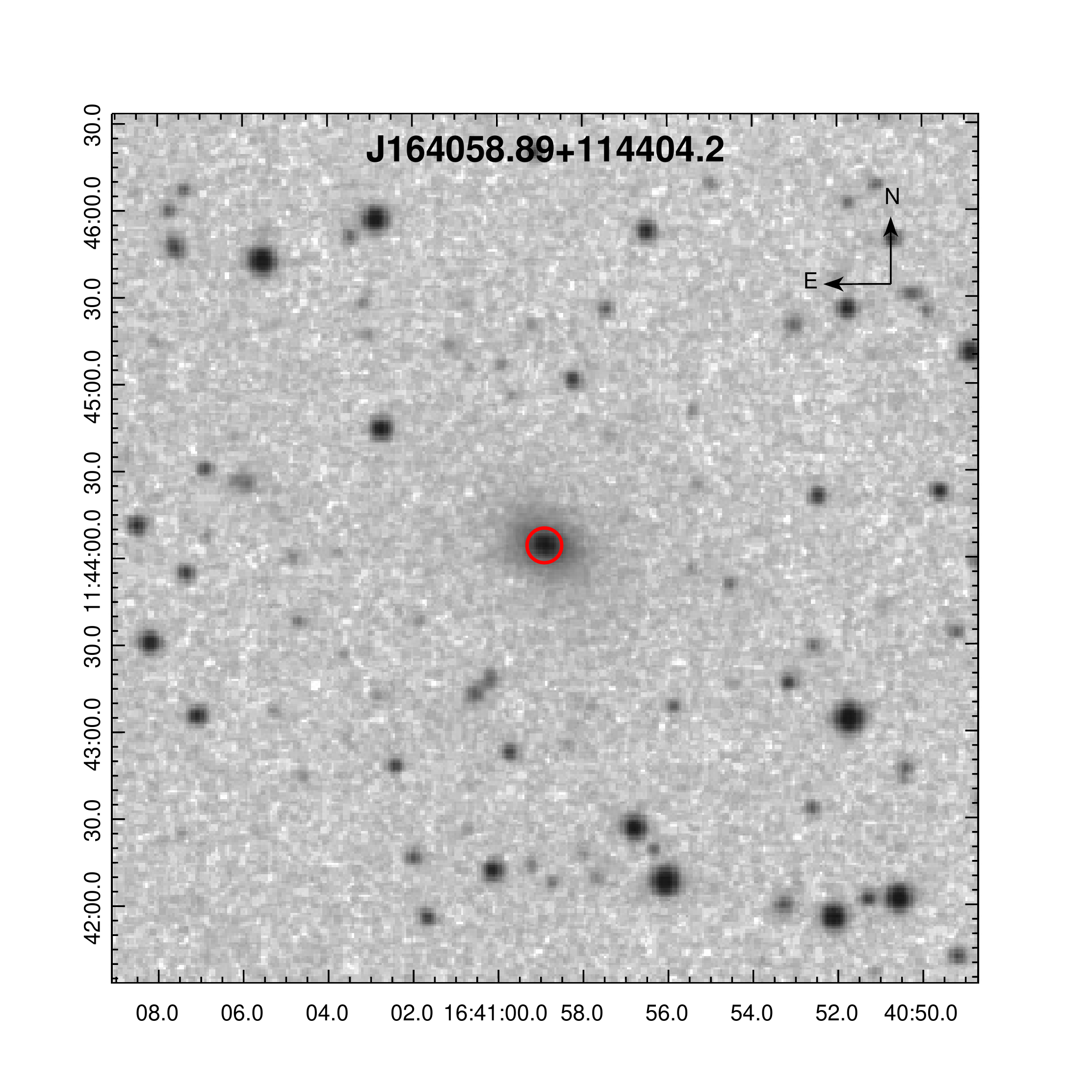} \\
\end{array}$
\end{center}
\caption{As in Figure~\ref{fig:J0837} but for WISE J164058.89+114404.2, the counterpart of 4FGL J1640.9+1143.}
\label{fig:J1640}
\end{figure*}

\begin{figure*}{}
\begin{center}$
\begin{array}{cc}
\includegraphics[width=\mywidth]{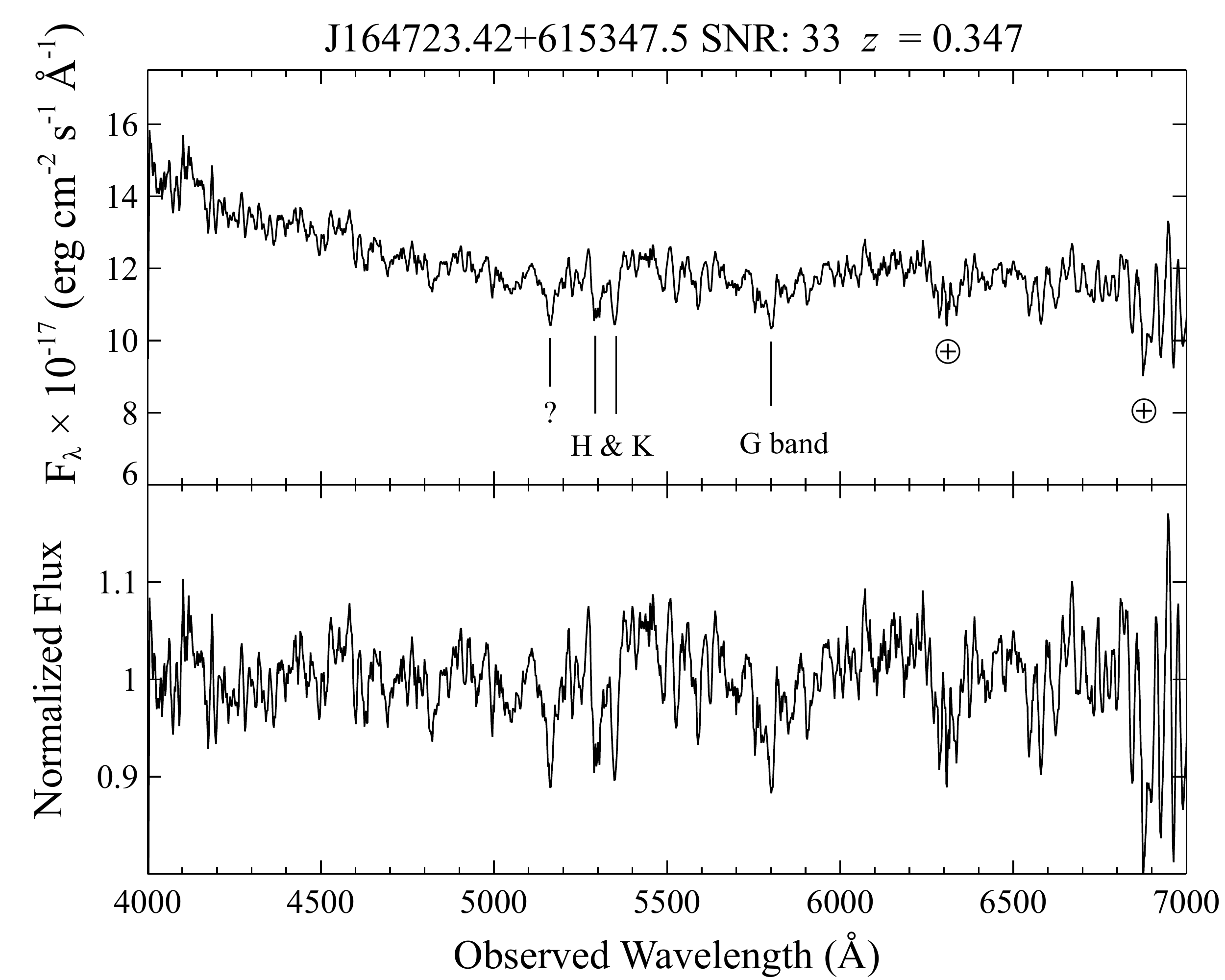} &
\includegraphics[clip=true, width=7cm]{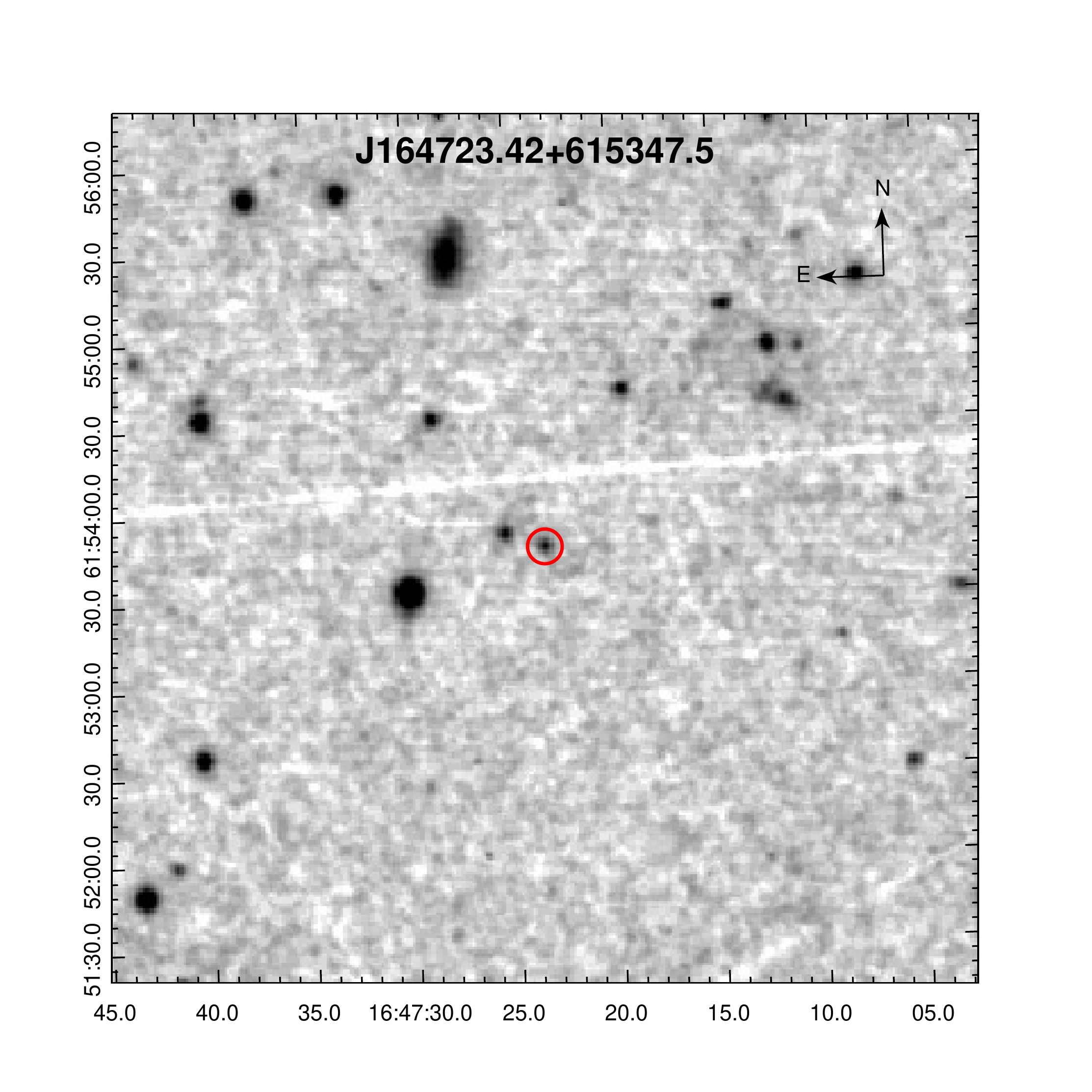} \\
\end{array}$
\end{center}
\caption{As in Figure~\ref{fig:J0837} but for WISE J164723.42+615347.5, the potential counterpart of 4FGL J1647.1+6149.}
\label{fig:J1647}
\end{figure*}

%%%%%%%%%%%
\begin{figure*}{}
\begin{center}$
\begin{array}{cc}
\includegraphics[width=\mywidth]{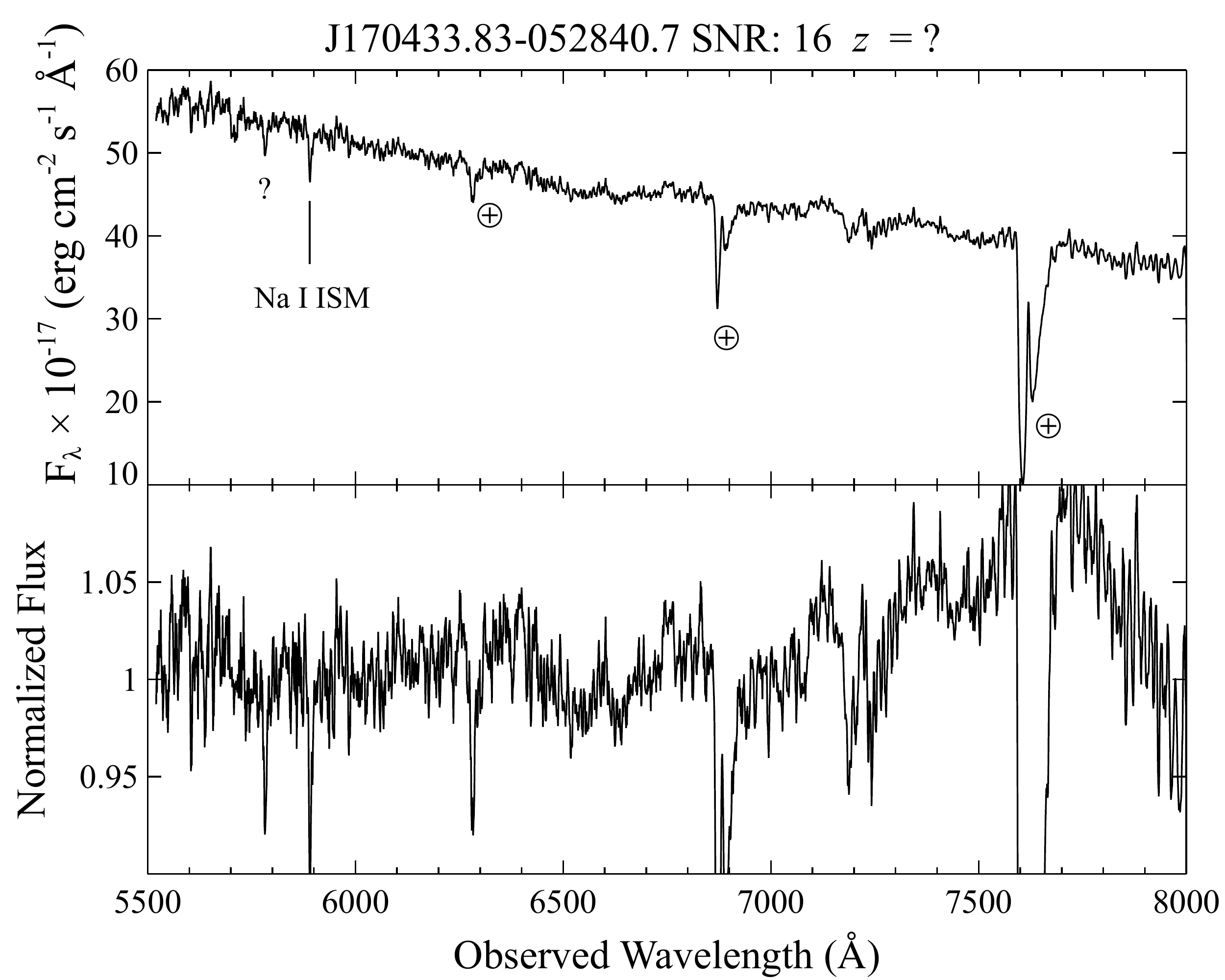} &
\includegraphics[clip=true, width=7cm]{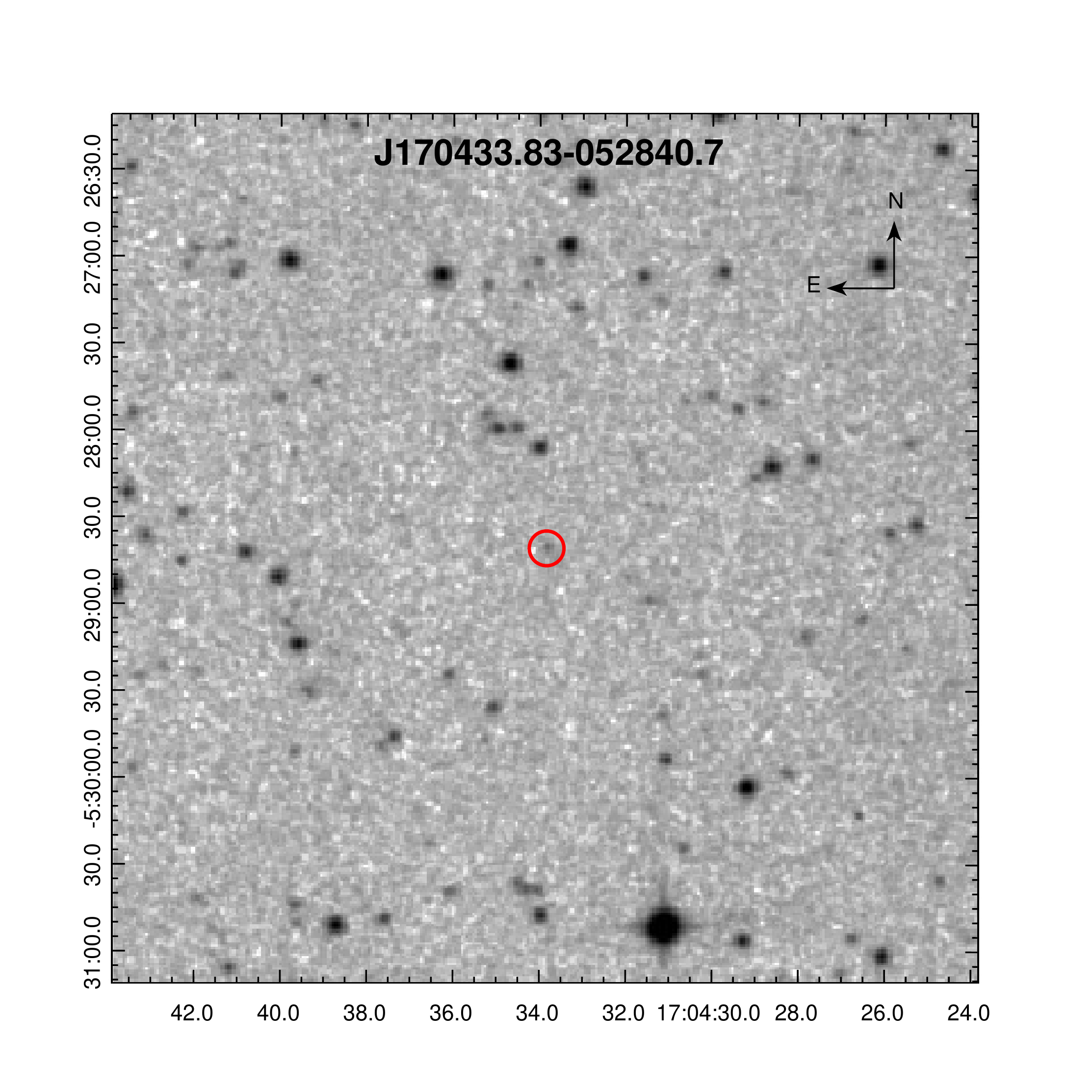} \\
\end{array}$
\end{center}
\caption{As in Figure~\ref{fig:J0837} but for WISE J170433.83-052840.7, the potential counterpart of 3FGL J1704.4-0528.}
\label{fig:J1704}
\end{figure*}

\begin{figure*}{}
\begin{center}$
\begin{array}{cc}
\includegraphics[width=\mywidth]{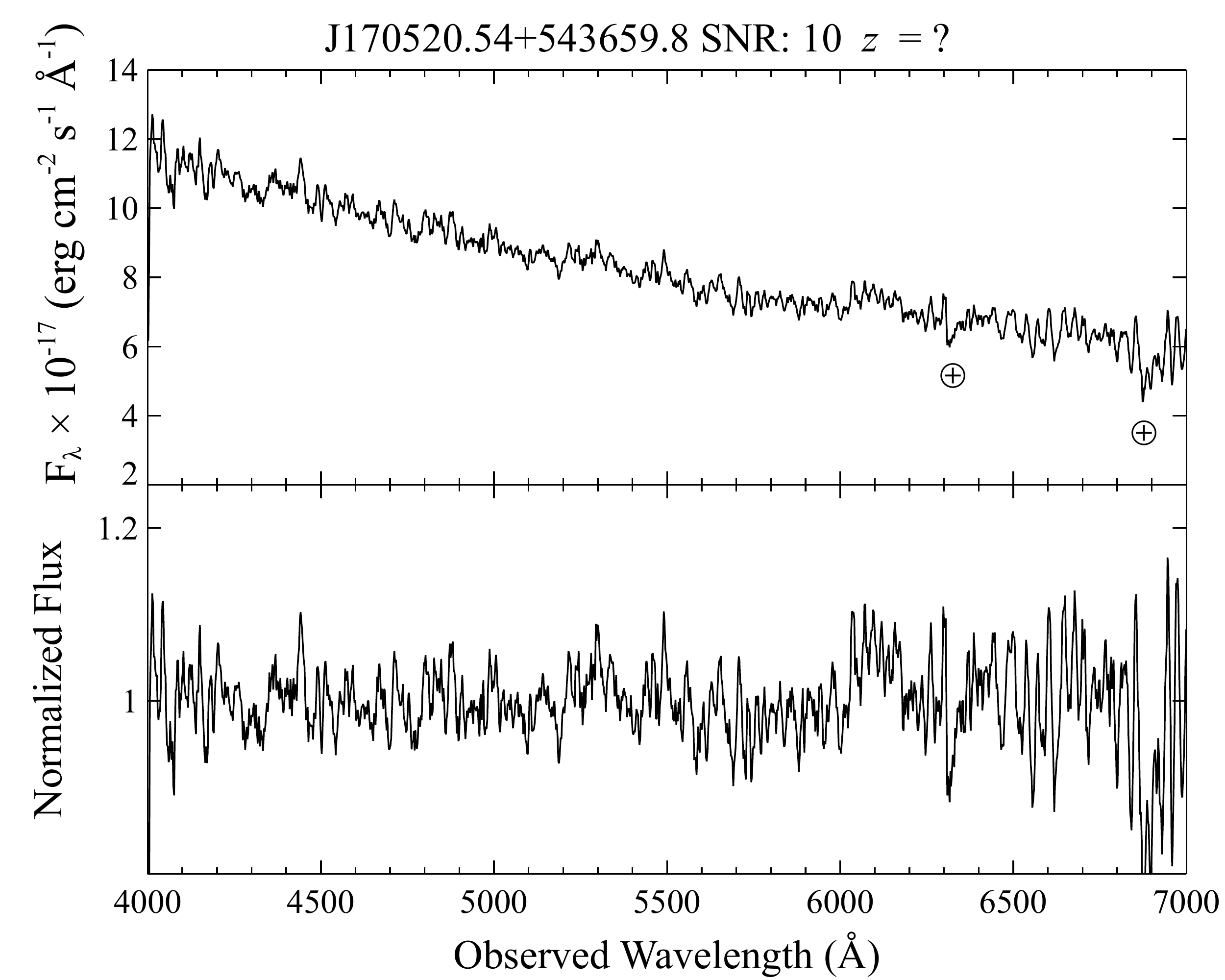} &
\includegraphics[clip=true, width=7cm]{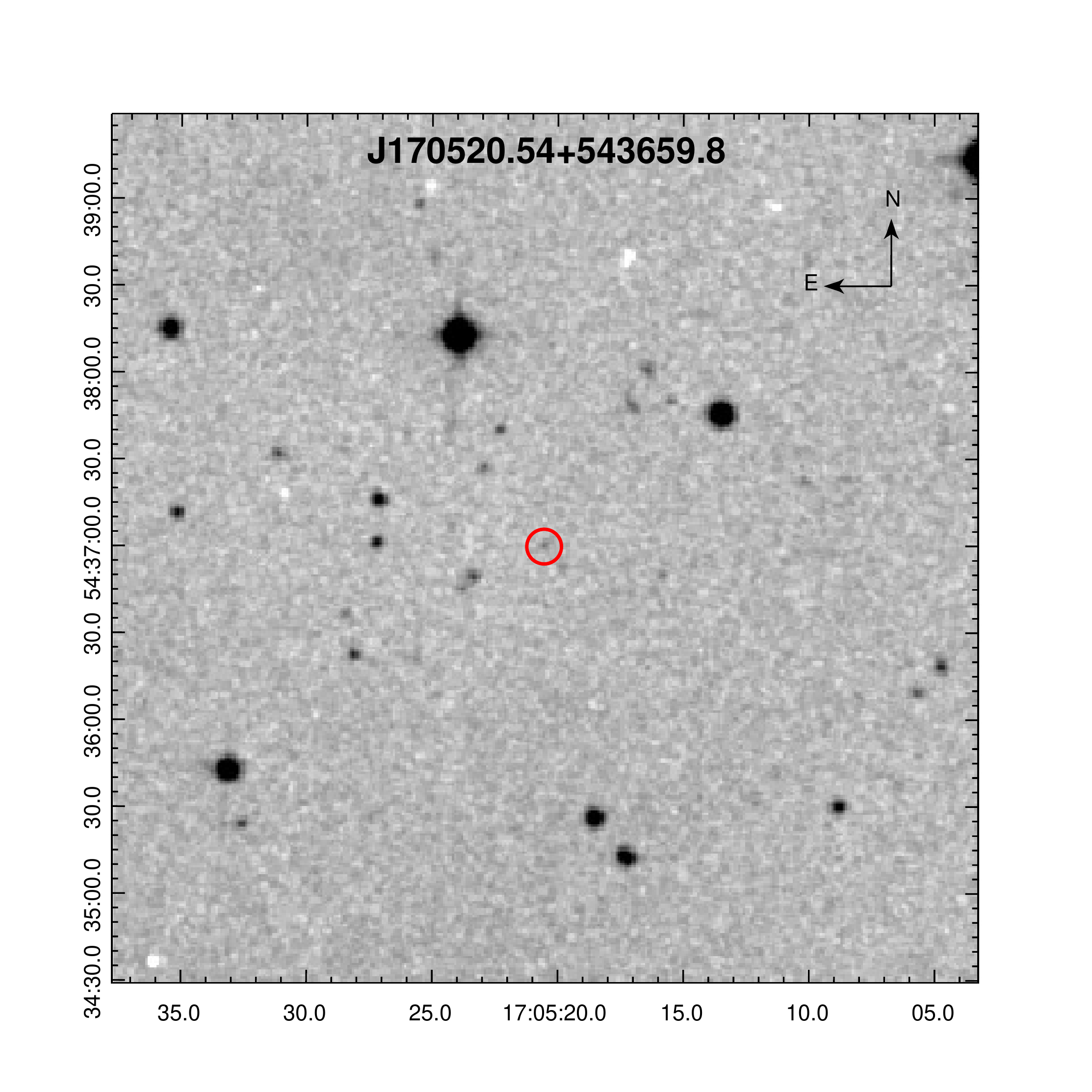} \\
\end{array}$
\end{center}
\caption{As in Figure~\ref{fig:J0837} but for WISE J170520.54+543659.8, the potential counterpart of 4FGL J1705.4+5436.}
\label{fig:J1705}
\end{figure*}

\begin{figure*}{}
\begin{center}$
\begin{array}{cc}
\includegraphics[width=\mywidth]{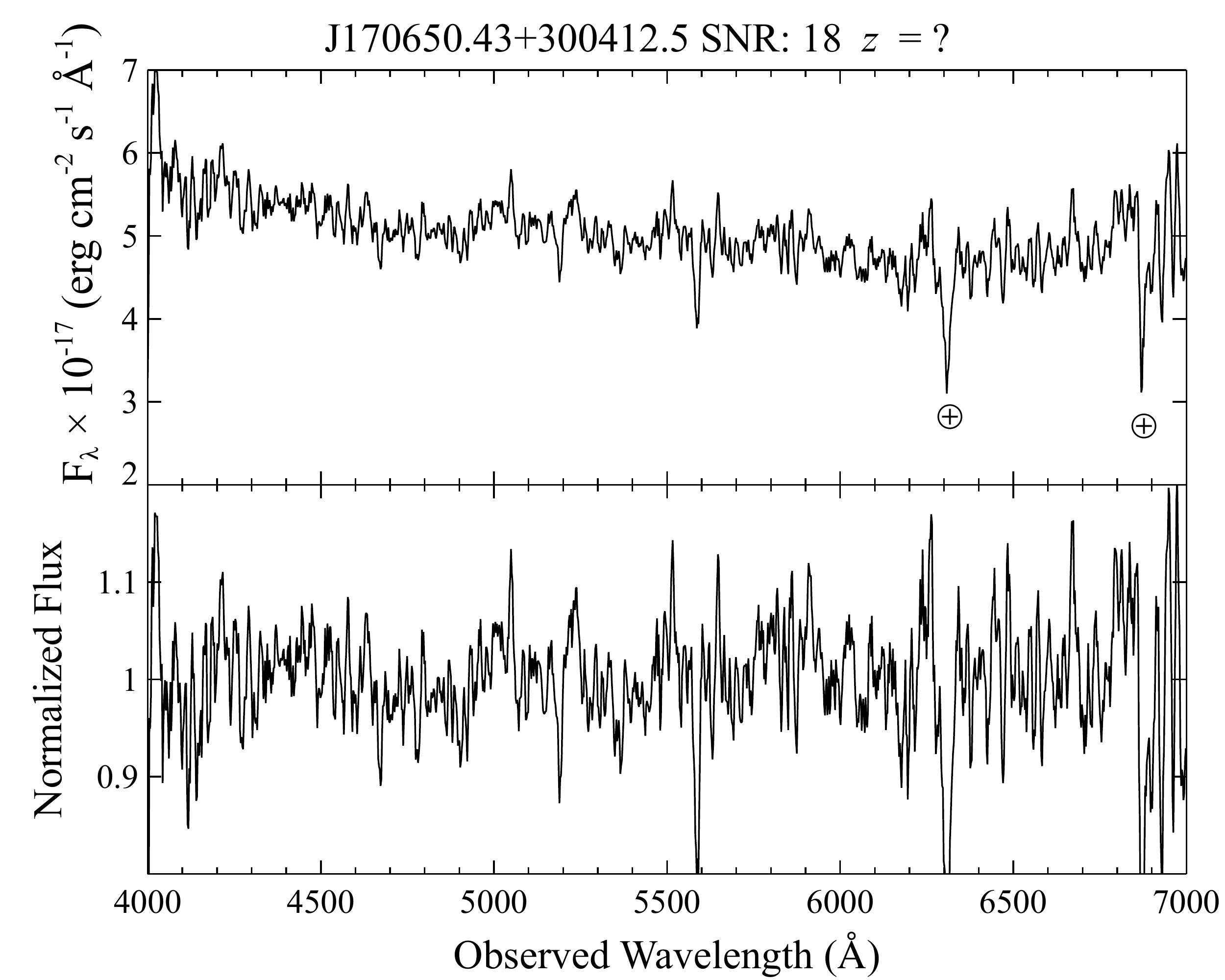} &
\includegraphics[clip=true, width=7cm]{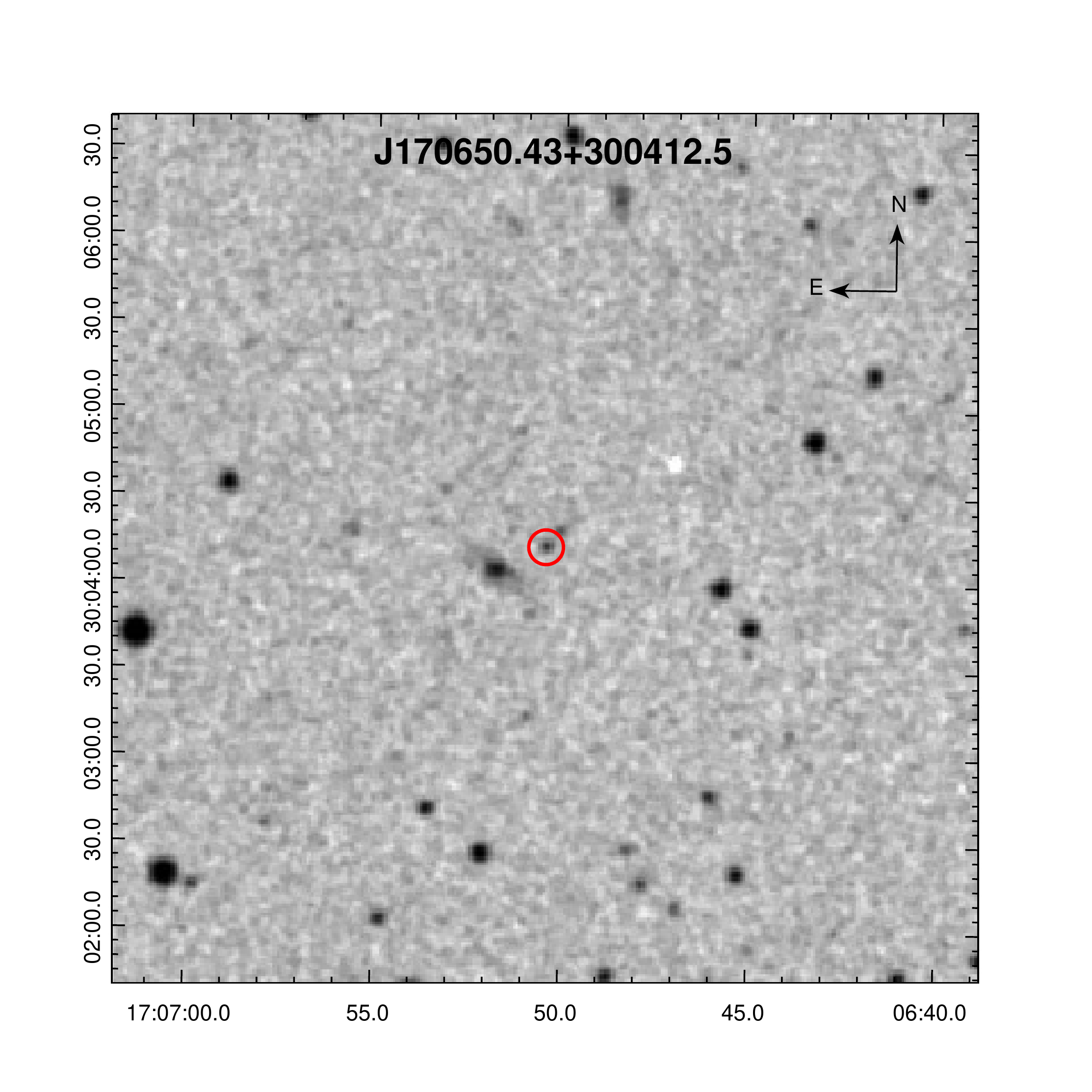} \\
\end{array}$
\end{center}
\caption{As in Figure~\ref{fig:J0837} but for WISE J170650.43+300412.5, the potential counterpart of 4FGL J1706.8+3004.}
\label{fig:J1706}
\end{figure*}

%%%%%%%%%%%
\begin{figure*}{}
\begin{center}$
\begin{array}{cc}
\includegraphics[width=\mywidth]{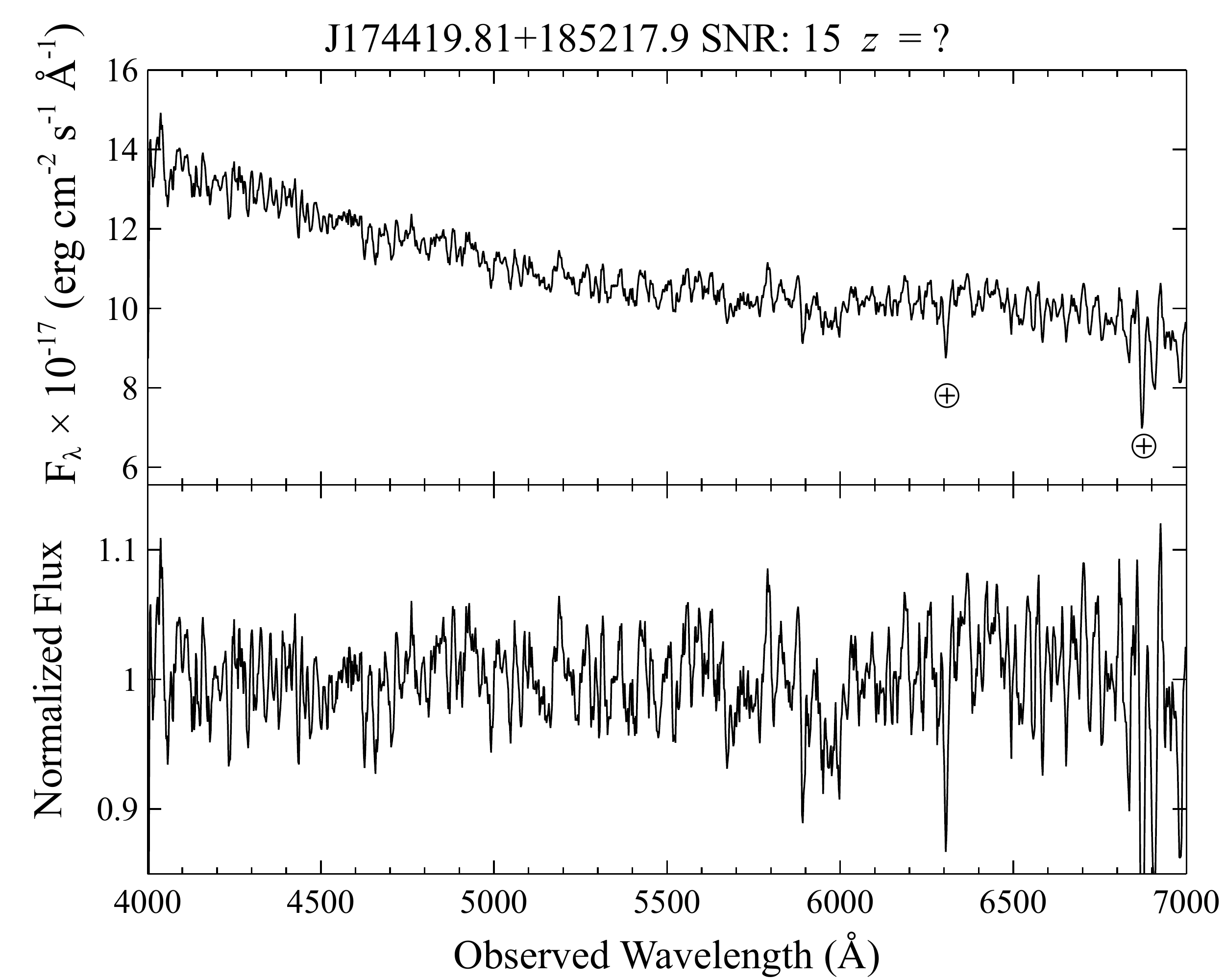} &
\includegraphics[clip=true, width=7cm]{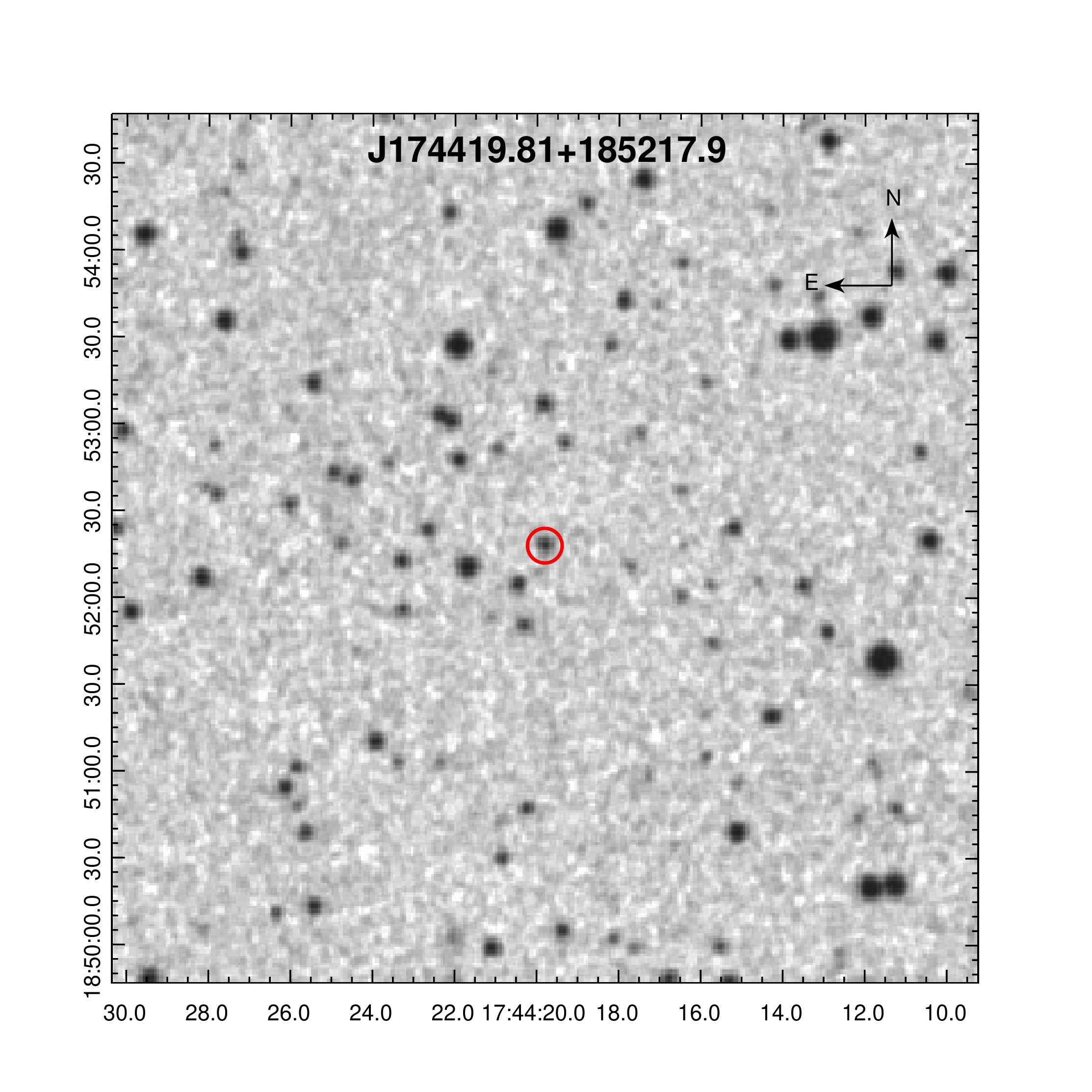} \\
\end{array}$
\end{center}
\caption{As in Figure~\ref{fig:J0837} but for WISE J174419.81+185217.9, the counterpart of 4FGL J1744.4+1851.}
\label{fig:J1744}
\end{figure*}

\begin{figure*}{}
\begin{center}$
\begin{array}{cc}
\includegraphics[width=\mywidth]{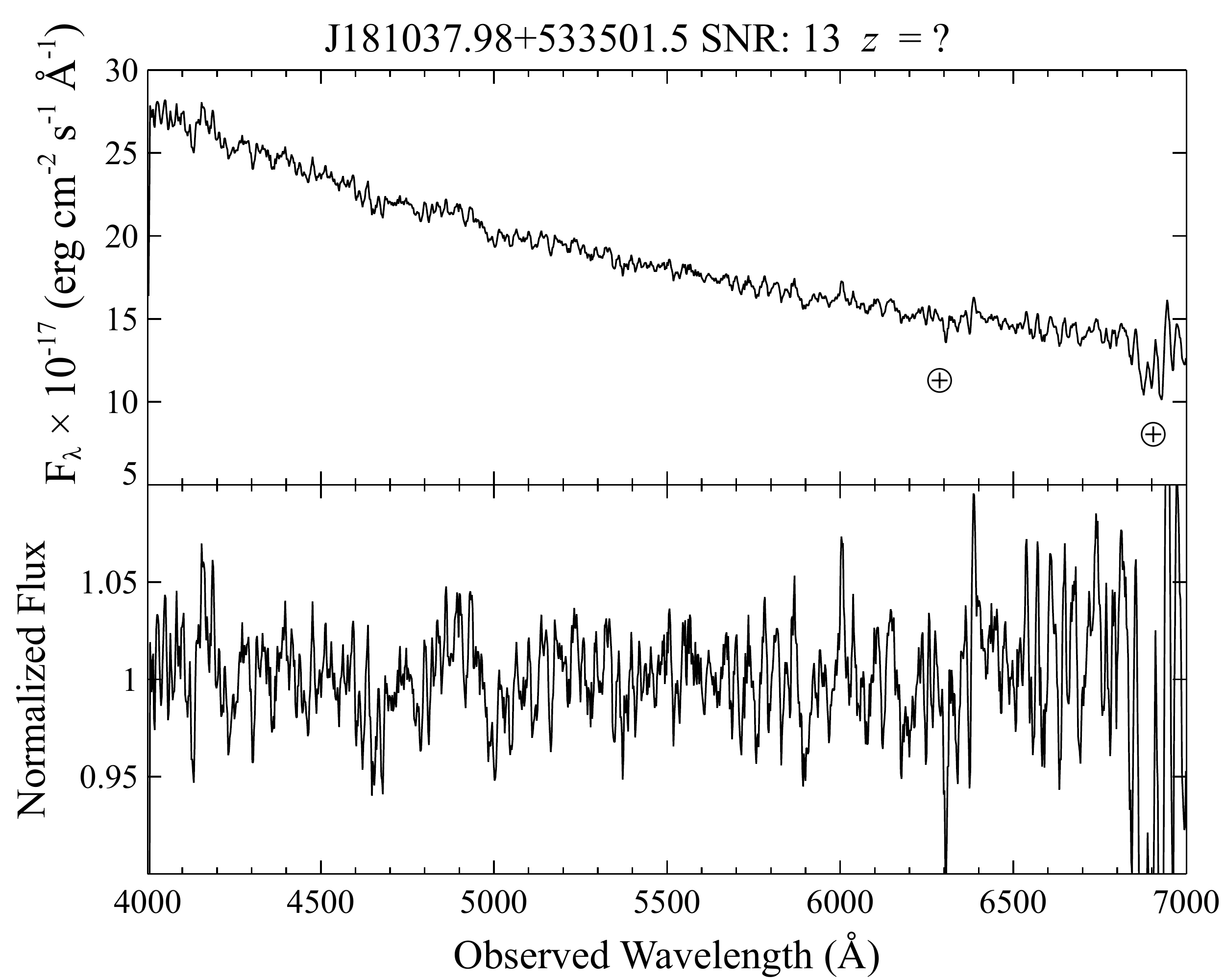} &
\includegraphics[clip=true, width=7cm]{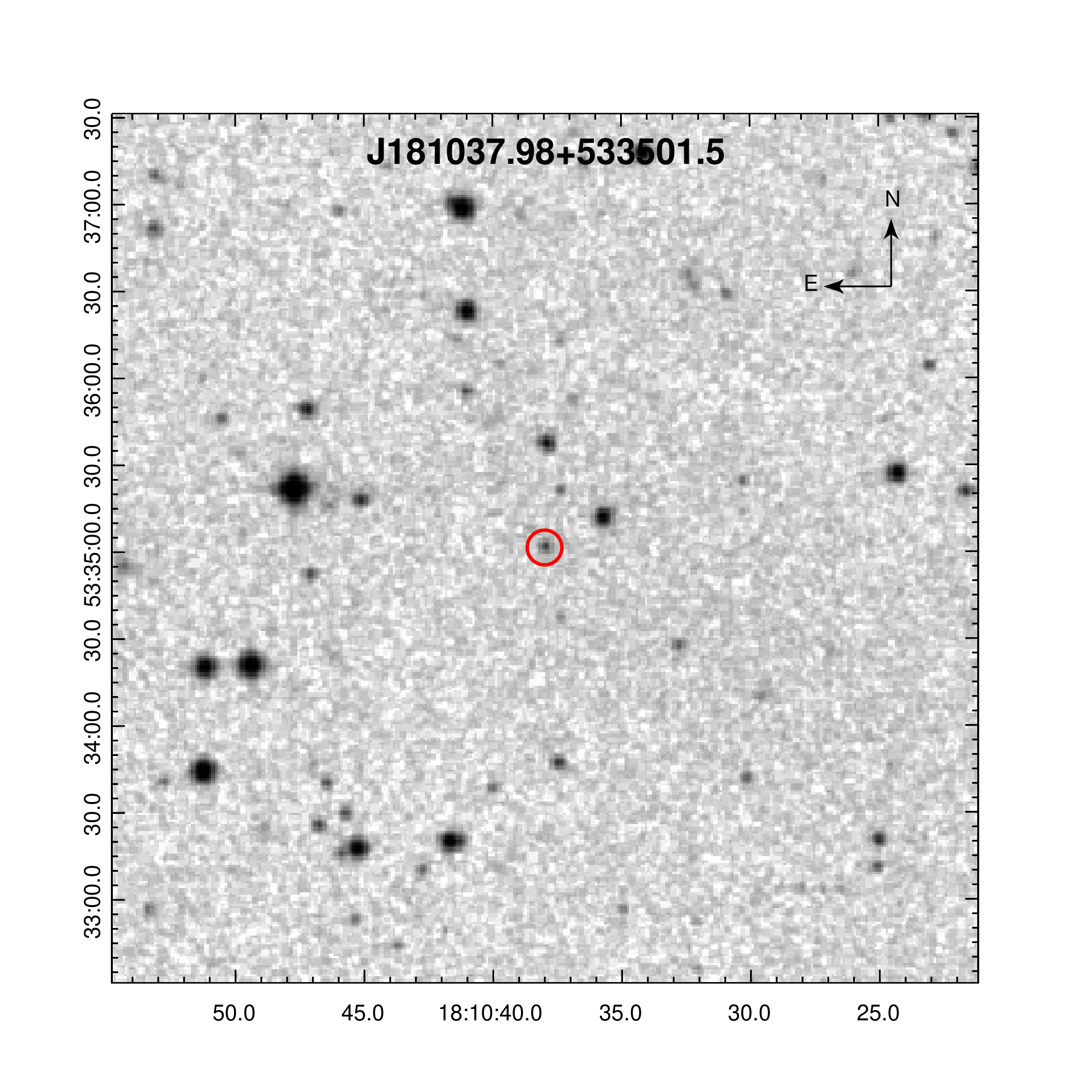} \\
\end{array}$
\end{center}
\caption{As in Figure~\ref{fig:J0837} but for WISE J181037.98+533501.5, the counterpart of 4FGL J1810.7+5335.}
\label{fig:J1810}
\end{figure*}

\begin{figure*}{}
\begin{center}$
\begin{array}{cc}
\includegraphics[width=\mywidth]{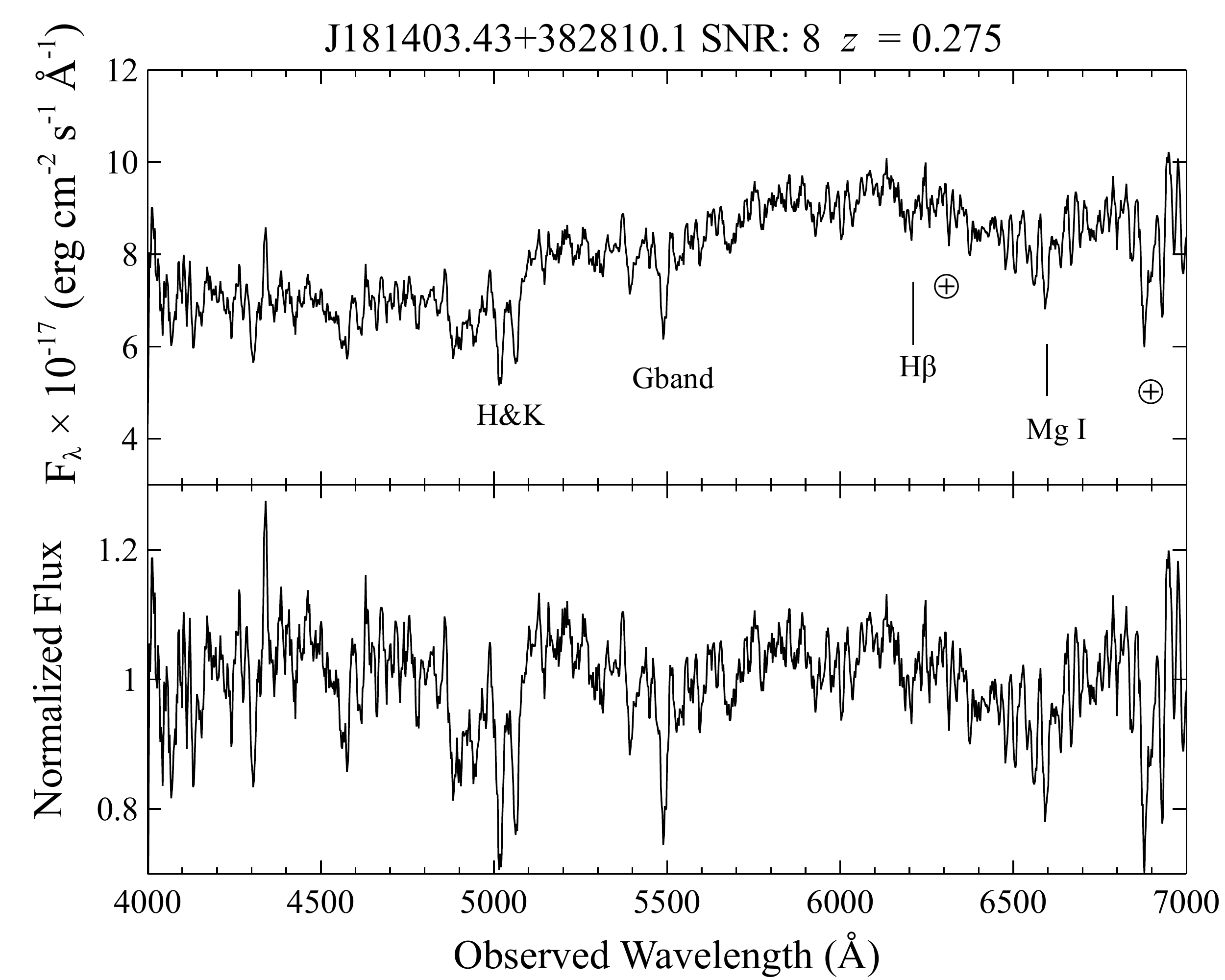} &
\includegraphics[clip=true, width=7cm]{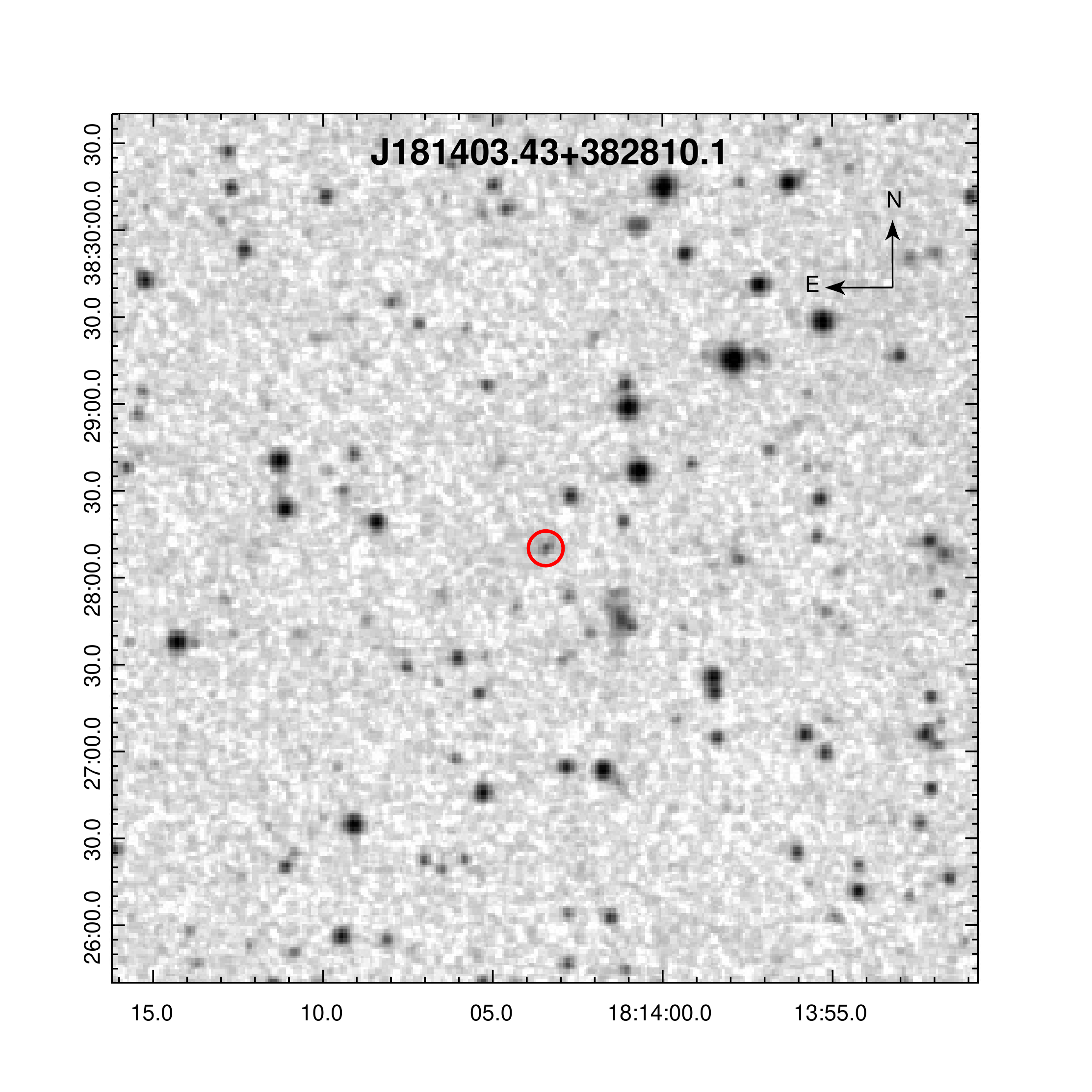} \\
\end{array}$
\end{center}
\caption{As in Figure~\ref{fig:J0837} but for WISE J181403.43+382810.1, the counterpart of 4FGL J1814.0+3828.}
\label{fig:J1814}
\end{figure*}

\begin{figure*}{}
\begin{center}$
\begin{array}{cc}
\includegraphics[width=\mywidth]{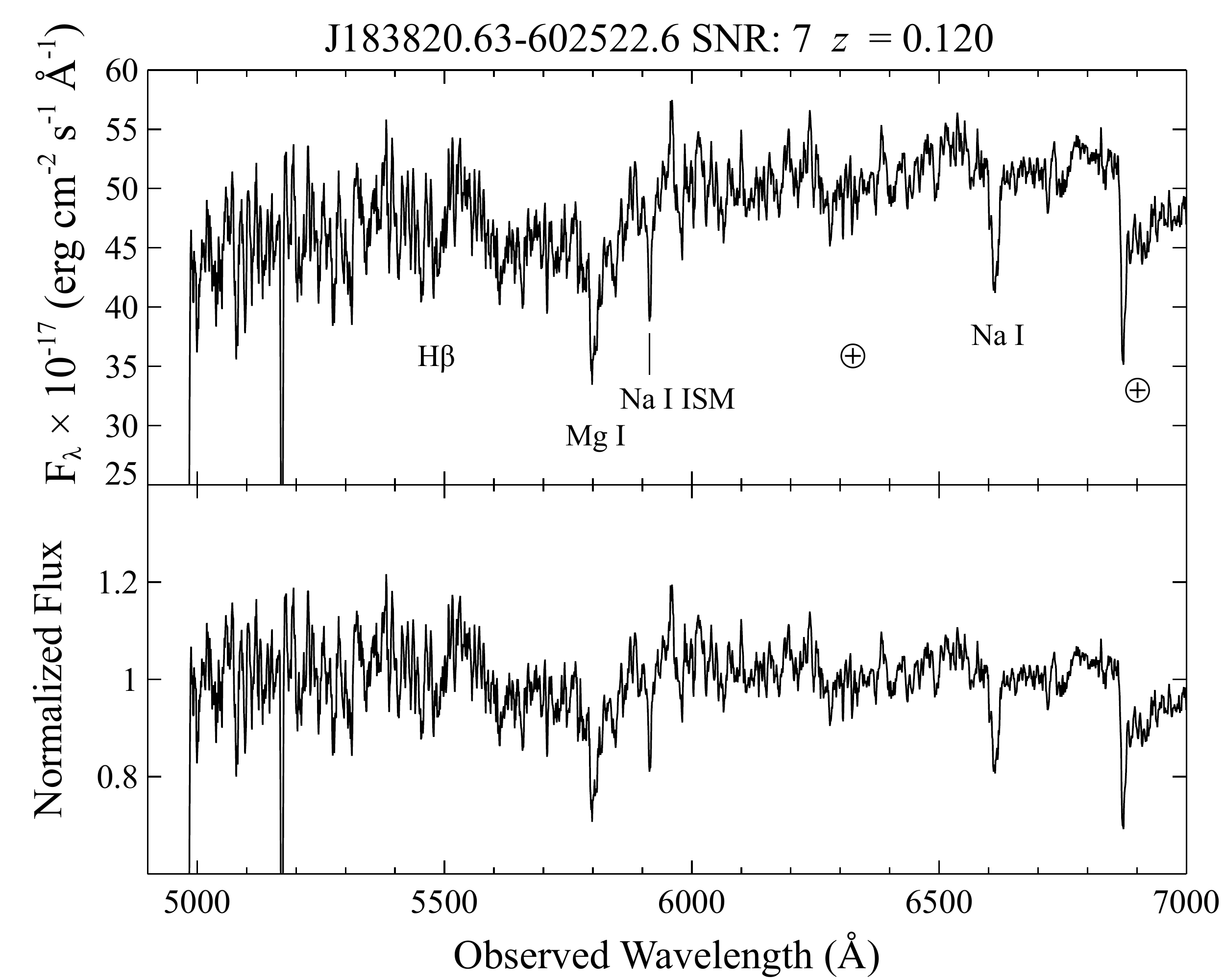} &
\includegraphics[clip=true, width=7cm]{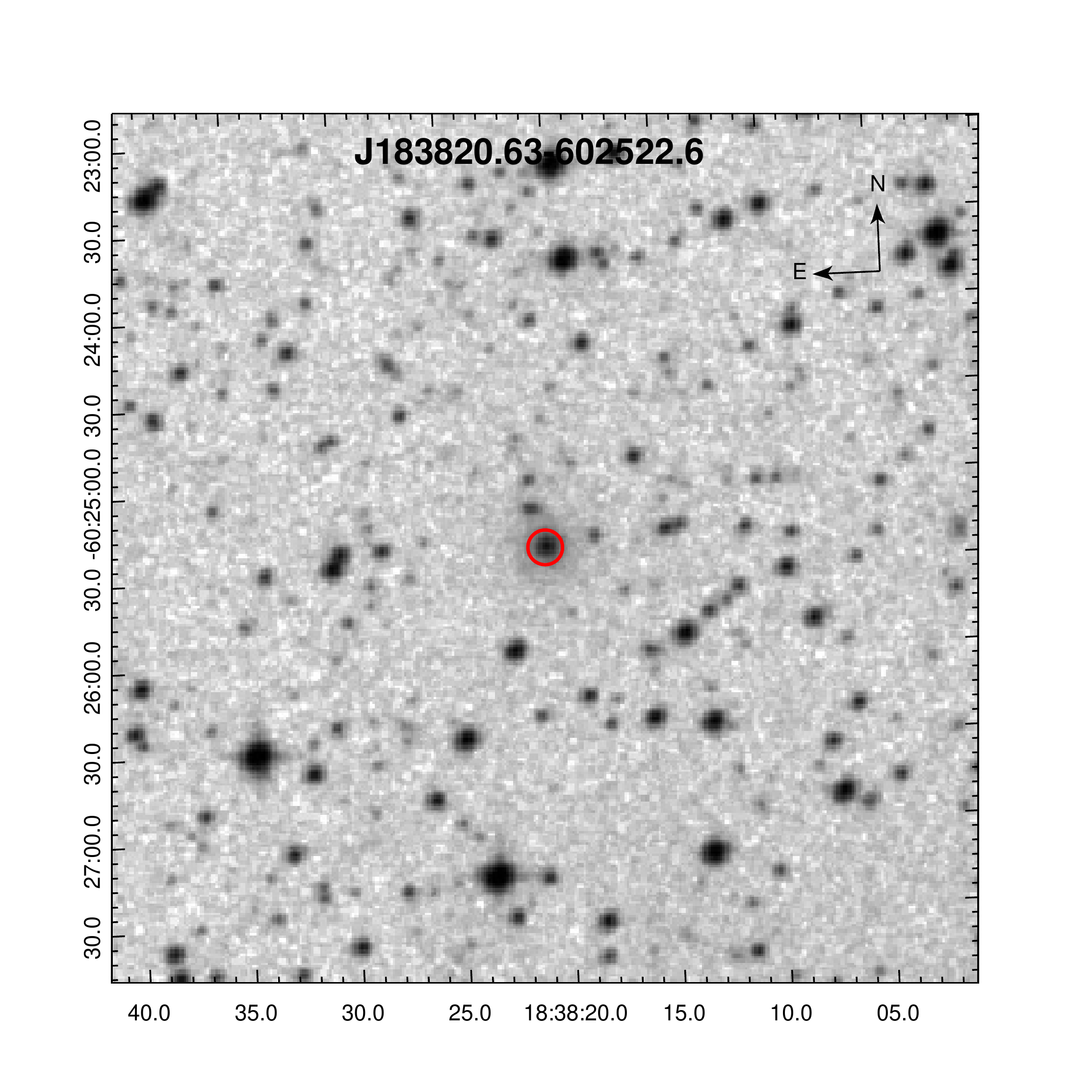} \\
\end{array}$
\end{center}
\caption{As in Figure~\ref{fig:J0837} but for WISE J183820.63-602522.6, the counterpart of 4FGL J1838.4-6023.}
\label{fig:J1838}
\end{figure*}

\begin{figure*}{}
\begin{center}$
\begin{array}{cc}
\includegraphics[width=\mywidth]{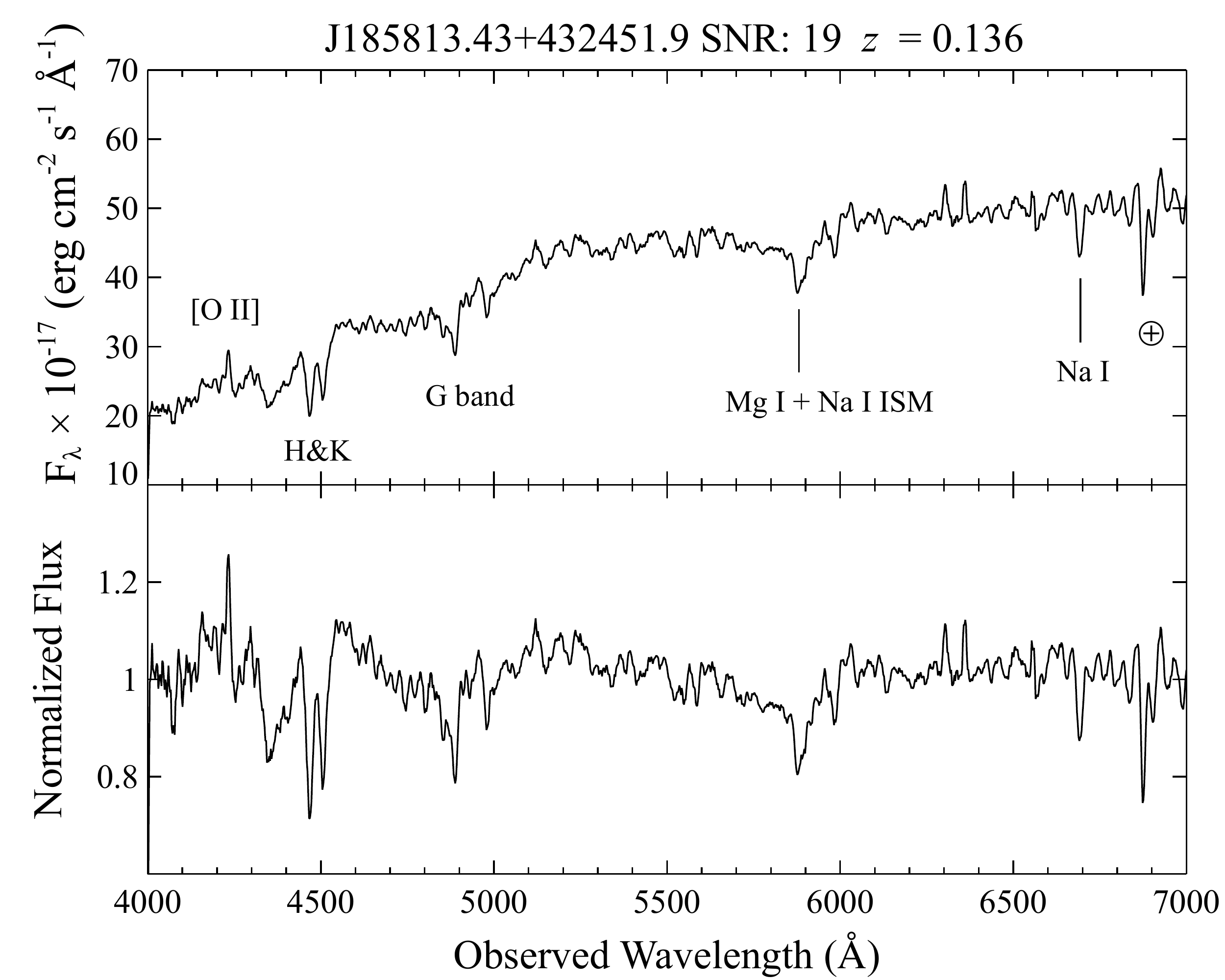} &
\includegraphics[clip=true, width=7cm]{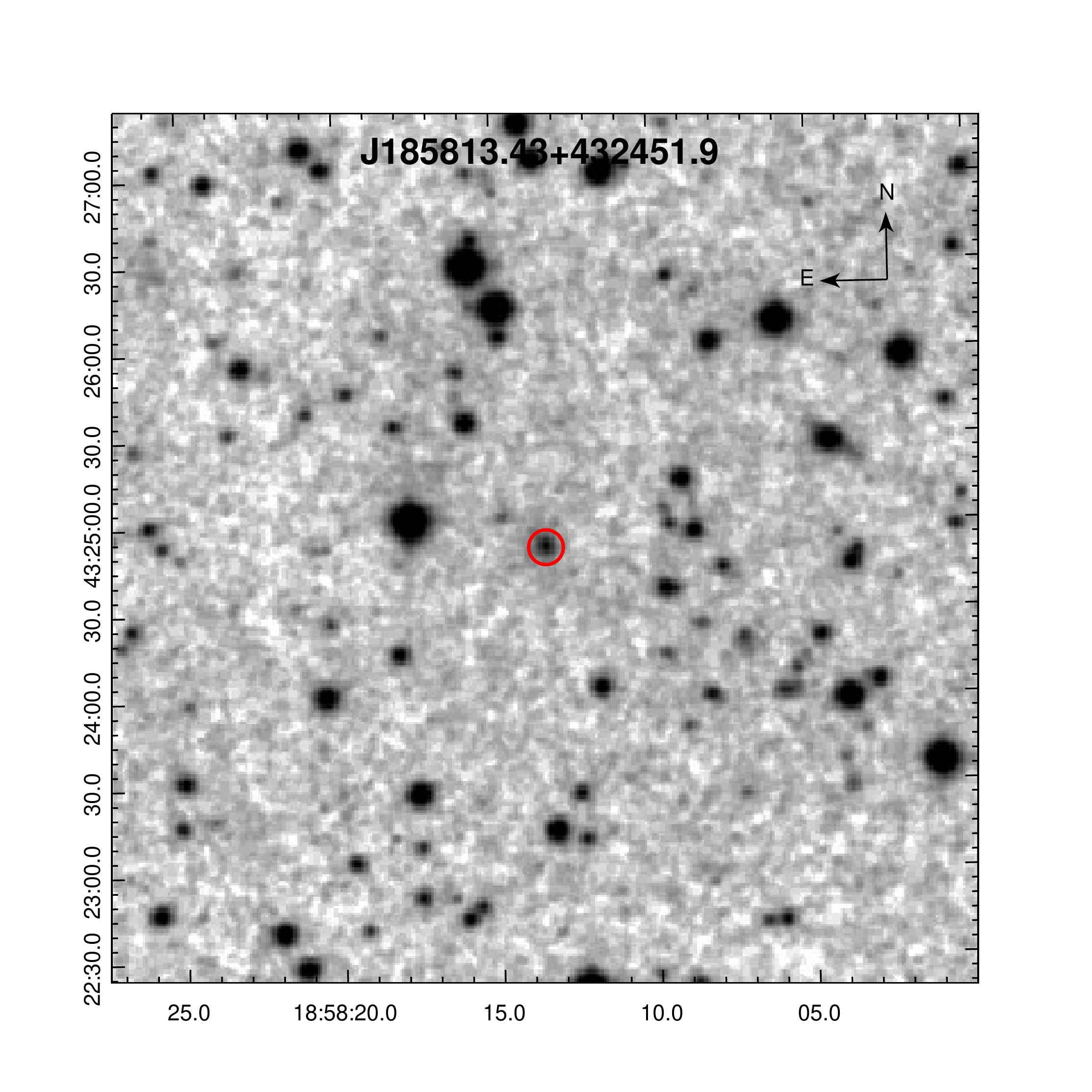} \\
\end{array}$
\end{center}
\caption{As in Figure~\ref{fig:J0837} but for WISE J185813.43+432451.9, the counterpart of 4FGL J1858.3+4321.}
\label{fig:J1858}
\end{figure*}

\begin{figure*}{}
\begin{center}$
\begin{array}{cc}
\includegraphics[width=\mywidth]{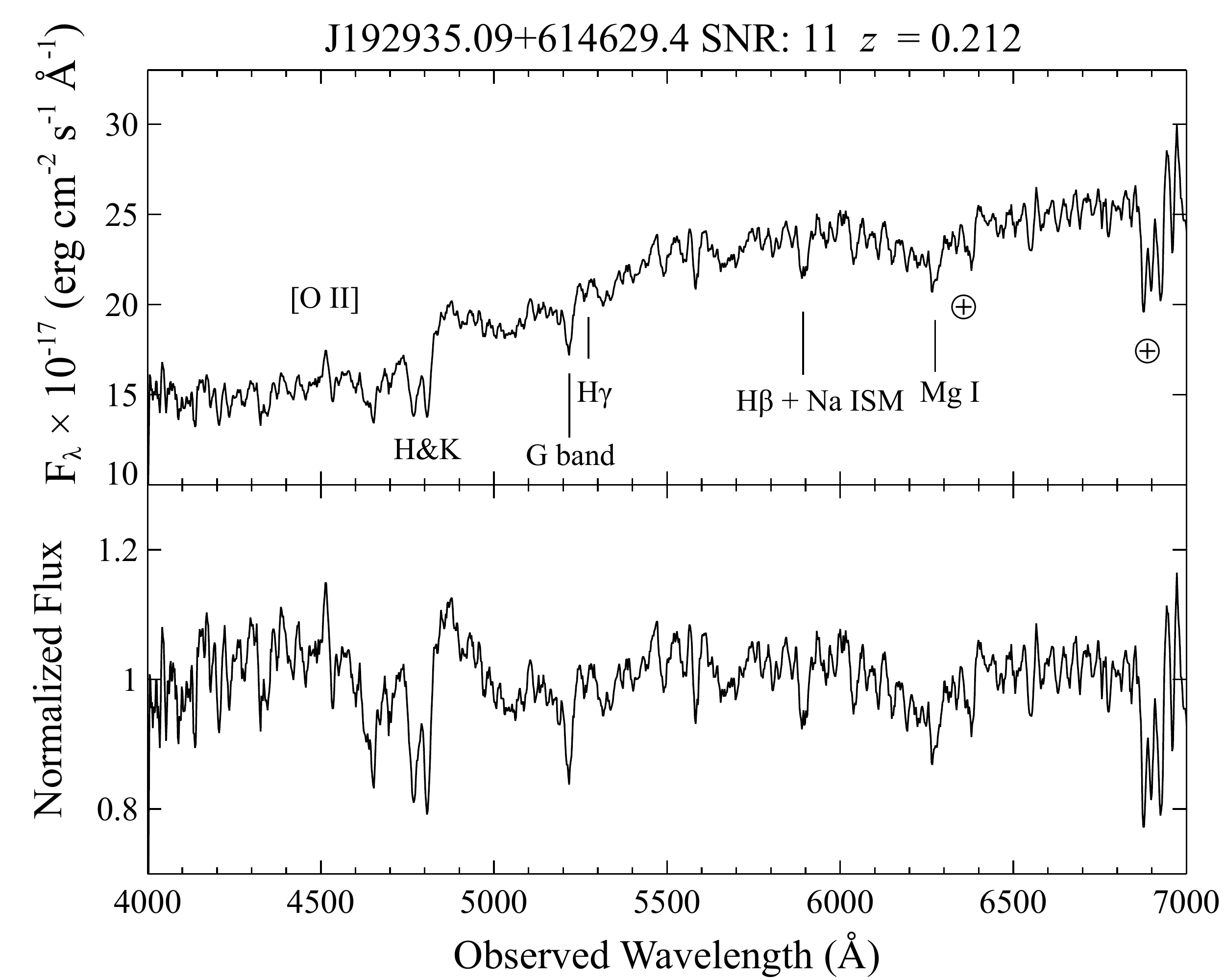} &
\includegraphics[clip=true, width=7cm]{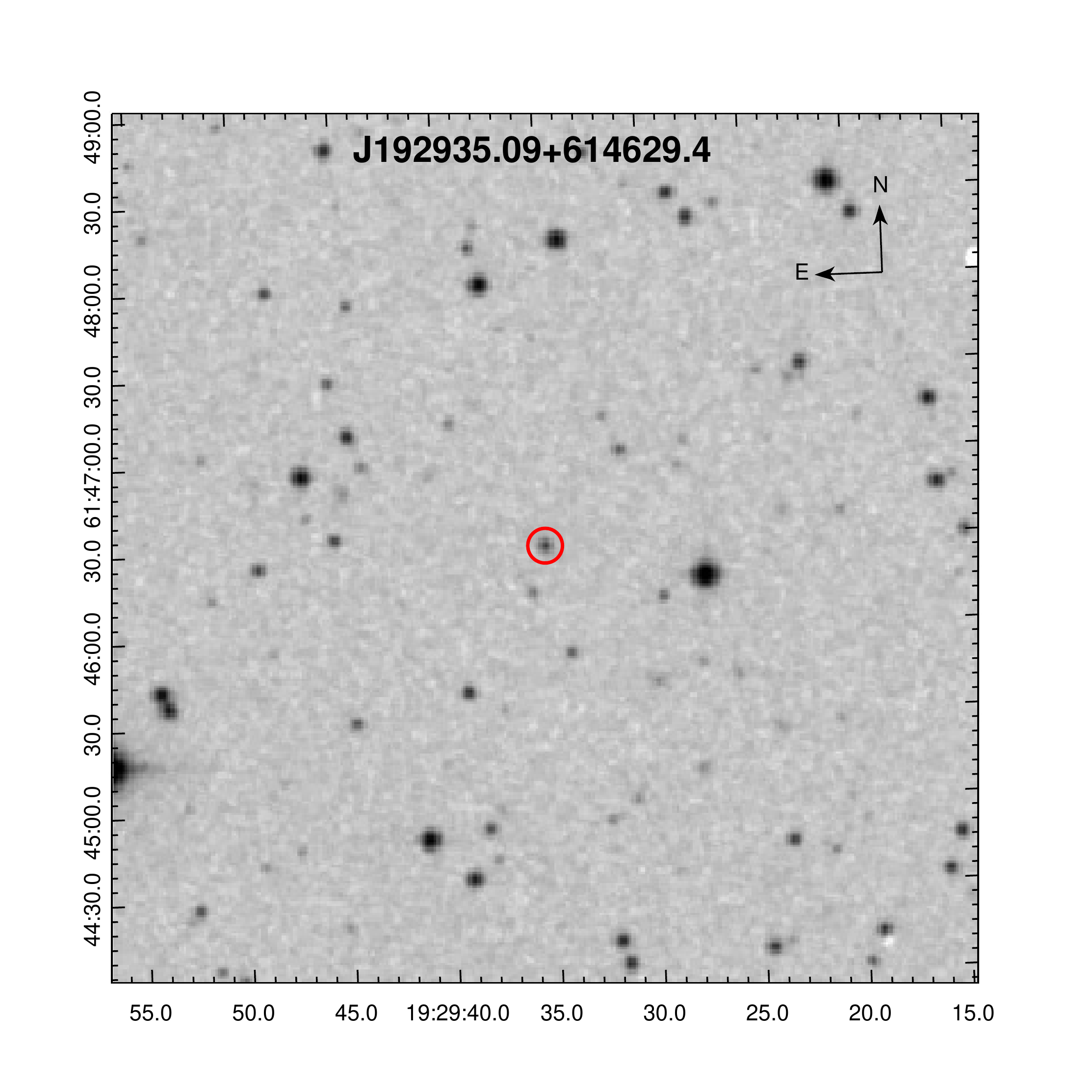} \\
\end{array}$
\end{center}
\caption{As in Figure~\ref{fig:J0837} but for WISE J192935.09+614629.4, the counterpart of 4FGL J1929.4+6146.}
\label{fig:J1929}
\end{figure*}

\begin{figure*}{}
\begin{center}$
\begin{array}{cc}
\includegraphics[width=\mywidth]{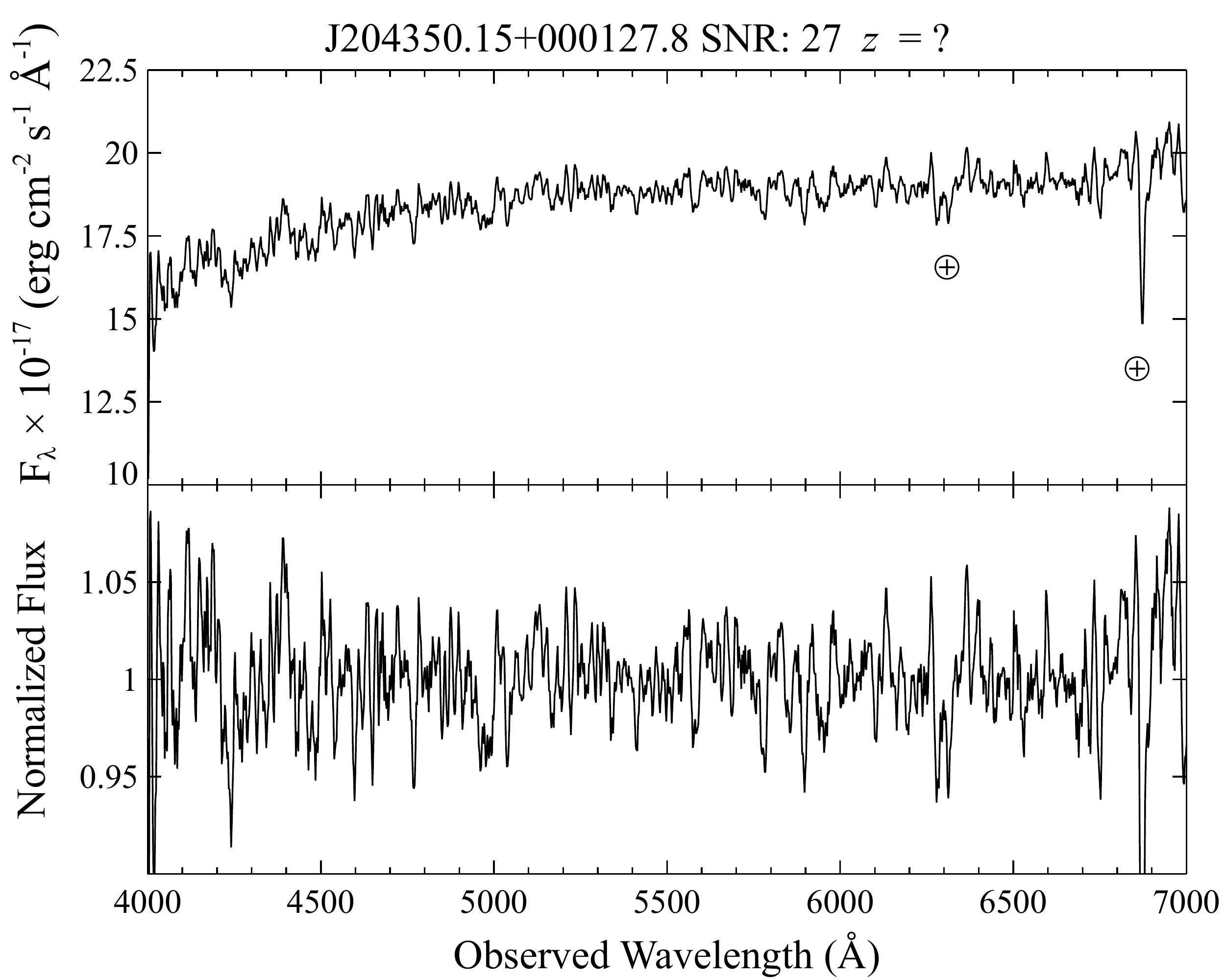} &
\includegraphics[clip=true, width=7cm]{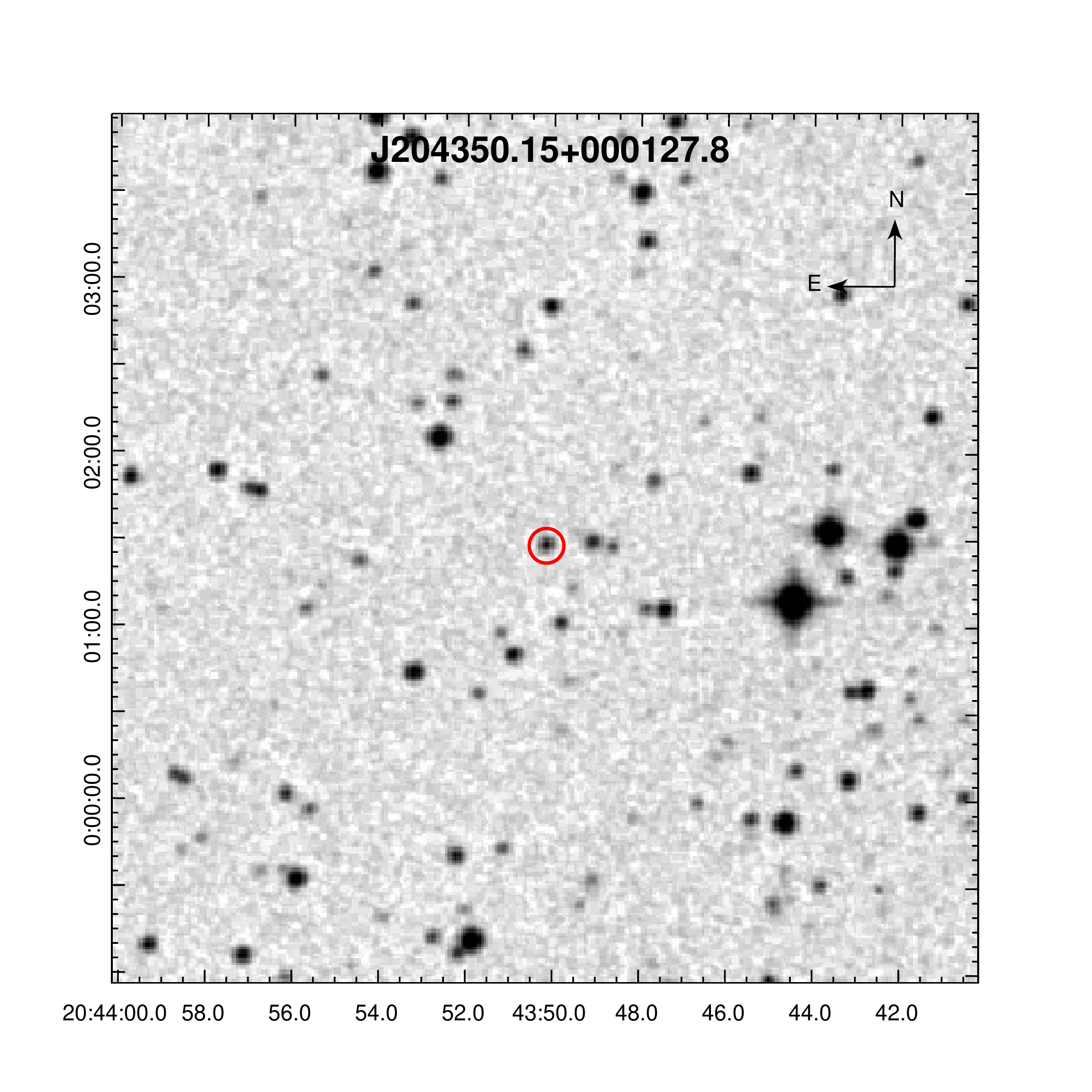} \\
\end{array}$
\end{center}
\caption{As in Figure~\ref{fig:J0837} but for WISE J204350.15+000127.8, the counterpart of 4FGL J2043.7+0000.}
\label{fig:J2043}
\end{figure*}

\begin{figure*}{}
\begin{center}$
\begin{array}{cc}
\includegraphics[width=\mywidth]{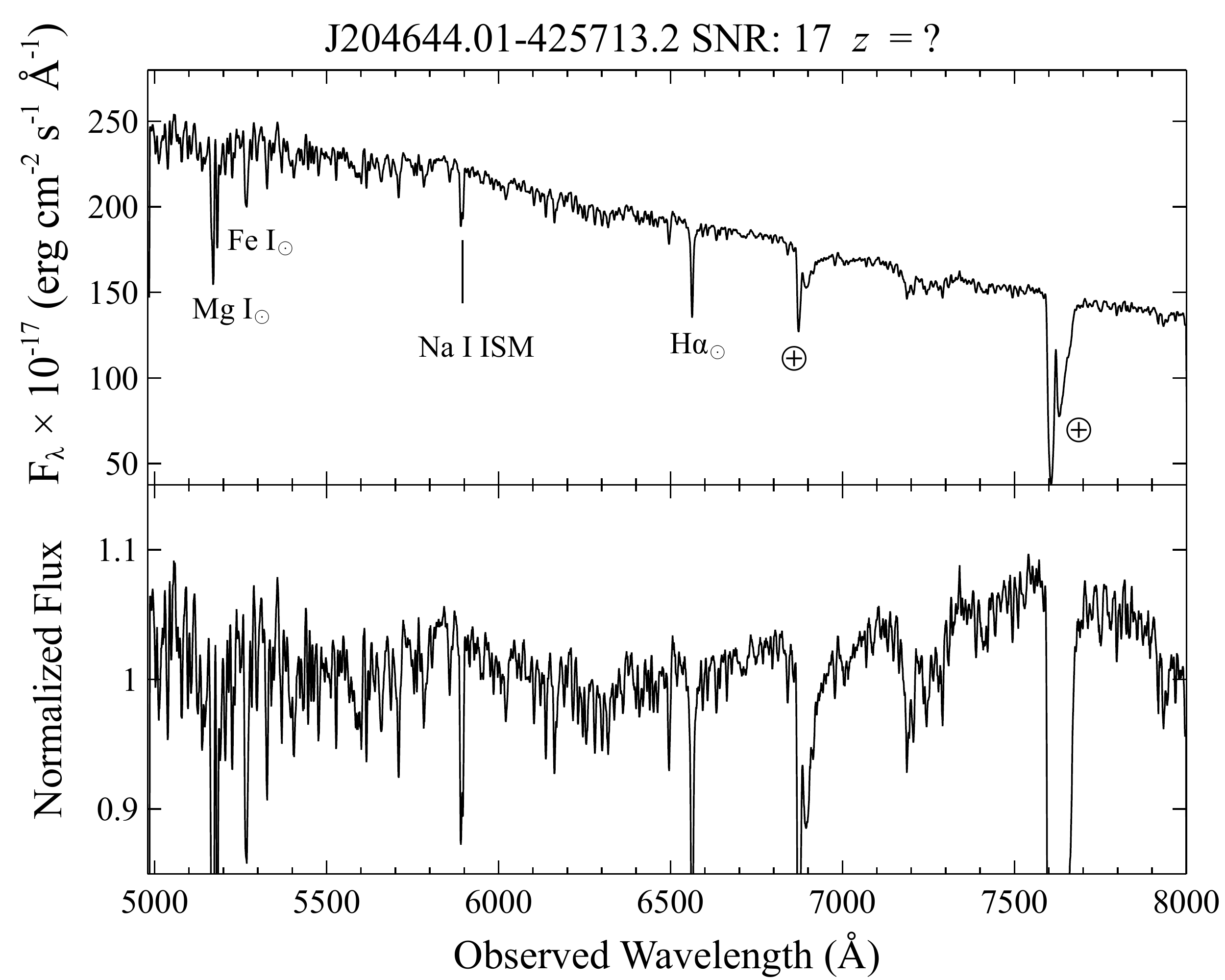} &
\includegraphics[clip=true, width=7cm]{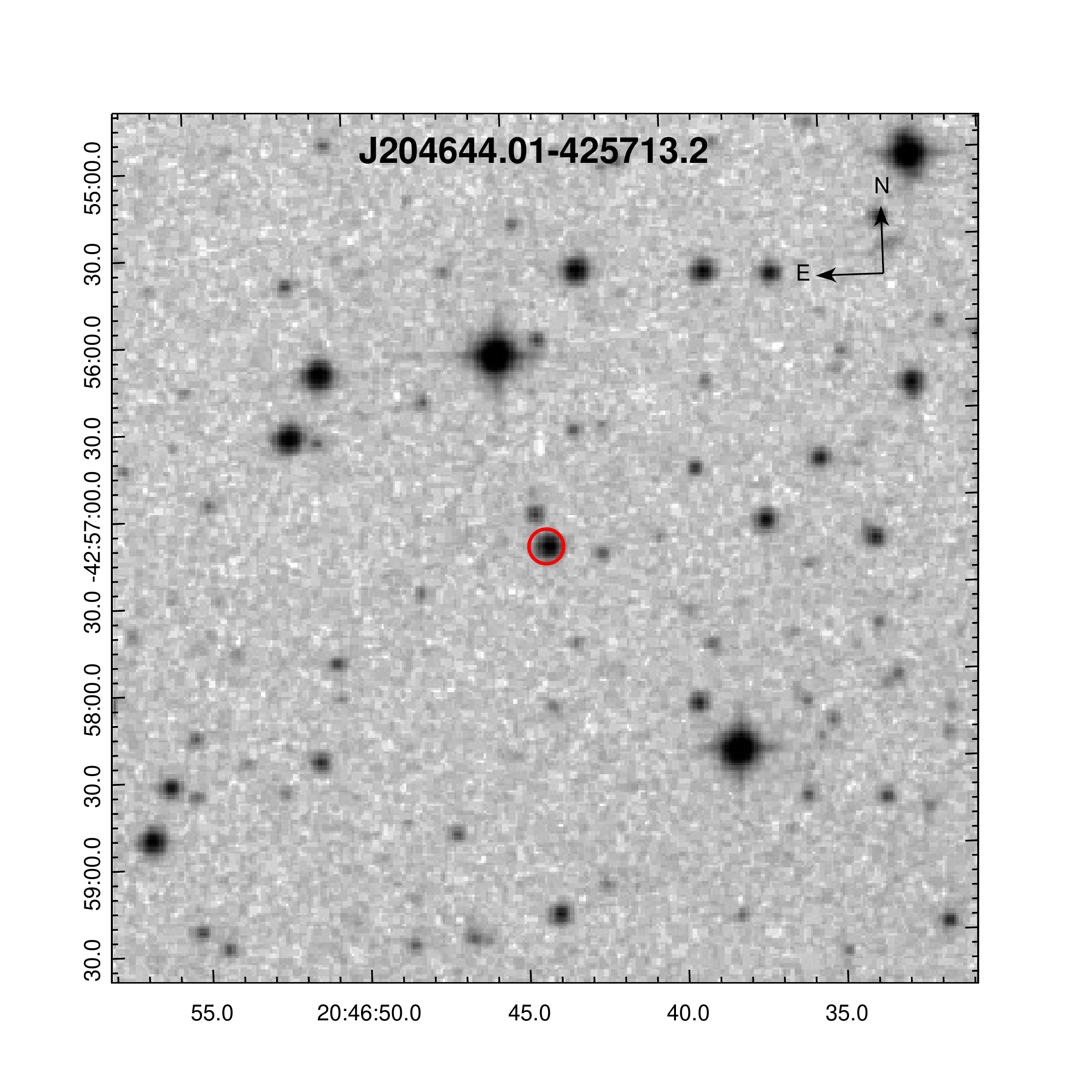} \\
\end{array}$
\end{center}
\caption{As in Figure~\ref{fig:J0837} but for WISE J204644.01-425713.2, the counterpart of 4FGL J2046.8-4258.}
\label{fig:J2046}
\end{figure*}

\begin{figure*}{}
\begin{center}$
\begin{array}{cc}
\includegraphics[width=\mywidth]{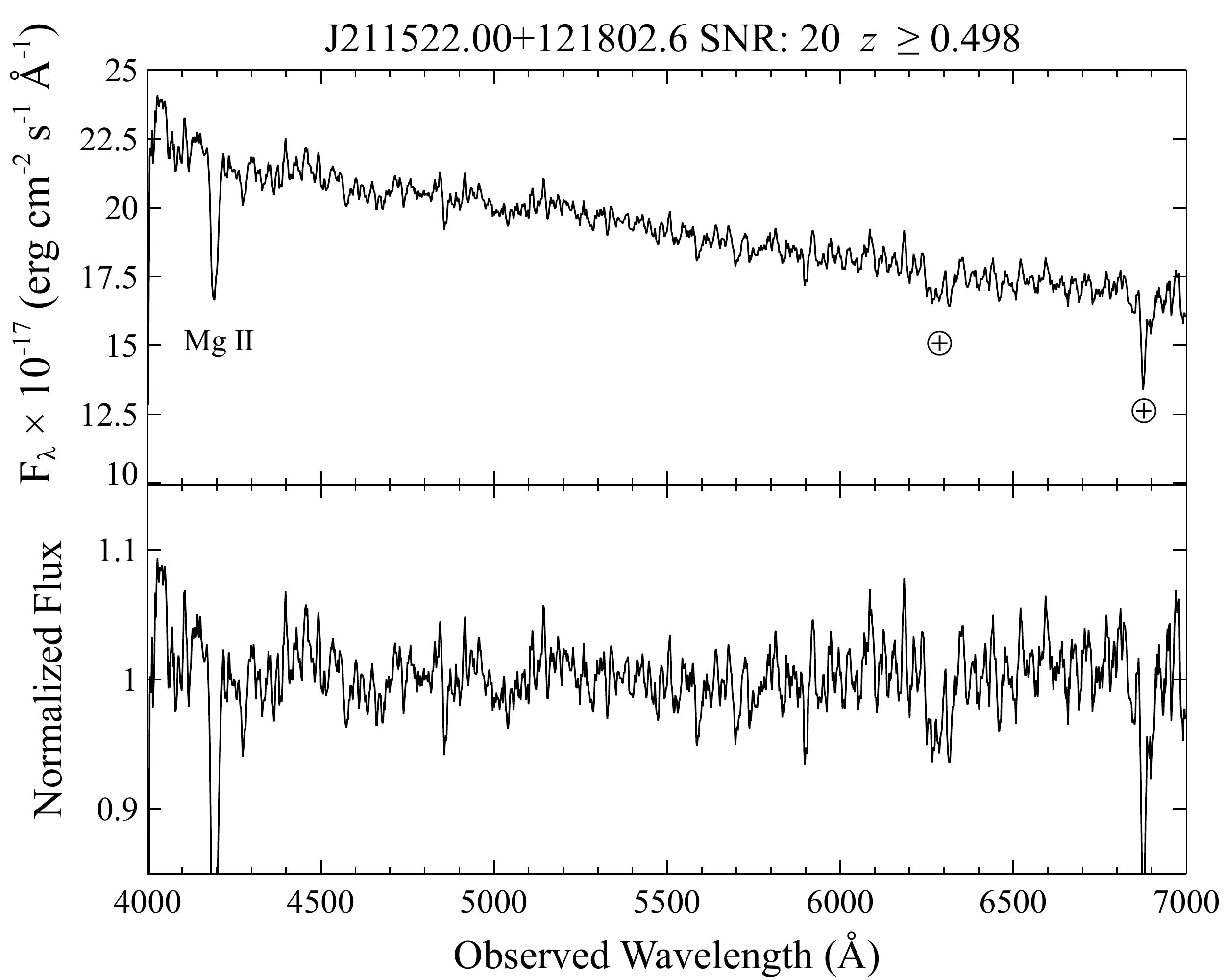} &
\includegraphics[clip=true, width=7cm]{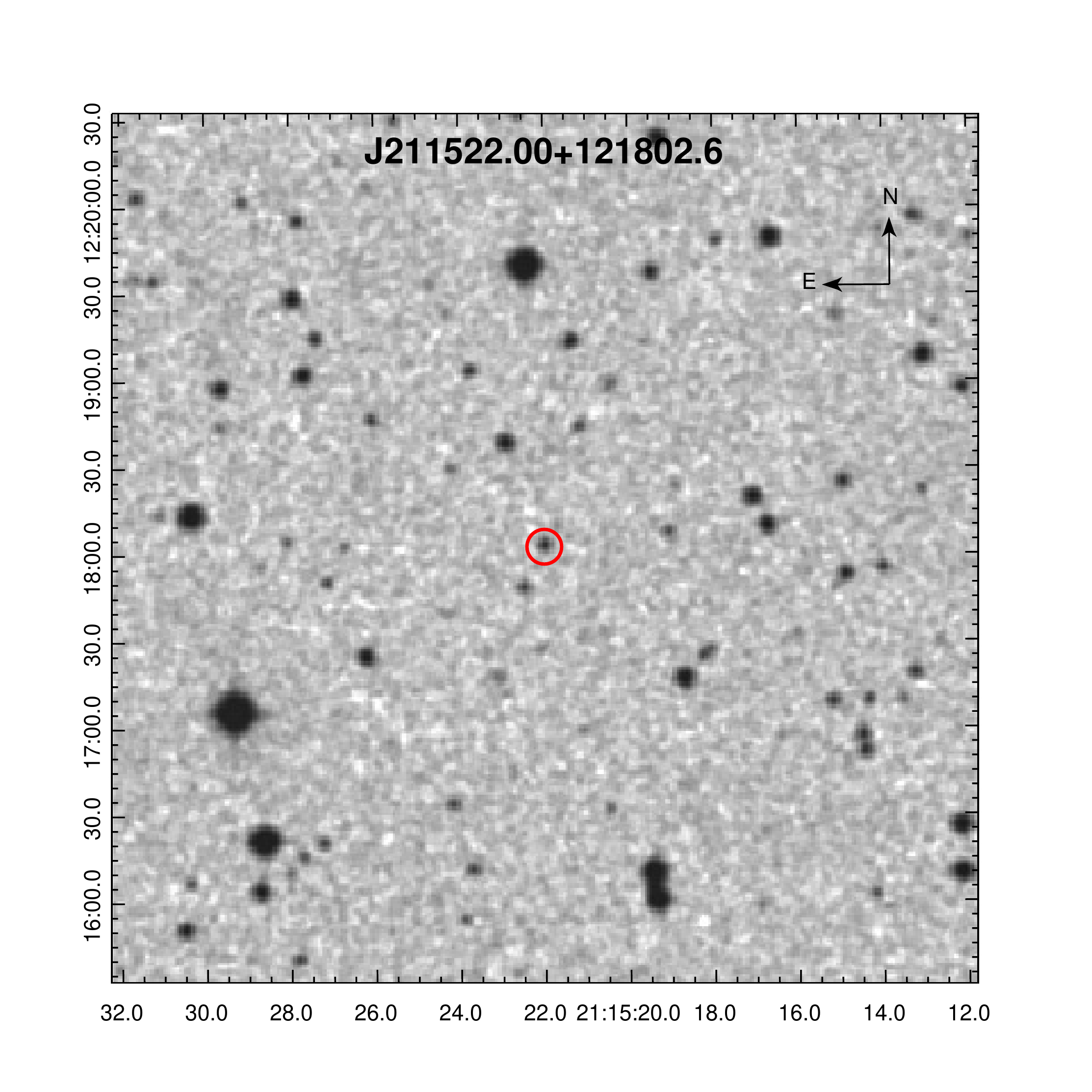} \\
\end{array}$
\end{center}
\caption{As in Figure~\ref{fig:J0837} but for WISE J211522.00+121802.6, the counterpart of 4FGL J2115.2+1218.}
\label{fig:J2115}
\end{figure*}

\begin{figure*}{}
\begin{center}$
\begin{array}{cc}
\includegraphics[width=\mywidth]{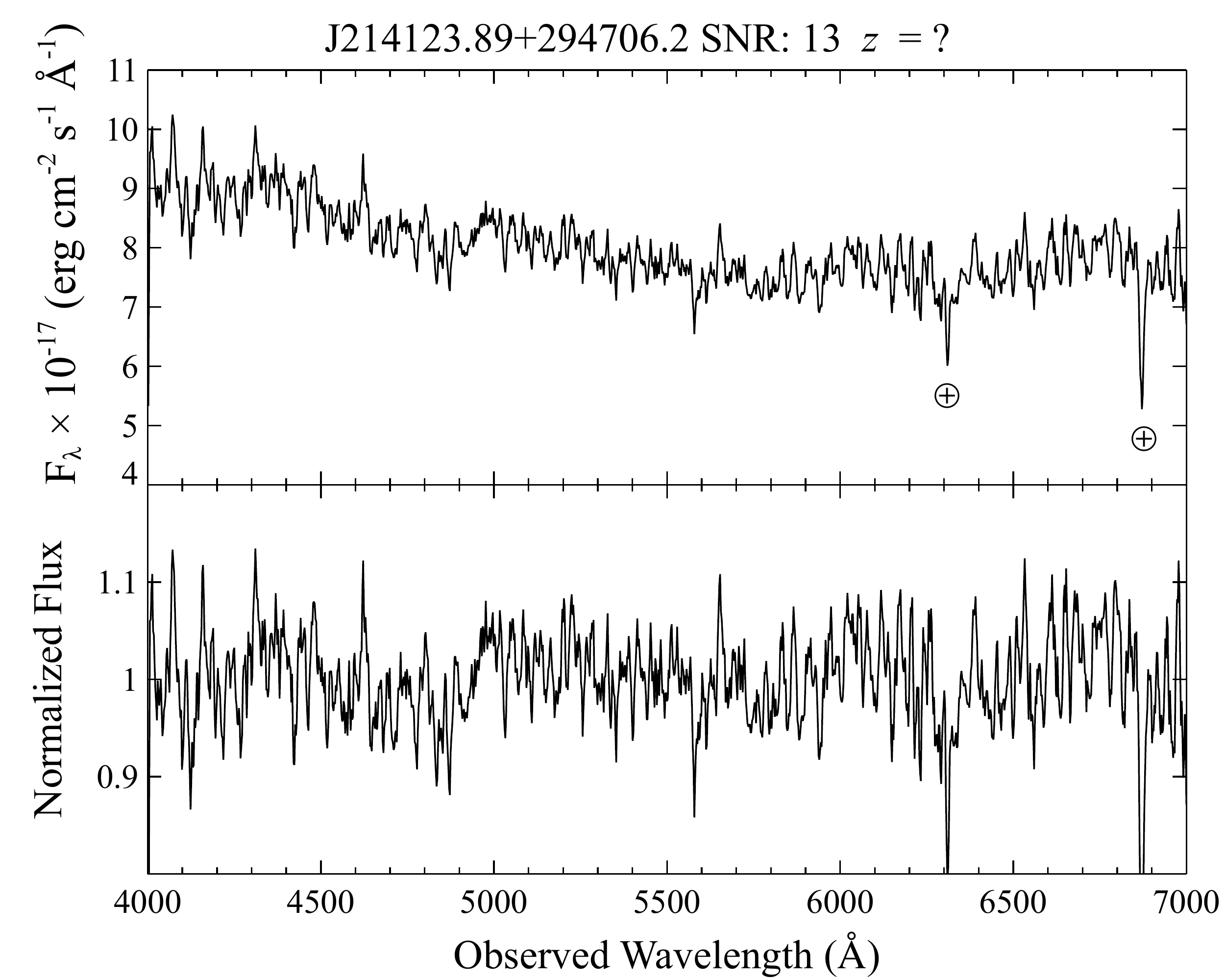} &
\includegraphics[clip=true, width=7cm]{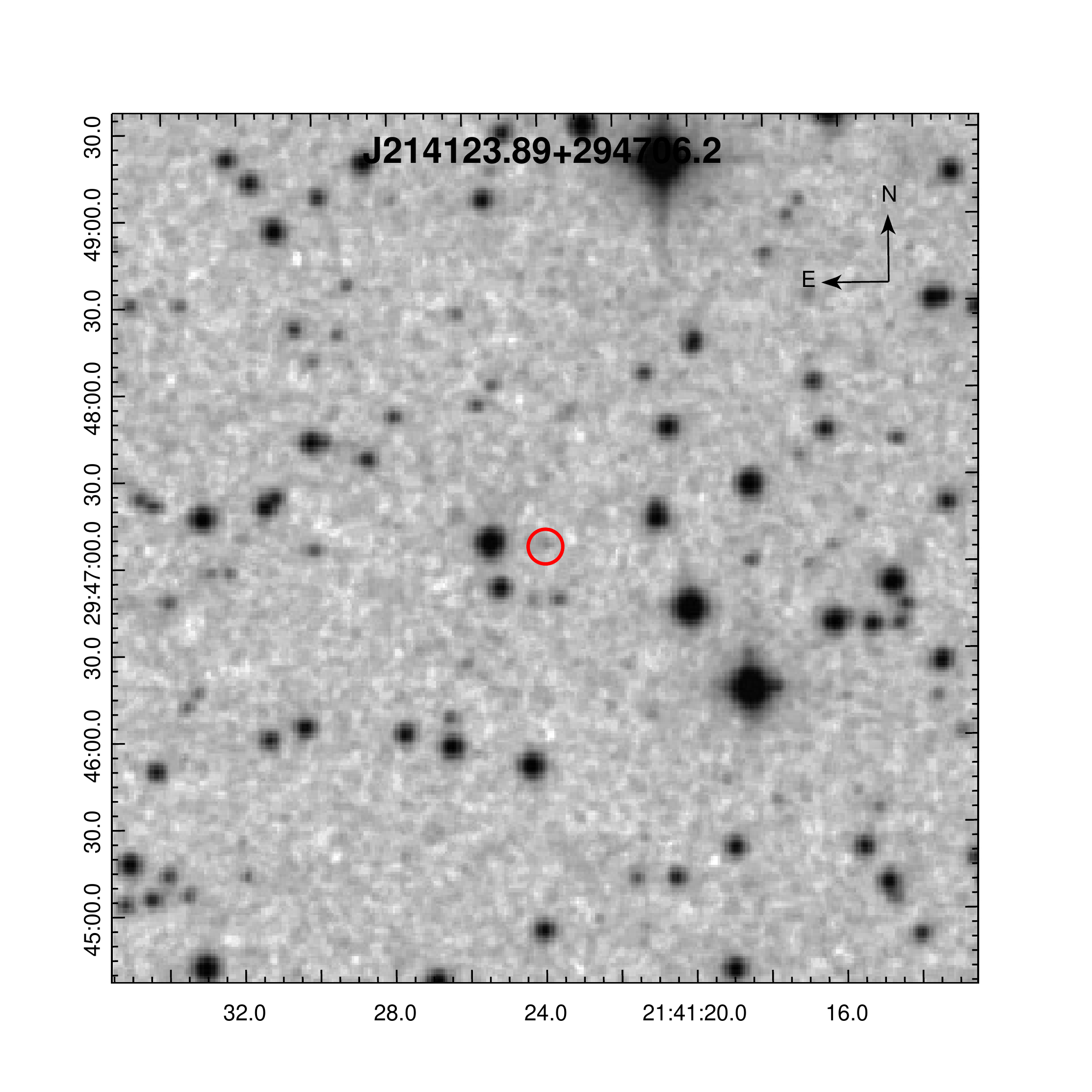} \\
\end{array}$
\end{center}
\caption{As in Figure~\ref{fig:J0837} but for WISE J214123.89+294706.2, the potential counterpart of 4FGL J2141.4+2947.}
\label{fig:J2141}
\end{figure*}

\begin{figure*}{}
\begin{center}$
\begin{array}{cc}
\includegraphics[width=\mywidth]{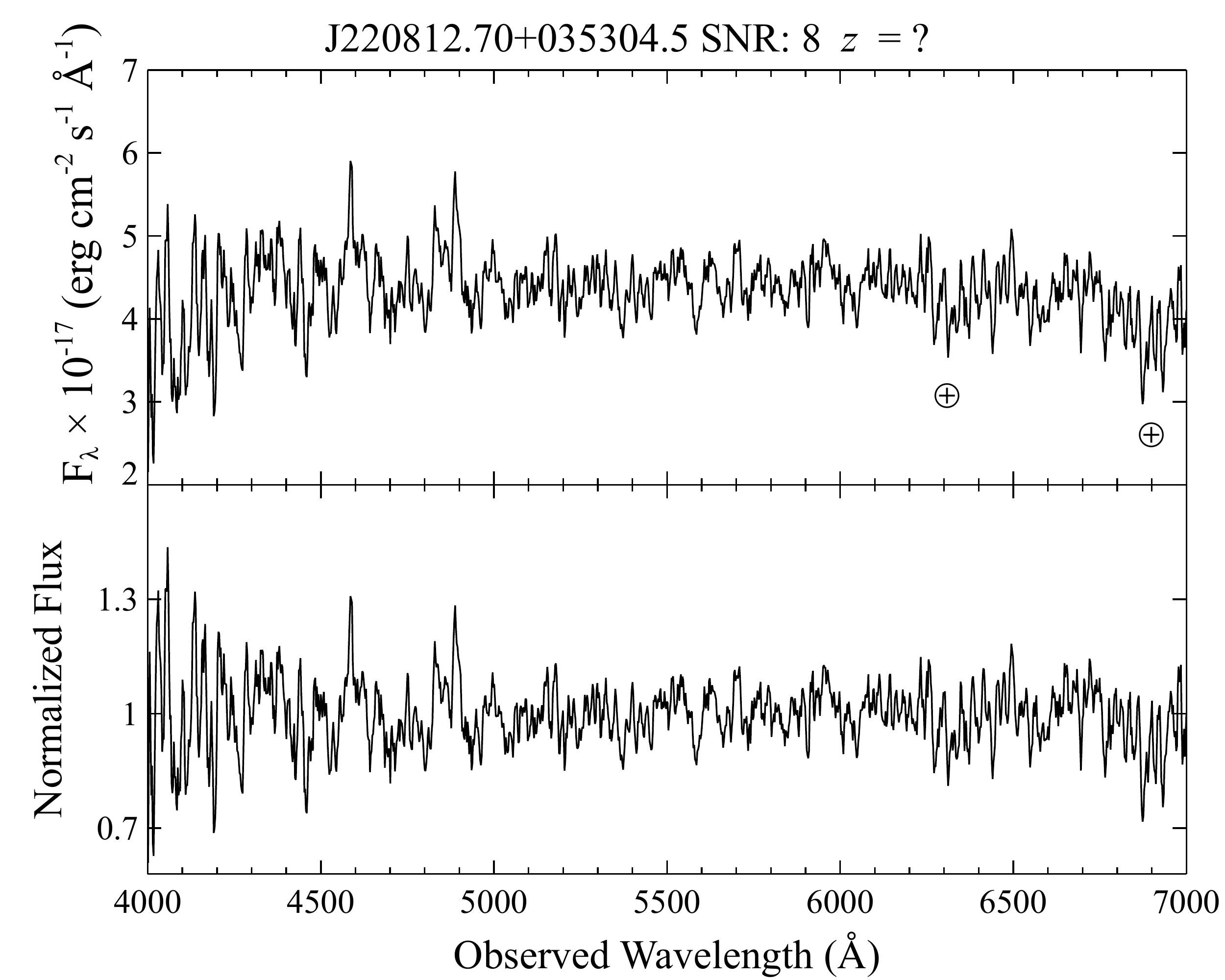} &
\includegraphics[clip=true, width=7cm]{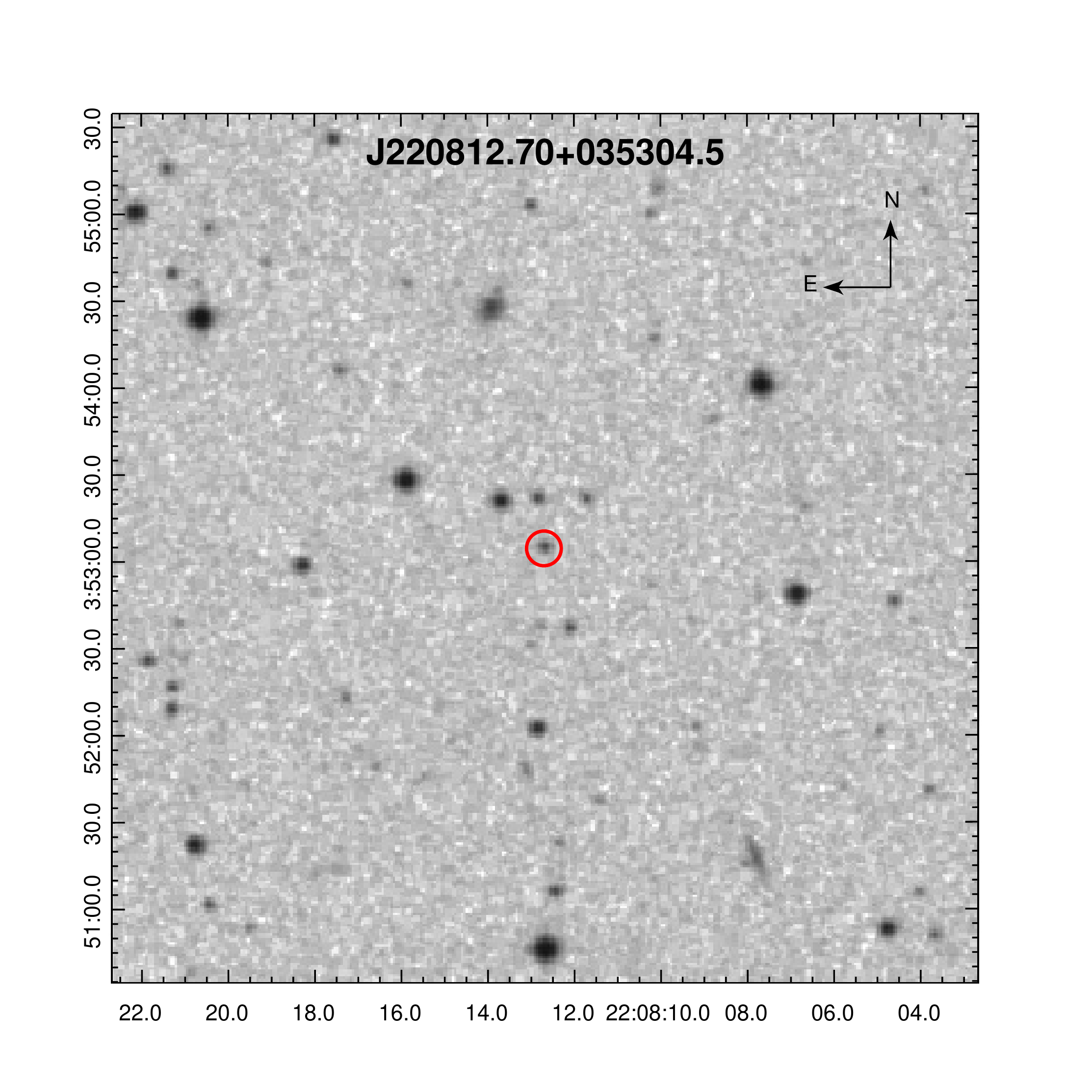} \\
\end{array}$
\end{center}
\caption{As in Figure~\ref{fig:J0837} but for WISE J220812.70+035304.5, the potential counterpart of 4FGL J2208.2+0350.}
\label{fig:J220}
\end{figure*}

\begin{figure*}{}
\begin{center}$
\begin{array}{cc}
\includegraphics[width=\mywidth]{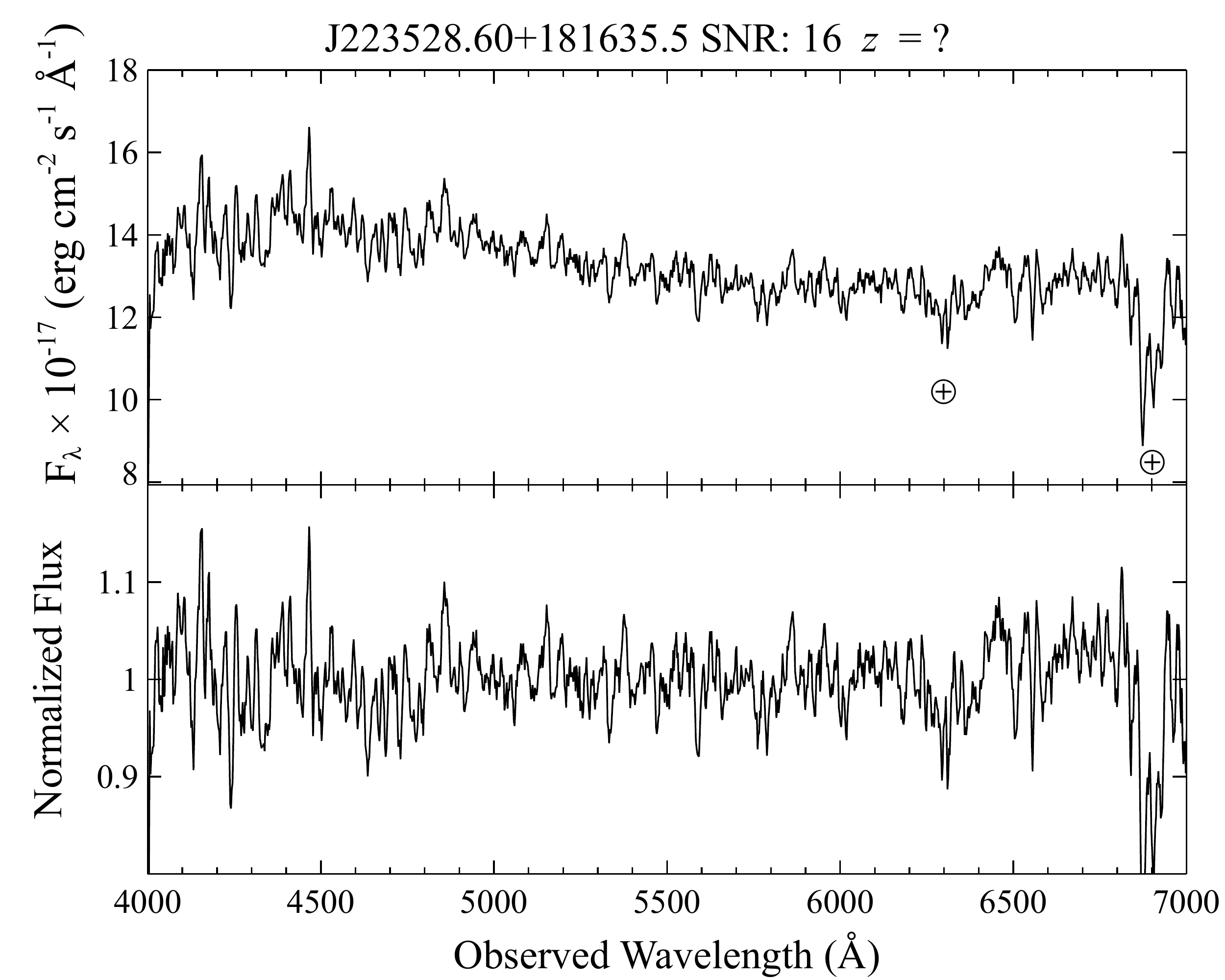} &
\includegraphics[clip=true, width=7cm]{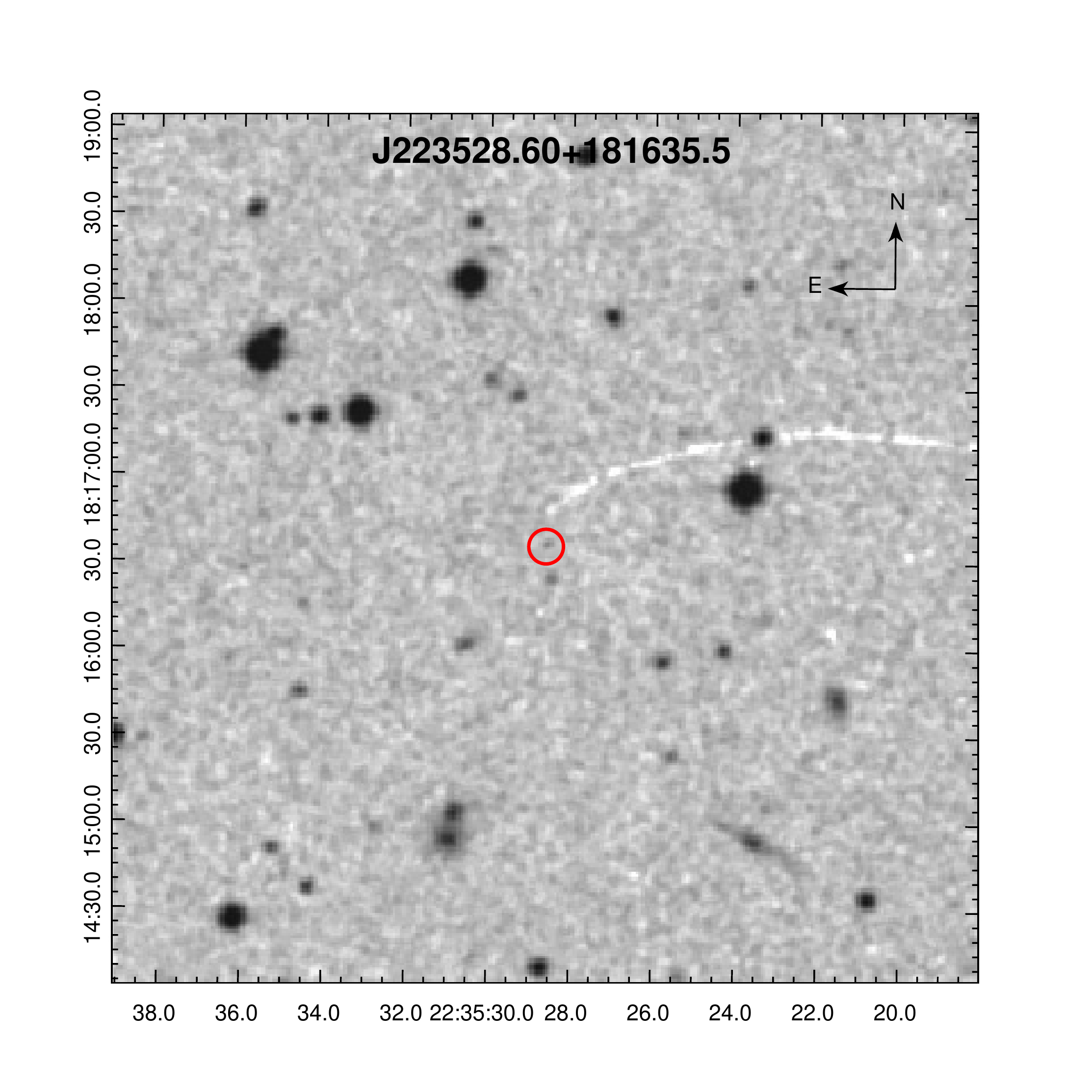} \\
\end{array}$
\end{center}
\caption{As in Figure~\ref{fig:J0837} but for WISE J223528.60+181635.5, the potential counterpart of 4FGL J2235.3+1818.}
\label{fig:J2235}
\end{figure*}

\clearpage
\section{Summary tables}
\label{sec:appB}
Here we report the first 12 lines of the two main tables summarizing {the} results (i) {achieved} thanks to our optical spectroscopic campaign including all 394 targets {(Table~\ref{tab:summary1}),} and (ii) {obtained from} the literature {consisting of} a total of 123 sources {(Table~\ref{tab:summary2})}. 

%Both tables report: the \textit{Fermi}-LAT name (4FGL {in} Table~\ref{tab:summary1} and 3FGL {in} Table~\ref{tab:summary2}), \textit{Fermi}-LAT classification provided {in the} catalog corresponding to the \fer-LAT name, the assigned low-energy counterpart {ultimately} associated in the corresponding \fer-LAT catalog, the name of the {associated} WISE counterpart, right ascension and declination (Equinox J2000) of the WISE counterpart, spectroscopic classification, redshift, references {to our campaign papers} (only for Table~\ref{tab:summary1}), telescopes used for acquiring their spectra, and references for targets that {were} also reported in the literature.

{\onecolumn
\begin{center}
\begin{landscape}
\begin{table*}[!ht] 
\tiny
\caption{Summary of {results from our} optical spectroscopic campaign.} \label{tab:summary1}
\begin{tabular}{lllllllrlll}
\hline
\fer-LAT & \fer-LAT &  Associated & WISE & R.A. (J2000) & Dec. (J2000) & Class & $z$ & Ref. & Telescope & Literature \\  
name & class &  source & counterpart & hh:mm:ss & dd:mm:ss & & & & & \\  
 {(1)} &  {(2)} &  {(3) }&  {(4)} &  {(5)} &  {(6)} &  {(7)} &  {(8)} &  {(9)} &  {(10)} &  {(11)} \\  
\noalign{\smallskip}
\hline 
\noalign{\smallskip}
  4FGL J0003.9-1149 & bll & PMN J0004-1148 & J000404.91-114858.3 & 00:04:04.91 & -11:48:58.33 & bzb & \dots & d & 6dF & \\
  4FGL J0009.7-3217 & rdg & IC 1531 & J000935.55-321636.8 & 00:09:35.56 & -32:16:36.85 & bzg & 0.0254 & d & 6dF & \\
  4FGL J0015.6+5551 & bll & GB6 J0015+5551 & J001540.13+555144.7 & 00:15:40.14 & +55:51:44.77 & bzb & \dots & b & KPNO & \\
  4FGL J0015.9+2440 & bcu & GB6 J0016+2440 & J001603.62+244014.7 & 00:16:03.62 & +24:40:14.77 & bzb & \dots & l & OAN-SPM & \\
  FL8Y J0024.1+2401 & ugs & \dots & J002406.10+240438.6 & 00:24:06.10 & +24:04:38.67 & bzb & 0.062? & o & SDSS & \\
  4FGL J0023.7-6820 & bcu & PKS 0021-686 & J002406.72-682054.5 & 00:24:06.72 & -68:20:54.50 & qso & 0.354 & f & SOAR & \\
  4FGL J0028.4+2001 & fsrq & TXS 0025+197 & J002829.81+200026.7 & 00:28:29.82 & +20:00:26.77 & bzq & 1.5517 & d & SDSS DR12 & \\
  4FGL J0037.9+2612 & bll & MG3 J003720+2613 & J003719.15+261312.6 & 00:37:19.15 & +26:13:12.60 & bzg & 0.1477 & k & SDSS & \\
  4FGL J0038.7-0204 & rdg & 3C 17 & J003820.53-020740.5 & 00:38:20.53 & -02:07:40.50 & bzq & 0.2204 & k & SDSS & \\
  4FGL J0040.4-2340 & bll & PMN J0040-2340 & J004024.90-234000.7 & 00:40:24.90 & -23:40:00.70 & bzg & 0.213 & f & SOAR & \\
  4FGL J0043.5-0442 & bll & 1RXS J004333.7-044257 & J004334.12-044300.6 & 00:43:34.12 & -04:43:00.67 & bzb & \dots & d & 6dF & \\
  4FGL J0043.7-1116 & bcu & 1RXS J004349.3-111612 & J004348.66-111607.2 & 00:43:48.66 & -11:16:07.23 & bzb & 0.264 & c & SOAR, NOT & h\\
\noalign{\smallskip}
\hline
\end{tabular}\\

\footnotesize{Column descriptions: 
{  (1) Fermi-LAT name;
  (2) Fermi-LAT classification;
  (3) Fermi-LAT low-energy counterpart association name; }
  (4) WISE counterpart name;
  (5) J2000 Right Ascension;
  (6) J2000 Declination;
  (7) {Spectroscopic class};
  (8) Redshift;
  (9) References {to our campaign papers}: a, \citep{cowperthwaite13}; b, \citep{crespo16a}; c, \citep{crespo16b}; d, \citep{crespo16c}; e, \citep{landoni15}; f, \citep{marchesini19}; g, \citep{massaro14}; h, \citep{massaro15a}; i, \citep{massaro15b}; j, \citep{massaro16}; k, \citep{demenezes19}; l, \citep{demenezes20}; m, \citep{paggi14}; n, \citep{pena17}; o, \citep{pena19}; p, This work; q, \citep{ricci15}.
  (10) Telescope;
  (11) References {to spectra that were also reported in the} literature: a \citep{desai19}; b, \citep{falco98}; c, \citep{jones09}; d, \citep{hewitt80}; e, \citep{klindt17}; f, \citep{lamura17}; g, \citep{marchesi18}; h, \citep{marchesini16}; i, \citep{marlow00}; j, \citep{marti04}; k, \citep{masetti13}; l, \citep{paiano17a}; m, \citep{paiano17b}; n, \citep{shaw13}; o, \citep{titov13}; p, \citep{tsarevsky05}; q, \citep{vermeulen95}; r, \citep{paiano19}.}
\end{table*}
\end{landscape}
\end{center}
}

{\onecolumn
\begin{landscape}

\begin{table*}[!hb] 
\tiny
\caption{Summary of literature search.} \label{tab:summary2}
\begin{center}
\begin{tabular}{lllllllllrll}
\hline
3FGL & 3FGL         &  4FGL & 4FGL & Associated & WISE         & R.A. (J2000) & Dec. (J2000) & Class & $z$ & Telescope & Literature \\  
name & class          &  name  & class &  name     &  name        & hh:mm:ss     & dd:mm:ss      &           &        &                  &                  \\  
 {(1)} &  {(2)}    &   {(3)} & { (4)} &  {(5)} &  {(6)}   &  {(7) }&  {(8)} &  {(9)} &  {(10)} &  {(11)} &  {(12)} \\  
\noalign{\smallskip}
\hline 
\noalign{\smallskip}
   J0004.2+0843 & ugs &  J0004.0+0840 & bcu & SDSS J000359.23+084138.1 & J000359.23+084138.1 & 00:03:59.23 & +08:41:38.15 & bzb & >1.503 & GTC & p\\
   J0006.2+0135 & ugs &  J0006.4+0135 & bcu & NVSS J000626+013611 & J000626.90+013610.6 & 00:06:26.90 & +01:36:10.70 & bzb & 0.787 & GTC & p\\
   J0008.0+4713 & bll &  J0008.0+4711 & bll & MG4 J000800+4712 & J000759.97+471207.7 & 00:07:59.98 & +47:12:07.75 & bzb & >1.659 & GTC & n\\
   J0031.3+0724 & bcu &  J0031.3+0726 & bll & NVSS J003119+072456 & J003119.71+072453.4 & 00:31:19.71 & +07:24:53.50 & bzb & >0.836 & KPNO & j\\
   J0040.3+4049 & bcu &  J0040.3+4050 & bll & B3 0037+405 & J004013.81+405004.5 & 00:40:13.82 & +40:50:04.54 & bzb & \dots & KPNO & j\\
   J0043.7-1117 & bcu &  J0043.7-1116 & bll & 1RXS J004349.3-111612 & J004348.66-111607.2 & 00:43:48.66 & -11:16:07.23 & bzb & \dots & NOT, Copernico & k and h\\
   J0045.2-3704 & bcu &  J0045.1-3706 & bcu & PKS 0042-373 & J004512.06-370548.5 & 00:45:12.06 & -37:05:48.54 & bzq & 1.033 & SALT & g\\
   J0049.0+4224 & ugs &  J0049.1+4223 & bcu & GALEXASC J004859.14+422351.4 & J004859.15+422351.1 & 00:48:59.16 & +42:23:51.12 & bzb & 0.302 & SDSS & o\\
   J0049.7+0237 & bll &  J0049.7+0237 & bll & PKS 0047+023 & J004943.23+023703.7 & 00:49:43.23 & +02:37:03.80 & bzb & >0.55 & GTC & n\\
   J0102.1+0943 & ugs &  J0102.4+0942 & bcu & 2MASS J01021713+0944098 & J010216.63+094411.1 & 01:02:17.10 & +09:44:09.50 & bzb & 0.42 & SDSS & o\\
   J0127.2+0325 & bcu &  J0127.2+0324 & bll & NVSS J012713+032259 & J012713.94+032300.6 & 01:27:13.95 & +03:23:00.64 & bzb & \dots & KPNO & j\\
   J0134.5+2638 & bcu &  J0134.5+2637 & fsrq & RX J0134.4+2638 & J013428.19+263843.0 & 01:34:28.20 & +26:38:43.01 & bzb & \dots & OAN-SPM, HET, and KPNO & s and j\\
\noalign{\smallskip}
\hline
\end{tabular}\\
\end{center}
\footnotesize{Column descriptions: 
  (1) 3FGL name;
  (2) 3FGL classification;
  (3) 4FGL name;
  (4) 4FGL classification;
  (5) {Fermi-LAT} low-energy counterpart association name;
  (6) WISE counterpart name;
  (7) J2000 Right Ascension;
  (8) J2000 Declination;
  (9) Spectroscopic class;
  (10) Redshift;
  (11) Telescope;
  (12) References: a, \citep{britzen07}; b, \citep{caccianiga02}; c, \citep{crespo16a}; d, \citep{crespo16b}; e, \citep{desai19}; f, \citep{hewitt80}; g, \citep{klindt17}; h, \citep{lamura17}; i, \citep{landoni18}; j, \citep{marchesi18}; k, \citep{marchesini16}; l, \citep{massaro15a}; m, \citep{paggi14}; n, \citep{paiano17a}; o, \citep{paiano17b}; p, \citep{paiano19}; q, \citep{peterson76}; r, \citep{ricci15}; s, \citep{shaw13}; t, \citep{wisotzki00}.}
\end{table*}

\end{landscape}

\end{document}